\documentclass[prc,aps,nofootinbib,superscriptaddress,showkeys,showpacs,twocolumn,10pt,floatfix]{revtex4-1}
\usepackage{epsfig}
\usepackage{graphicx}
\usepackage[small]{subfigure}
\usepackage{setspace}
\usepackage[T1]{fontenc}
\usepackage[utf8]{inputenc}
\usepackage[english]{babel}
\usepackage{mathrsfs}
\usepackage{amssymb}
\usepackage{amsmath}
\usepackage{amsfonts}
\usepackage{siunitx}
\usepackage{amsopn}
\usepackage{mathtools}
\usepackage{float}
\usepackage{chemformula}
\usepackage{comment} 
\usepackage{tikz}
\usepackage{booktabs}
\usepackage{multirow}
\usepackage{verbatim}
\usepackage{ulem}
\usepackage{soul}
\usetikzlibrary{arrows}

\begin{document}
\title{Multi-channel experimental and theoretical constraints \\ for the $^{116}$Cd($^{20}$Ne,$^{20}$F)$^{116}$In charge exchange reaction at 306 MeV}
\author{S. Burrello} \affiliation{Technische Universit\"{a}t Darmstadt, Fachbereich Physik, Institut f\"{u}r Kernphysik, Darmstadt, Germany} \affiliation{Departamento de F\'{i}sica Atomica, Molecular y Nuclear, Facultad de F\'{i}sica, Universidad de Sevilla, Sevilla, Spain}
\author{S. Calabrese} \affiliation{Istituto Nazionale di Fisica Nucleare, Laboratori Nazionali del Sud, Catania, Italy} \affiliation{Dipartimento di Fisica e Astronomia  ``Ettore Majorana'',  Universit\`{a} di Catania, Catania, Italy}
\author{F. Cappuzzello} \affiliation{Istituto Nazionale di Fisica Nucleare, Laboratori Nazionali del Sud, Catania, Italy} \affiliation{Dipartimento di Fisica e Astronomia  ``Ettore Majorana'',  Universit\`{a} di Catania, Catania, Italy}
\author{D. Carbone} \affiliation{Istituto Nazionale di Fisica Nucleare, Laboratori Nazionali del Sud, Catania, Italy}
\author{M. Cavallaro} \affiliation{Istituto Nazionale di Fisica Nucleare, Laboratori Nazionali del Sud, Catania, Italy}
\author{M. Colonna} \affiliation{Istituto Nazionale di Fisica Nucleare, Laboratori Nazionali del Sud, Catania, Italy}
\author{J. A. Lay} \affiliation{Departamento de F\'{i}sica Atomica, Molecular y Nuclear, Facultad de F\'{i}sica, Universidad de Sevilla, Sevilla, Spain} \affiliation{Instituto Carlos I de F\'{i}sica Teórica y Computacional, Universidad de Sevilla, Sevilla, Spain}
\author{H. Lenske} \affiliation{Institut f\"ur Theoretische Physik, Justus-Liebig-Universit\"at Giessen, Giessen, Germany}
\author{C. Agodi} \affiliation{Istituto Nazionale di Fisica Nucleare, Laboratori Nazionali del Sud, Catania, Italy}
\author{J. L. Ferreira} \affiliation{Instituto de Física, Universidade Federal Fluminense, Niter\'{o}i, Brazil}
\author{S. Firat} \affiliation{Department of Physics, Akdeniz University, Antalya, Turkey}
\author{A. Hacisalihoglu} \affiliation{Institute of Natural Sciences, Karadeniz Teknik Universitesi, Trabzon, Turkey}
\author{L. La Fauci} \affiliation{Istituto Nazionale di Fisica Nucleare, Laboratori Nazionali del Sud, Catania, Italy} \affiliation{Dipartimento di Fisica e Astronomia  ``Ettore Majorana'',  Universit\`{a} di Catania, Catania, Italy}
\author{A. Spatafora} \affiliation{Istituto Nazionale di Fisica Nucleare, Laboratori Nazionali del Sud, Catania, Italy} \affiliation{Dipartimento di Fisica e Astronomia  ``Ettore Majorana'',  Universit\`{a} di Catania, Catania, Italy}
\author{L. Acosta} \affiliation{Instituto de F\'{i}sica, Universidad Nacional Autónoma de M\'{e}xico, Mexico City, Mexico} \affiliation{INFN, Sezione di Catania, Catania, Italy}
\author{J. I. Bellone} \affiliation{Istituto Nazionale di Fisica Nucleare, Laboratori Nazionali del Sud, Catania, Italy}
\author{T. Borello-Lewin} \affiliation{Instituto de Fisica, Universidade de S\~{a}o Paulo, S\~{a}o Paulo, Brazil}
\author{I. Boztosun} \affiliation{Department of Physics, Akdeniz University, Antalya, Turkey}
\author{G. A. Brischetto} \affiliation{Istituto Nazionale di Fisica Nucleare, Laboratori Nazionali del Sud, Catania, Italy} \affiliation{Dipartimento di Fisica e Astronomia  ``Ettore Majorana'',  Universit\`{a} di Catania, Catania, Italy}
\author{D. Calvo} \affiliation{INFN, Sezione di Torino, Torino, Italy}
\author{E. R. Ch\'{a}vez-Lomel\'{i}} \affiliation{Instituto de F\'{i}sica, Universidad Nacional Autónoma de M\'{e}xico, Mexico City, Mexico}
\author{I. Ciraldo} \affiliation{Istituto Nazionale di Fisica Nucleare, Laboratori Nazionali del Sud, Catania, Italy} \affiliation{Dipartimento di Fisica e Astronomia  ``Ettore Majorana'',  Universit\`{a} di Catania, Catania, Italy} 
\author{M. Cutuli} \affiliation{Istituto Nazionale di Fisica Nucleare, Laboratori Nazionali del Sud, Catania, Italy} \affiliation{Dipartimento di Fisica e Astronomia  ``Ettore Majorana'',  Universit\`{a} di Catania, Catania, Italy}
\author{F. Delaunay} \affiliation{Istituto Nazionale di Fisica Nucleare, Laboratori Nazionali del Sud, Catania, Italy} \affiliation{Dipartimento di Fisica e Astronomia  ``Ettore Majorana'',  Universit\`{a} di Catania, Catania, Italy} \affiliation{LPC Caen, Normandie Universit\'{e}, ENSICAEN, UNICAEN, CNRS/IN2P3, Caen, France}
\author{P. Finocchiaro} \affiliation{Istituto Nazionale di Fisica Nucleare, Laboratori Nazionali del Sud, Catania, Italy}
\author{M. Fisichella} \affiliation{Istituto Nazionale di Fisica Nucleare, Laboratori Nazionali del Sud, Catania, Italy}
\author{A. Foti} \affiliation{INFN, Sezione di Catania, Catania, Italy}
\author{F. Iazzi} \affiliation{INFN, Sezione di Torino, Torino, Italy} \affiliation{DISAT, Politecnico di Torino, Torino, Italy}
\author{G. Lanzalone} \affiliation{Istituto Nazionale di Fisica Nucleare, Laboratori Nazionali del Sud, Catania, Italy} \affiliation{Facolt\`{a} di Ingegneria e Architettura, Università di Enna ``Kore'', Enna, Italy}
\author{R. Linares} \affiliation{Instituto de Física, Universidade Federal Fluminense, Niter\'{o}i, Brazil} 
\author{J. Lubian} \affiliation{Instituto de Física, Universidade Federal Fluminense, Niter\'{o}i, Brazil} 
\author{M. Moralles} \affiliation{Instituto de Pesquisas Energeticas e Nucleares IPEN/CNEN, S\~{a}o Paulo, Brazil}
\author{J. R. B. Oliveira} \affiliation{Instituto de Fisica, Universidade de S\~{a}o Paulo, S\~{a}o Paulo, Brazil}
\author{A. Pakou} \affiliation{Department of Physics, University of Ioannina and Hellenic Institute of Nuclear Physics, Ioannina, Greece}
\author{L. Pandola} \affiliation{Istituto Nazionale di Fisica Nucleare, Laboratori Nazionali del Sud, Catania, Italy}
\author{H. Petrascu} \affiliation{IFIN-HH, Bucarest, Romania}
\author{F. Pinna} \affiliation{INFN, Sezione di Torino, Torino, Italy} \affiliation{DISAT, Politecnico di Torino, Torino, Italy}
\author{G. Russo} \affiliation{INFN, Sezione di Catania, Catania, Italy}
\author{O. Sgouros} \affiliation{Istituto Nazionale di Fisica Nucleare, Laboratori Nazionali del Sud, Catania, Italy}
\author{S. O. Solakci} \affiliation{Department of Physics, Akdeniz University, Antalya, Turkey}
\author{V. Soukeras} \affiliation{Istituto Nazionale di Fisica Nucleare, Laboratori Nazionali del Sud, Catania, Italy}
\author{G. Souliotis} \affiliation{Department of Chemistry, University of Athens and Hellenic Institute of Nuclear Physics, Athens, Greece}
\author{D. Torresi} \affiliation{Istituto Nazionale di Fisica Nucleare, Laboratori Nazionali del Sud, Catania, Italy}
\author{S. Tudisco} \affiliation{Istituto Nazionale di Fisica Nucleare, Laboratori Nazionali del Sud, Catania, Italy}
\author{A. Yildirin} \affiliation{Department of Physics, Akdeniz University, Antalya, Turkey}
\author{V. A. B. Zagatto} \affiliation{Instituto de Física, Universidade Federal Fluminense, Niter\'{o}i, Brazil}
\collaboration{for the NUMEN collaboration} \noaffiliation

\date{\today}

\begin{abstract}
\begin{description}
\item[Background] Charge exchange (CE) reactions offer a major opportunity to excite nuclear isovector modes, providing important clues about the nuclear interaction in the medium. Moreover, double charge exchange (DCE) reactions are proving to be a tempting tool to access nuclear transition matrix elements (NME) related to double beta-decay processes. The latter are also of crucial importance to extract neutrino properties from the half-life of the hypothetical neutrinoless double beta decay and to search for physics beyond the standard model. 
\item[Purpose] Through a multi-channel experimental analysis and a consistent theoretical approach of the $^{116}$Cd($^{20}$Ne,$^{20}$F)$^{116}$In single charge exchange (SCE) reaction at 306 MeV, we aim at %providing more stringent constraints %\textcolor{magenta}{involving the} 
%on the nuclear models involved for the calculations of the NME of the system $^{116}$Cd, in light of their interest for beta decay processes. Our goal is, in particular, to disentangle 
disentangling from the experimental cross section the contribution of the competing mechanisms, %given by 
associated with second or higher order sequential transfer and/or inelastic processes.
\item[Method] We measured excitation energy spectra and absolute cross sections for elastic + inelastic, one-proton transfer and SCE channels, using the MAGNEX large acceptance magnetic spectrometer to detect the ejectiles. For the first two channels, we also extracted the experimental cross section angular distributions. The experimental data are compared with theoretical predictions obtained by performing two-step distorted wave Born approximation and coupled reaction channel calculations. We employ spectroscopic amplitudes for single-particle transitions derived within a large-scale shell model approach and different optical potentials for modeling the initial and the final state interactions.
\item[Results] The present study significantly mitigates the possible model dependence %with respect to the ingredients usually adopted 
existing in the description of these complex reaction mechanisms, thanks to the satisfactory reproduction of several channels at once. In particular, our work demonstrates that the two-step transfer mechanisms produce a non negligible contribution to the total cross section of the $^{116}$Cd($^{20}$Ne,$^{20}$F)$^{116}$In reaction channel, although a relevant fraction is still missing, % within our picture, being provided 
%proving to be provided by 
being ascribable to the direct SCE mechanism, which is not addressed here. 
\item[Conclusions] %The competition between the direct SCE mechanism and the competing processes analyzed in this work deserves further investigation. Nevertheless, o
Our %study
analysis provides %suggests that 
a careful estimation of the %competing 
sequential transfer %or 
%and inelastic 
processes which are competing with the direct SCE mechanism for the heavy ion reaction under investigation. 
The study suggests that the %is mandatory to extract structure information concerning the NME from the SCE reaction cross section. The 
direct SCE should play %a primary 
an important role among the mechanisms populating the final channel. % of the reaction under investigation. 
Nevertheless, the analysis of the higher order processes considered here is mandatory to isolate the direct SCE process contribution and approach structure information on the corresponding NME from the reaction cross section. %Therefore, %the study suggests that
The description of the latter process and the competition between the %direct mechanism and the higher order processes analyzed in this work deserves
two mechanisms deserves further investigation. %to complete the picture. %Nevertheless, the latter are expected to be suppressed when further increasing the number of steps involved and they should be thus safely neglected when investigating DCE reactions. 
\end{description}
\end{abstract}
\date{\today}
\maketitle

\section{Introduction}
Charge exchange (CE) transitions induced by %a strong external field 
the nuclear interaction have raised a widespread theoretical and experimental interest over the last decades, being regarded as a major source of information on nuclear isovector excitations~\cite{Osterfeld, Ichimura06, Frekers13}. On the other hand, %nuclear charge exchange (CE) 
CE reactions offer a quite appealing opportunity to excite these modes~\cite{Brendel:1988mwl, AndersonPRC1991, blomgren, YakoPLB2005, FrekersPPNP2006, FujitaPRC2007, FujitaPPNP2011, Douma2020}, providing crucial clues about the nucleon-nucleon (NN) interaction in the medium, whose behavior in the spin-isospin channels is still not fully understood. Indeed, CE reactions are nowadays %on the top of agenda 
the object of recent intensive experimental campaigns~\cite{EPJA2018, EPJA2015, Agodi1, Agodi2, sasanoPRC2012, Kisamori2016, Miki2017}.  % around the world~\cite{EPJA2018,Kisamori2016,Miki2017}. 

In the case of high-energy reactions with light projectiles, %whereas 
where the reaction mechanism is rather well known, single CE (SCE) reactions have been widely investigated to probe spin-isospin properties of nuclei. Moreover, $\beta$ decay was used as calibration whenever possible, in view of the close relationship between $\beta$-decay %probabilities 
strengths and CE cross sections at low momentum transfer~\cite{Taddeucci87,Alford}. 

A renewed interest has also %however 
recently emerged in studying CE reactions with heavy ions, which are used to address a wide spectrum of research topics, from work on quasi--elastic SCE excitations with low and intermediate energy ion beams, to investigations of the subnuclear sector by  excitations of nucleon resonance. e.g. $\Delta_{33}(1232)$, at relativistic beam energies, as discussed in the recent review article Ref.\cite{LenPPNP2019}. 

Over the years, a plethora of heavy ion SCE data has been collected and used for spectroscopic work. The study of the charge-converting components of the projectile-target nucleon--nucleon (NN) interactions has been in many cases the driving force behind experimental and theoretical research. In this context, some progress on the theoretical description of the complex reaction dynamics has been made over the past %year
years~\cite{Lenske18, LenPPNP2019, CavFRO2021}. %, in the case of heavy-ion SCE reactions~\cite{Lenske18,LenPPNP2019,CavFRO2021}.
Not to the least, the interest in that kind of interaction is motivated
by the similarity of the nuclear transition operators to those acting in nuclear beta--decay, as elucidated in Refs.~\cite{Lenske18, LenPPNP2019, BelPLB2020, LenskeUniverse}. 

Even the large research potential of heavy ion double charge exchange (DCE) reactions was recognized quite early in a first (unsuccessful) attempt to study the double Gamow--Teller (DGT) mode \cite{blomgren}.
More recently, experimental studies are also investigating the DCE reactions between heavy ions to identify the connections of these processes not only with DGT or double
Fermi transitions but also with double-$\beta$ ($2\beta$) decay~\cite{SanPRC2018, BelPLB2020}.
%CE reactions have long been recognized as well established tools for spectroscopic studies, especially when nuclear states which may be hardly excited otherwise are concerned~\cite{}. 
%However, in 
In %addition to the new spectroscopic information one can extract from these studies, especially when nuclear states which may be hardly excited otherwise are concerned, 
this context, the NUMEN and NURE projects at LNS-INFN have proposed an innovative way to exploit DCE reactions to access nuclear transition matrix elements (NME) related to those involved in $2\beta$ decay~\cite{EPJA2018}. A very accurate knowledge of %spin-isospin 
the NME is indeed instrumental to extract the properties of neutrinos from the half-life of the hypothetical neutrinoless double beta decay (0$\nu$2$\beta$), that is, to search for physics beyond the standard model~\cite{EngRPP2017, ShimizuPRL2018}. 
Besides, extraction of nuclear structure information from the total DCE reaction cross section is helpful to %put 
add new and more stringent constraints on the nuclear models~\cite{EngRPP2017, GamPRL2020}.
%All these features make 
These goals pose an %the study of DCE reactions a particularly 
exciting challenge on the study of DCE reactions %and 
that are stimulating recent developments, %which are trying to 
aimed to describe the (virtually unexplored) underlying reaction mechanism~\cite{SanPRC2018, BelPLB2020}. 
Before being able to address those spectroscopic issues on a quantitative level, the experimental and theoretical research program of the NUMEN and NURE projects demands thus a full understanding of the reaction mechanism of the underlying SCE and DCE reactions at beam energies in the region of 10 to 20 AMeV, which is located well above the Coulomb barrier in practical experiments.

Heavy ion collisions above the Coulomb barrier are however generally characterized by a large number of different populated channels. For example, it is well known that nuclear charge exchange processes proceed in principle by two distinct, but interfering mechanisms. Direct CE is a collisional process mediated by the isovector nucleon--nucleon (NN) interactions, acting between projectile and target. However, an unwanted but unavoidable %complications 
complication in the heavy ion induced CE reactions %are admixtures 
is the admixture %from more complicated 
of multi-step reaction mechanisms, typically given by sequential transfer of nucleons or inelastic processes of second or higher order~\cite{CapLenske04}. The two-step transfer mechanism is indeed sensitive to the nucleon-nucleus mean-field potential and cannot probe the NN interaction responsible for Fermi and Gamow-Teller isospin-flip transitions, which are observed in a CE reaction~\cite{LayBurrello}. For SCE reactions, %The
the competition of one-step direct charge exchange  and two-step sequential processes was discussed in detail, e.g., in Refs.~\cite{Oertzen, Brendel:1988mwl, Winfield, Len89}. 
The probably first fully microscopic theoretical investigations of heavy ion SCE reaction showed that, at energies close to the Coulomb barrier, transfer CE is by far the dominant reaction mechanism \cite{Brendel:1988mwl}, but with increasing beam energy the strength of direct SCE rapidly increases and finally dominates the SCE cross sections \cite{Len89}.

In Ref.~\cite{Len89} it was also shown that %calculations based on the assumption of 
direct CE calculations give differential cross section angular distributions with a steeper decrease at backward angles than the experimental data, so indicating the increased importance of higher-order processes at large momentum transfer. Other pioneering studies~\cite{DasPRC1986} showed that, under suitable conditions, the sequential transfer of proton and neutron pairs could even constitute the dominant process populating the final channel of a DCE reaction, at least when not very forward angles were considered. 

The second order character and the dominance of mean--field dynamics lead to a quite pronounced dependence of transfer SCE on incident energies, quantum numbers and matching conditions of the involved single particle orbitals, and the structure and multipolarities of the initial, intermediate, and final states, see e.g. Ref. \cite{Len89}. In other words, although the reaction mechanism follows general rules, the transfer SCE yield may depend critically on the reacting nuclei and the kinematical conditions %However, the % validity of the result 
%importance of sequential nucleon transfer is generally dependent on the kinematical conditions %, which may favour the transfer processes 
and a full understanding of the competition between the different mechanisms calls for further investigation. To complicate matters, one should take into account that also projectile and target inelastic excitations could play a relevant role in the description of the reaction mechanism, producing strong coupling effects among the various reaction channels. Therefore, if one wants to extract information on the NME, it %reveals 
is mandatory to disentangle from the experimental CE cross section the contribution of each competing process. %The two-step transfer mechanism is for instance sensitive to the nucleon-nucleus mean-field potential and cannot probe the NN interaction responsible of Fermi and Gamow-Teller isospin-flip transitions, which are observed in a CE reaction~\cite{LayBurrello}. 

On the other hand, the theoretical description of the complex mechanisms involved requires several ingredients, whose reliability cannot be generally guaranteed when applied beyond the domain in which they have been determined. %Recent studies have for instance shown that the nucleus-nucleus potential has to be specifically modeled for the projectile-target system under investigation~\cite{SpaPRC19, FauPRC21, CavFRO2021, Akis, CarUNI2021}. 
For instance, the nucleus-nucleus potential should be in principle specifically modeled for the projectile-target system under investigation. However, recent studies have shown that, in the energy region of our interest, a satisfactory reproduction of the scattering cross sections is achieved for a wide range of masses, under suitable general prescriptions~\cite{SpaPRC19, FauPRC21, CavFRO2021, Akis, CarUNI2021}.  It is nonetheless important to validate this strategy for each system under investigation and, in this context, the experimental measurement and analysis of the elastic and inelastic scattering reveals thus crucial to constrain also the SCE and DCE calculations. 

Because of the complexity in diagonalizing the Hamiltonian of systems involving open shell medium and heavy nuclei, other requirements are then also needed to perform the nuclear structure calculations. In particular, further constraints are required for the model %sub-space 
space and interaction combination which is frequently adopted to deduce the single-particle transitions characterizing one- and two-nucleon transfer processes~\cite{Diana, Jonas, Akis}. A comprehensive analysis should take the limitations of the model spaces into account and assess whether the global interaction parameters are suitably applicable to the system under investigation.

A possible gateway to overcome this issue, which has revealed successfull in some recent works~\cite{CavFRO2021, Akis}, is trying to perform measurements of %all the competitive processes 
several reaction mechanisms under the same experimental conditions and to describe them within an unified theoretical reaction scheme. In the present work, we propose thus to perform a multi-channel experimental analysis, to be compared with a combined structure and reaction theoretical study, with the aim to assess the reliability of the ingredients adopted in the description of the involved mechanisms. 

As part of the intensive experimental campaigns proposed by the NUMEN project, we focus on the $^{116}$Cd($^{20}$Ne,$^{20}$F)$^{116}$In SCE reaction at 306 MeV. 
$^{116}$Cd was chosen as a target not to the least because that nucleus is one of the candidates for the pursued $0\nu2\beta$ decay \cite{sasanoPRC2012}. Keeping that goal in mind, there are other, lighter nuclei under scrutiny as $2\beta$ decay–candidates. However, once reactions on $^{116}$Cd are understood, then SCE – and finally DCE – reactions on lighter target nuclei can be analyzed accordingly by the same methods.
%\st{This reaction deserves indeed a special attention, being the involved $^{116}$Cd nucleus a candidate for the pursued $0\nu2\beta$ decay.}

%The goal of this paper is twofold: first of all,  we aim at achieving a comparison with a more comprehensive set of experimental observables than has been done so far; then, with refined model ingredients, we evaluate all the competing contributions, with respect to the direct process, feeding the SCE outgoing channel. 
The goal of this paper is twofold: first of all, the work serves to show the experimental feasibility of such a demanding reaction for spectroscopic purposes. Secondly, on the theory side we aim at the evaluation of the various competing transfer contributions feeding the outgoing SCE channel by a comparison with a more comprehensive set of experimental observables beyond what had been done so
far.
%In other words, t
%The goal of this work is then to evaluate all the competing contributions, %with the final \textcolor{magenta}{aim-->goal} 
%to get a complete description of CE reactions, accounting for all the mechanisms feeding the same outgoing channels and to achieve a comparison with a more comprehensive set of experimental observables than it has been done so far. %In such a way, we aim also at getting a more quantitative understanding of the rich phenomenology that involves the nuclear spin and spin-isospin properties, which is still missing, despite the significant progresses achieved in last decades.

%For their study, a multi-channel approach should be thus recommended if one wants to assess the reliability of the ingredients usually adopted in the description of the complex reaction mechanisms involved. 

The paper is organized as follows.  %\textcolor{magenta}{
In Sect. II we illustrate the theoretical framework adopted, trying to underline the main differences existing among the various reaction schemes considered. In Sect. III the experimental set-up and the data reduction are described. In Sect. IV we discuss the results concerning the analysis of the elastic scattering, the one-proton transfer reaction and the sequential-nucleon transfer processes. In the last section, conclusions are drawn and future perspectives are indicated.  %}  

\section{Theoretical framework} 
We consider an ion-ion SCE reaction according to the scheme
\begin{equation}
^a_zx + ^A_ZX \to ^a_{z\pm 1}y + ^A_{Z\mp 1}Y
\end{equation}
leading from the entrance channel
$\alpha = \{x;X\}$ to the exit channel $\beta = \{y; Y\}$.  In the outgoing nuclei, states of multipolarity $J^\pi$ are excited, including Gamow-Teller modes of unnatural parity $\pi=(-)^{J+1}$ and Fermi--modes of natural parity, $\pi=(-)^J$, respectively. During the reaction,
the mass partition is retained but the charge partition is changed, either by a balanced redistribution
of protons and neutrons in a sequence of single particle transfer reactions or by direct single charge exchange. 

Direct SCE is mediated by the exchange of (virtual) isovector mesons between nucleons in projectile and target. It is a process of first order in the projectile-target isovector interaction, described well by Distorted Wave Born Approximation (DWBA) methods. Hence, in direct SCE the initial and final states are connected in an especially simple manner by transition operators of the same spin--isospin structure as known from nuclear beta--decay ~\cite{Lenske18, LenPPNP2019}. From the point of view of nuclear structure physics, the direct SCE reaction mechanism is the by far preferred mechanism because of the unfiltered access to nuclear isovector spectroscopy.

However, the same exit channels may also be populated in a considerably more complicated manner by transfer SCE. That alternative process requires a sequence of proton--neutron exchange processes between projectile and target nuclei, finally merging into a configuration which is part of the wave function of the final state. Obviously, that process requires the appropriate stepwise rearrangement of neutrons
and protons. In a reaction like the one here investigated, we encounter the neutron pick up--proton stripping sequence
\begin{equation}
^a_zx + ^A_ZX \to ^{a+1}_{z}x' + ^{A-1}_{Z}X'\to ^a_{z- 1}y + ^A_{Z+ 1}Y
\label{eq:np1}
\end{equation}
and the complementary proton stripping--neutron pickup sequence
\begin{equation}
^a_zx + ^A_ZX \to ^{a-1}_{z-1}y' + ^{A+1}_{Z+1}Y'\to ^a_{z- 1}y + ^A_{Z+ 1}Y,
\label{eq:np2}
\end{equation}
both being at least second order reactions in the respective mean--field potentials defining the transfer interactions.
From a nuclear structure point of view, transfer SCE is probing the overlap of the final projectile and target configurations with specific proton and neutron configurations. Since these configurations are populated stepwise in a sequence of pick up and stripping reactions, also the single particle spectroscopy in the intermediately reached $A\pm 1$ and $a\mp 1$ nuclei, respectively, plays an important role. Moreover, in order to exhaust the full single particle strengths, the transfer steps should in principle scan over the whole spectral distributions for each single particle spin and parity. 

Also admixtures of core--excited configurations may affect significantly the spectroscopic distributions especially in the intermediately populated odd--even and even--odd nuclei. A complete theoretical analysis might require to take into account also the coupling of the various channels to inelastically excited states in projectile-- and target--like nuclei. The proper handling of these complexities in the reaction mechanism requires methods going beyond standard DWBA.

At this stage we do not consider
direct SCE processes further, mainly due to yet to be resolved conceptual uncertainties with respect to the consistency of the theoretical models. The widely used standard approach to the nuclear response function of direct SCE is the Quasiparticle Random Phase Approximation (QRPA). QRPA by definition focuses on the nuclear 1 particle-1 hole response in a quite complete manner, even over very large energy intervals as entering e.g. as intermediate states into the theory of DCE reactions~\cite{BelPLB2020, LenskeUniverse}. The spectroscopy of the transfer channels, however, is described by many--body shell model methods, which emphasize the multi--configuration dynamics in a limited number of the nuclear valence shells. Thus, in order to exclude arbitrary effects due to  possibly incommensurable theoretical input, at this point we refrain from combining the two distinct sectors, leaving that demanding task to later work.

It is worthwhile to notice that the strong
absorption in the overlap region enforces the localization of the interaction regions mostly
to the surface or even tail regions of the density distributions of the colliding ions. Different
direct reaction frameworks are available to tackle
such processes.
Firstly, the DWBA framework has proven in many cases
to describe successfully direct reactions as single nucleon
transfer reactions\cite{Sat83, TimoPPNP2019, MontPRC2011, GasPRC2018, Akis}. For a reaction $\alpha\to \beta$ the DWBA amplitude is given as a matrix element of distorted waves $\chi^{(\pm)}_{\alpha\beta}$  and a form factor or reaction kernel
\begin{equation} 
M_{\alpha\beta}=\langle \chi^{(-)}_\beta|F_{\alpha\beta}|\chi^{(+)}_\alpha\rangle ,
\end{equation}
where the nuclear structure information is contained in the (non--local) reaction kernel
\begin{equation}
F_{\alpha\beta}(\mathbf{r}_\alpha,\mathbf{r}_\beta)=\langle yY|U_{\alpha\beta}|xX\rangle 
\end{equation}
depending on the channel coordinates $\mathbf{r}_{\alpha,\beta}$. The transition potential $U_{\alpha\beta}$ accounts for the nuclear dynamics of the reaction. For single nucleon transfer reactions, $U_{\alpha\beta}$ is defined in terms of the binding potentials of the transferred nucleon plus eventually necessary non--orthogonality terms, for details see e.g. \cite{Tamura:1974,Sat83}. In the case of a second order reaction like transfer SCE, $U_{\alpha\beta}$ is by itself a second order operator, accounting for the population of the intermediate states and the corresponding distorted wave channel propagators, see~\cite{BelPLB2020, LenskeUniverse}. 

A widely used successful approach is to separate the transition potentials into (products of) spectroscopic amplitudes, determined either by nuclear theory or phenomenologically by fits to data, and reduced form factors carrying the information on single particle wave functions and potentials. The calculations discussed below are performed with realistic single particle wave functions and potentials describing separation energies and nuclear radii. Spectroscopic amplitudes are taken from many--body shell model calculations.

The DWBA approximation assumes, however, that the
probability of the process under study is small with respect
to the elastic scattering and requires that the final
state is reached directly, i.e. without any intermediate
excitations of excited states lying in the same mass partitions.
Such requirement is not always fulfilled, especially in
the case of strongly excited rotational or vibrational modes. As a result,
when coupling effects among the various reaction channels
are strong, the DWBA approach fails to reproduce
the experimental data. For such cases, the reaction should be described by solving a set of coupled wave equations.
For the direct solution of such coupled channels (CC) problems standard numerical methods are available, but restricted
to couplings within the same mass partition. Iterative methods are used if transfer channels are to be included. A widely
used approach is the
Coupled Channels Born Approximation (CCBA).   
On top of the CCBA solutions the coupling of initial and final SCE channels via sequential transfer is added 
iteratively by the Coupled Reaction Channel (CRC) formalism. For details we refer again to the literature \cite{Tamura:1974,Sat83,KeePRC2020}. The various formal approaches are realized numerically in the Exact Finite Range (EFR) computer code FRESCO~\cite{Thompson} which treats properly the afore mentions non--localities of transfer form factors. FRESCO was used for all reaction calculations. 

In the following sections, we first study elastic scattering. As discussed in Ref.~\cite{LenPPNP2019}, the order of magnitude of the quasi--elastic CE and transfer cross sections is determined by the distorted waves. Their properties are fixed by the optical potentials used to describe elastic scattering. Hence, a proper description of elastic cross sections and total reaction cross sections is essential for a realistic transfer yield. Therefore, below some space is devoted to elastic scattering and the derivation of optical potentials. Turning to the transfer SCE, we start with studying the population of several
low-lying excited states reached in the first step of the full SCE sequence.  All
of them are relevant intermediate channels of the nucleon
transfer processes which coherently contribute to the full
SCE reaction. Finally, the transitions to the final SCE states are considered by investigating the combined sequential
one neutron/one proton transfer rearrangements.
Further details
concerning each reaction framework, the description of
the ion-ion interaction as well as about the structure ingredients
required within our scheme, will be provided especially in
section~\ref{sec:results}.

\section{Experiment and data reduction}
The experiment was performed at the INFN-LNS laboratory in Catania. A $^{20}$Ne$^{4+}$ beam was accelerated at 306 MeV incident energy by the K800 Superconducting Cyclotron, fully stripped by crossing a thin carbon foil %\textcolor{magenta}{
located at the accelerator exit %} 
and then transported to the scattering chamber. The targets were 96\% isotopically enriched $^{116}$Cd foils produced by rolling at the LNS target laboratory of 
thickness (1370 $\pm$ 70), (1080 $\pm$ 60) and (1330 $\pm$ 70) $\mu$g/cm$^{2}$ in the case of one-proton transfer, SCE and scattering measurements, respectively. The first two were coupled to natural C foils with thicknesses of %thick 
(990 $\pm$ 50) and (900 $\pm$ 45) $\mu$g/cm$^{2}$, while the third one to a CH$_{2}$ %of 
(950 $\pm$ 45) $\mu$g/cm$^{2}$ %of thickness. 
thick. Such post-strippers foils %secondary targets 
were introduced %as post-strippers  
\cite{CAVALLARO2020334,shima1992} to conveniently %\textcolor{magenta}{re}
readjust the charge state distribution of the ions emerging from the cadmium targets \cite{cavallaro2019}. However, due to the %different kinematic conditions
kinematics of the explored reactions, the data shown in the following are not affected, %\textcolor{magenta}{,} 
in the analyzed regions, %\textcolor{magenta}{,} 
by the contributions due to reactions on post-stripper materials. \newline
A copper Faraday cup of 0.8 cm entrance diameter and 3 cm %of 
depth was used in order to stop the beam and collect its charge. It was mounted 15 cm downstream of the target and equipped with an electron suppressor polarized at -200 V to ensure, even with beam currents of few enA as the ones typically available in the described %\textcolor{magenta}{
experimental %} 
conditions \cite{EPJA2018}, a charge collection accuracy better than 10\%. \newline
The reaction ejectiles were momentum analyzed by the MAGNEX spectrometer \cite{magnex_review}, %\textcolor{magenta}{
whose %} 
experimental setup was optimized to select, in separated runs, specific reaction channel:
\begin{itemize}
\item For the $^{116}$Cd($^{20}$Ne,$^{20}$Ne)$^{116}$Cd elastic and inelastic scattering measurement, the spectrometer optical axis was set at three different angular settings, %$\overline{\theta}_{lab}$
$\theta_{opt}$ = 8°, 13° and 20° in the laboratory frame, spanning the 3° $< \theta_{lab} <$ 26° angular range. All the runs were performed adopting the full angular acceptance ($\Omega \sim$ 50 msr). %In the run at 8°, in particular, the beam current was decreased to reduce the count-rate at the focal plane detector. \textcolor{blue}{MEMO: il set 8deg lama chiusa è stato scartato nell'analisi}
\item For the $^{116}$Cd($^{20}$Ne,$^{19}$F)$^{117}$In one-proton transfer measurement the spectrometer optical axis was placed at %$\overline{\theta}_{lab}$ 
$\theta_{opt}$ = 8°, corresponding to the 3°$< \theta_{lab} <$ 14° angular range. The MAGNEX angular acceptance was slightly decreased, determining a solid angle %value 
of $\sim$ 45 msr.   
\item For the $^{116}$Cd($^{20}$Ne,$^{20}$F)$^{116}$In single charge exchange the spectrometer optical axis was oriented at %$\overline{\theta}_{lab}$ 
$\theta_{opt}$= 9°, thus exploring the angular range 4° $< \theta_{lab} <$ 15°. In this run the vertical angular acceptance was considerably reduced in order to limit the overall event-rate at the focal plane detector, decreasing the covered solid angle to $\sim$ 1.3 msr. 
\end{itemize}

The ejectile identification as well as the data reduction techniques are described in details in Refs. \cite{cappuzzello2010, CAVALLARO2020334}. %The latter is based on a fully differential algebraic method \cite{a},
They require the accurate measurements of the horizontal and vertical positions and angles provided by the MAGNEX focal plane detector \cite{cavallaro2012, TORRESI2021164918} and the %faithful 
knowledge of the high-order transport map of the particles throughout the spectrometer \cite{magnex_review}. 
\newline
The differential cross sections were extracted for all the measured %\st{measured} \textcolor{magenta}{investigated} 
reaction channels according to the technique described in Ref. \cite{magnex_review}. %taking into account the overall MAGNEX efficiency \cite{a}.
The error bars reported in the corresponding energy and angular distributions account for the statistical contribution, the uncertainty due to the solid angle evaluation %\textcolor{magenta}{
and the fit procedure when performed. %}.
%\red{An overall} \gre{A fully correlated} \blu{Intendete dire che l'incertezza è "fully correlated"? (ossia... tutti i punti si muovono coerentemente in su/giù della stessa quantità?)} 
A systematic uncertainty of $\sim$10\%, not shown in the plots, 
is common to all the experimental data,
originating from target thickness measurement and
Faraday cup charge collection uncertainties.
%is common to the one-proton transfer and SCE experimental data, originating from target thickness measurement and Faraday cup charge collection uncertainties. In the case of elastic scattering measurement, the data were normalized to the Rutherford cross section at very forward angles, adopting a scale factor equal to 1.6, which is larger than the systematic error and that could be associated to efficiency losses in the focal plane detector, due to the corresponding high counting rate. \newline

\section{Results}
\label{sec:results}

%\subsection{\st{Quasi-elastic} \textcolor{magenta}{Scattering?} channel}
\subsection{Scattering channel}
A crucial issue for a quantitative understanding of heavy ion reactions is the proper treatment of projectile-target interaction, thus motivating elastic and inelastic scattering studies.

\begin{figure}[tbp!]
\includegraphics[width=\columnwidth]{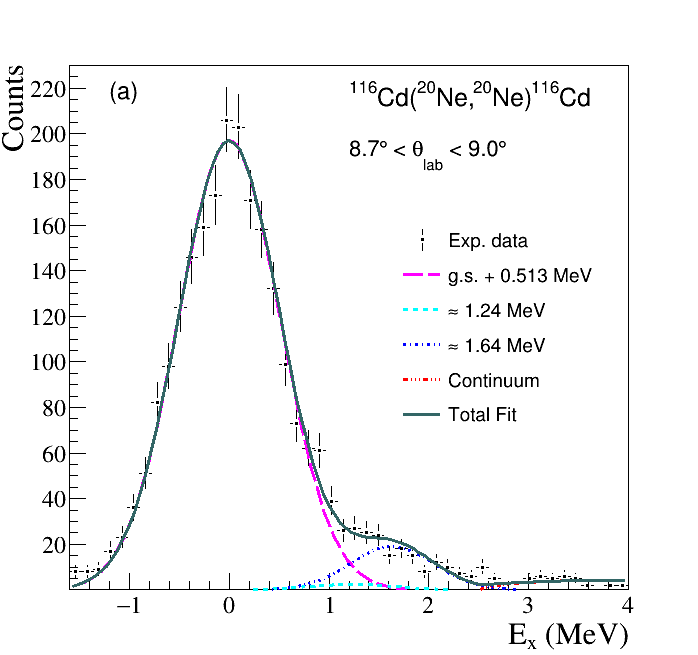}
\includegraphics[width=\columnwidth]{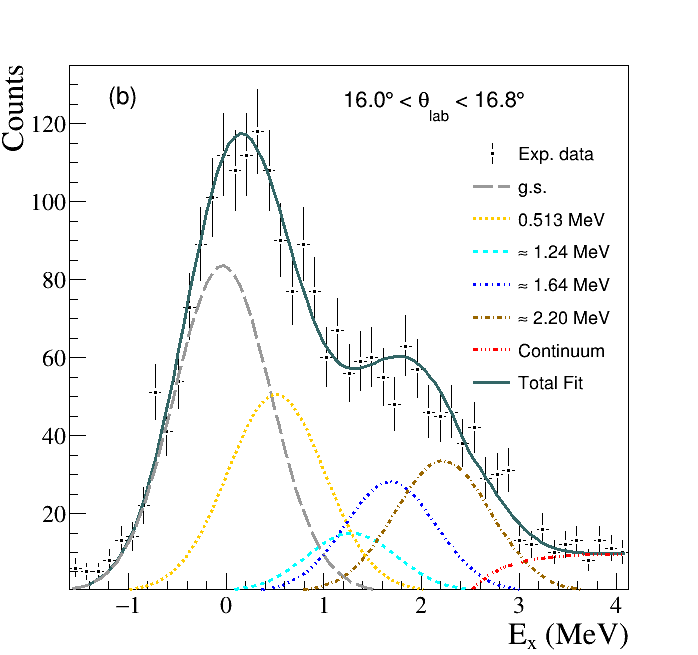}
\caption{Excitation energy spectra of the scattering channel %\textcolor{magenta}{
in two different angular regions. Panel (a): Energy spectrum
%} 
in the angular range 8.7° $<  \theta_{lab} <$ 9.0°. The magenta curve represents the energetically unresolved ground and first excited states of the target. The cyan and blue Gaussian functions are introduced to reproduce the small bump associated to the unresolved excitation of the first low-lying states of $^{116}$Cd and $^{20}$Ne (see the text). The red curve reproduces the continuos spectrum in the excitation energy region above $\sim$2.5 MeV. The sum of all the mentioned contributions is shown as the green line. 
%\textcolor{magenta}{
Panel (b): Energy spectrum in the angular range 16.0° $<  \theta_{lab} <$ 16.8°. The ground and the first $^{116}$Cd excited states are fitted separately as the grey and yellow curves. The brown curve represents the additional Gaussian functions peaked around 2.2 MeV, not present in the forward angle spectrum of panel (a). For all the other contributions as well as for the total sum function the same color code as in panel (a) is adopted.
%}
}
\label{fig:elastic_spectrum}
\end{figure}

\begin{figure}[tbp!]
\includegraphics[width=.85\columnwidth]{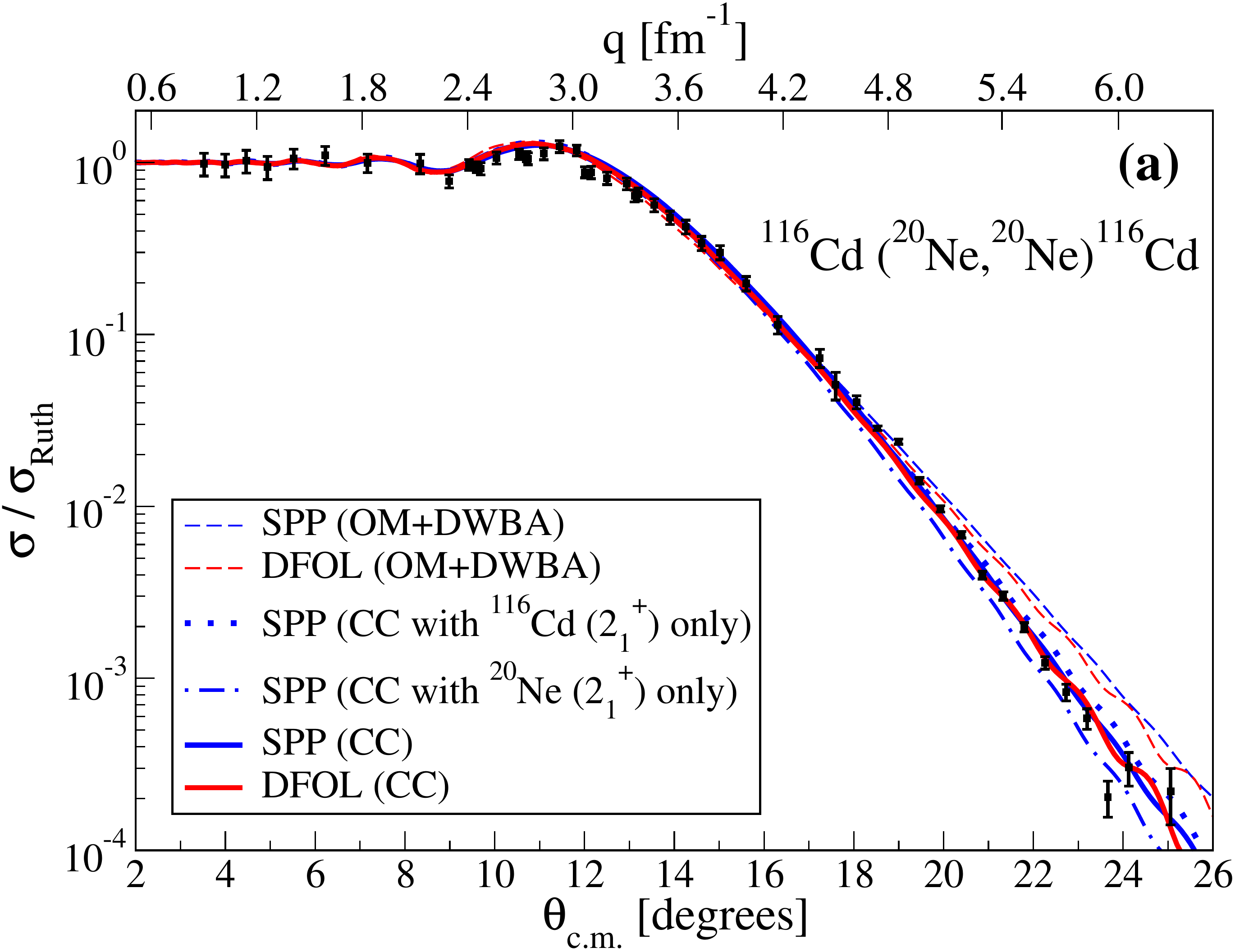} \includegraphics[width=.85\columnwidth]{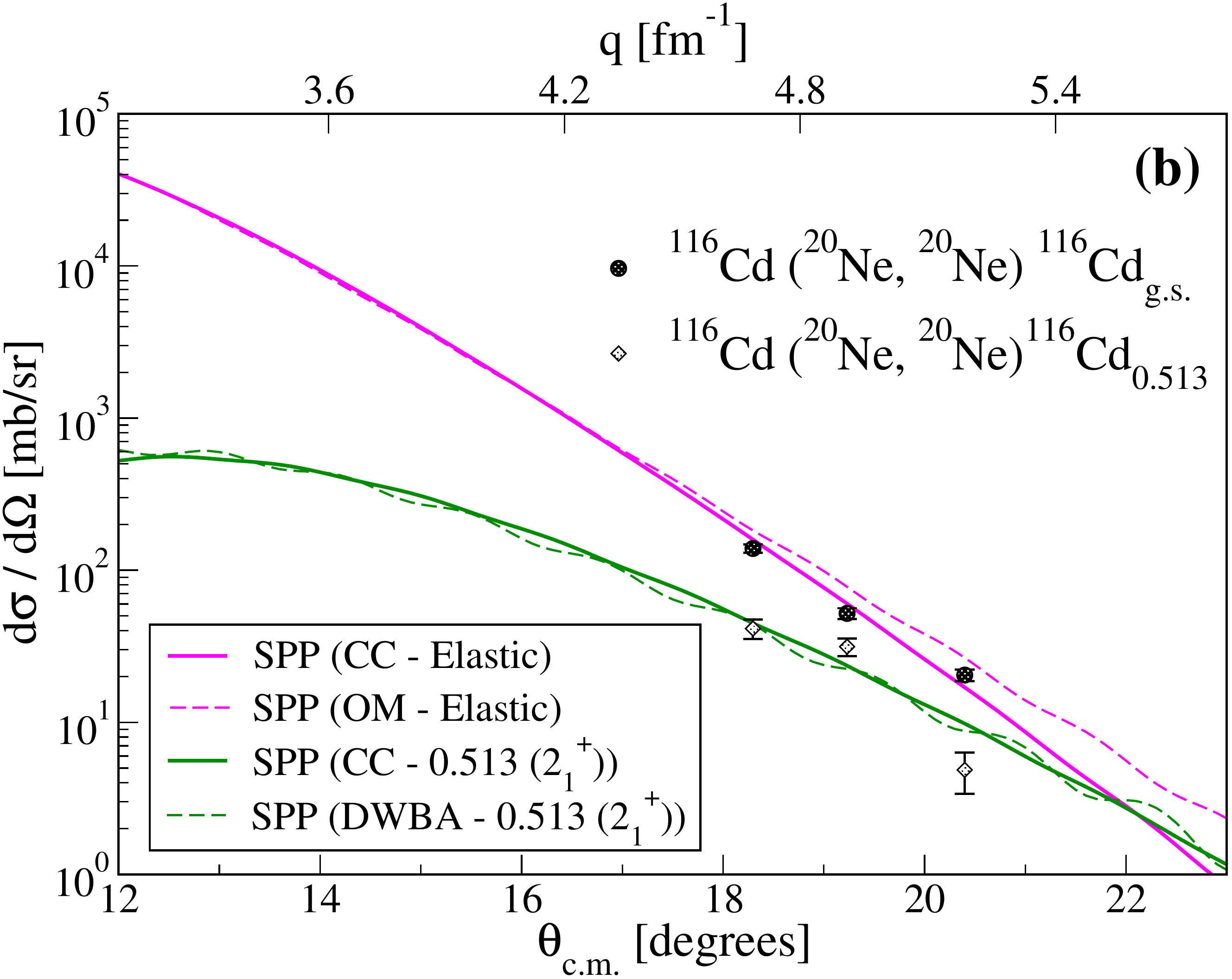} \includegraphics[width=.85\columnwidth]{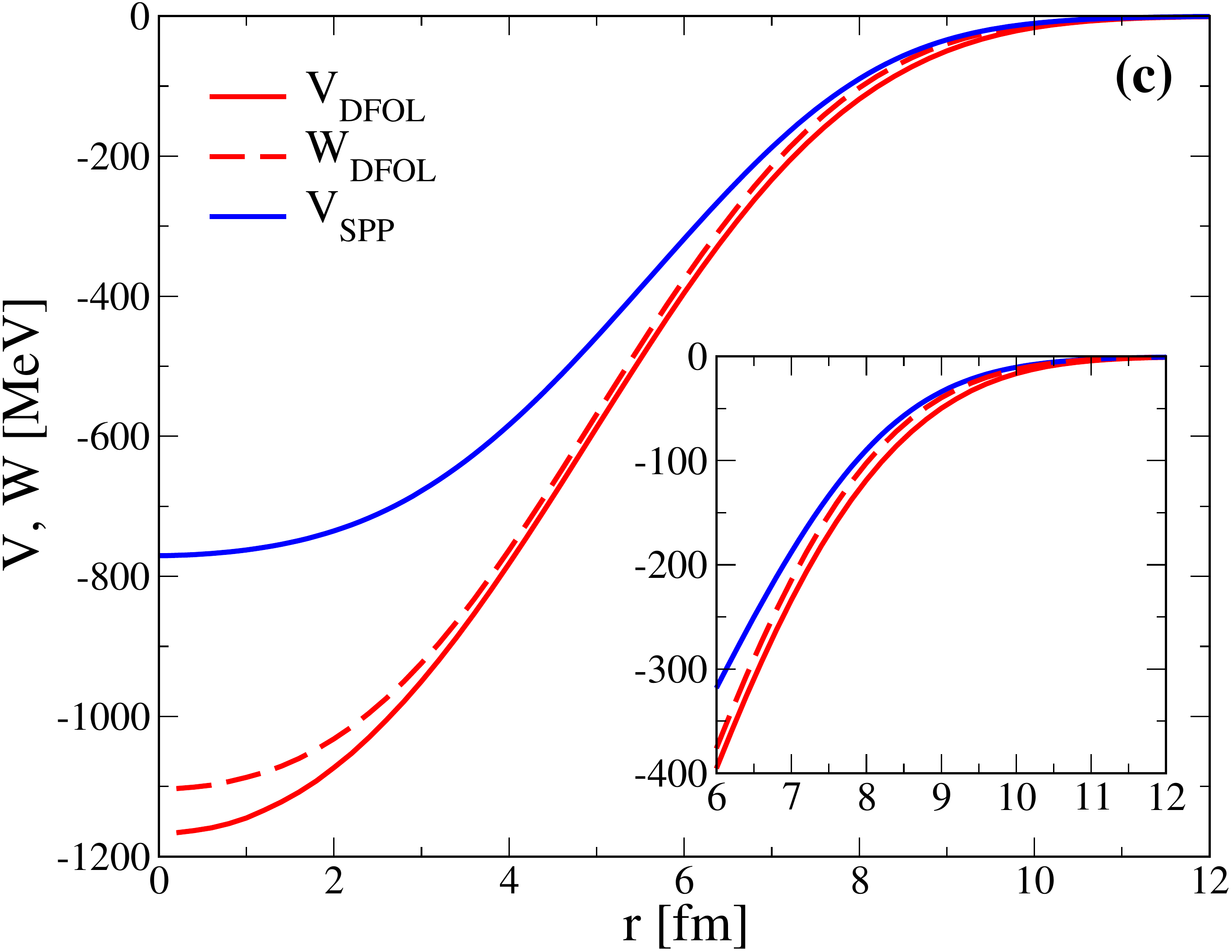}
\caption{Panel (a): Experimental angular distribution of differential quasi-elastic scattering cross section in the $\sigma / \sigma_{\textup{Ruth}}$ representation. Theoretical results obtained for the DFOL (red lines) and SPP (blue lines) potentials, within two different reaction schemes, namely OM (elastic) + DWBA (inelastic), in dashed lines, and CC, in full lines. Both experimental values and theoretical calculations also include the contribution of the inelastic scattering to the (2$_1^+$) target state. The dotted and dashed-dotted lines show the results obtained with SPP by alternatively switching on the contributions of the (2$_1^+$) target and projectile states, respectively. Panel (b): Experimental angular distribution of differential cross section for elastic and  (2$_1^+$) $^{116}$Cd inelastic excitation, as separately considered and compared with OM/DWBA or CC calculations, with SPP potential. Panel (c): Radial dependence of the real V (red full line) and imaginary W (red dashed line) part for DFOL and the real part V for SPP (blue full line). The inset shows also a zoom on the peripheral region.}
\label{fig:elastic}
\end{figure}

%\textcolor{magenta}{
The excitation energy spectrum of the ($^{20}$Ne, $^{20}$Ne) scattering channel is shown in Fig. \ref{fig:elastic_spectrum}, being $E_x$ = $Q_0$ - $Q$ with $Q_0$ the ground-state-to-ground state Q-value. 
%\textcolor{magenta}{T}he 
The limited experimental energy resolution ($\delta E \sim$ 1.1 MeV full width half maximum, mainly due to straggling effects in the used target) %\textcolor{magenta}{
and the available statistics do %} 
not allow to discriminate the elastic transition and the first excited state of the system, the ($2^+_1$) state of the $^{116}$Cd expected at 0.513 MeV, %\textcolor{magenta}{
except for a restricted angular region (15° $\lesssim \theta_{lab} \lesssim$ 18°). %}. 
%\red{
%Thus, the dominant Gaussian function of Fig. \ref{fig:elastic_spectrum}\textcolor{magenta}{(a)} represents the sum of the %two corresponding 
%contributions of \textcolor{magenta}{such} states}
In Fig. \ref{fig:elastic_spectrum} (a), the sum of the contributions of such states is represented by the dominant Gaussian function,
%\textcolor{magenta}{
whereas in Fig. \ref{fig:elastic_spectrum}(b) the two transitions are considered separately. %}. 
%\textcolor{magenta}{A}dditional
Additional Gaussian functions peaked around 1.24 and 1.64 MeV, %\textcolor{magenta}{are shown in Figs. \ref{fig:elastic_spectrum}(a, b) and were} 
introduced to better reproduce the bump visible in such energy region, are shown in Figs. \ref{fig:elastic_spectrum}(a) and (b). These distributions are expected to correspond to the sum of the cadmium two-phonon quadrupole states ($2^+$, $4^+$, $0^+$) at 1.213, 1.219, 1.282 MeV, respectively, and the (2$^+$) $^{20}$Ne and $^{116}$Cd states at 1.634 and 1.642 MeV.
However,  they can also collect the contributions from other possible state like the $^{116}$Cd (0$^+$) predicted at 1.382 MeV and, partially, from the cadmium ones foreseen around 2 MeV. %\textcolor{magenta}{
The excitation energy spectrum of Fig. \ref{fig:elastic_spectrum}(b), corresponding to the backward angular range, presents a further Gaussian distribution introduced to better model the counting shape observed around 2.2 MeV, which is attributable to the latest quoted excited states of the target as well as to the simultaneous excitation of the first target and projectile transitions. %} 
Due to the low yields, %\textcolor{magenta}{, 
the high number of the involved possible states %} 
and the limited energy resolution available which does not allow to distinguish among the different possible transitions,  the distributions lying beyond 1 MeV will not be further considered in the following. Finally, the experimental spectrum beyond %\textcolor{magenta}{the discussed structures} 
the commented structures resulting in a continuous and suppressed shape is fitted by a smooth curve. 
The experimental differential cross section of the quasi-elastic transition, which includes the ground state (g.s.) and the (2$^{+}_1$) state of $^{116}$Cd at 0.513 MeV, is shown in Fig. \ref{fig:elastic} (a) %\textcolor{magenta}{(a)}. 
and compared to the results of the theoretical calculations,
The experimental elastic and (2$^{+}_1$) $^{116}$Cd inelastic differential cross sections, as extracted in the restricted angular range indicated above, with the corresponding calculations, are reported separately in Fig. \ref{fig:elastic} (b).
%}
Figures \ref{fig:elastic} (a) and (b) also report the scale of transferred linear momentum $q$. As seen from the upper abscissa of Fig. \ref{fig:elastic} (a), %\textcolor{magenta}{(a)}, 
the data cover a large range of linear momentum transfers, extending from $ q\sim 0.6$ fm$^{-1}$ to  $q\sim 6$ fm$^{-1} $. Thus, heavy ion reactions of this kind are probing nuclear  properties in much detail. %\textcolor{magenta}{In Fig. \ref{fig:elastic}(b), moreover, the experimental elastic and (2$^{+}_1$)  $^{116}$Cd inelastic differential cross sections, extracted in the restricted angular range as previously discussed, are reported separately with the corresponding calculations.}
%}

In the present work, the ion-ion interaction is described by a complex optical potential $U$, whose radial dependence is microscopically derived by doubly folding the one-body g.s. projectile and target nuclear densities, which are parameterized by two-parameters Fermi-Dirac distributions, with an effective NN interaction. In the numerical calculations of this work, two different double folding optical potentials will be tested.

On the one hand, we consider the S\~{a}o Paulo Potential (SPP)~\cite{PerPLB09}, which has already demonstrated to be successful in describing the elastic scattering and peripheral reaction channels for several heavy-ion reactions involving light~\cite{PerPLB2012, FonPRC2019}, medium~\cite{SpaPRC19, PerNPA2009, ZagPRC2018} and heavy~\cite{AlvNPA2003} mass targets %for several heavy-ion systems 
in a wide energy region. %\textcolor{magenta}{

%Moreover, it has no adjustable parameter, %}, %Within the context of the systematics for the densities, which 
%being 
The general SPP parameterization, being based on theoretical calculations with the Dirac-Hartree-Bogoliubov model and also on experimental results for charge distributions, has no adjustable parameter~\cite{PerPLB09}. %the SPP has no adjustable parameter, although 
%\textcolor{magenta}{
However, %}, 
small deviations around the average values might be expected due to the effects of the structure of the nuclei. 

On the other hand, %in analogy to what has been done in some previous works~\cite{CapLenske04}, 
another optical potential, which was used in some other works~\cite{CapLenske04, Lenske18, LenskeUniverse} and labeled as DFOL, is employed. The latter makes use of the complex NN T-matrix derived by Franey and Love~\cite{FraneyLove}, including both the isoscalar and isovector components and it is extrapolated down to the present energy region. %Indeed, despite the issues which emerge in the extrapolation method at low beam energies, below $T_{lab} = 50$ MeV the potential values do not change much and the strong increase which could be observed in the total NN cross section is mainly ascribable to a phase space effect.  %As a 
%At difference with respect to  
Differently from the SPP case, where the imaginary part is obtained from a simple scaling of the real one, for the DFOL optical potential the imaginary potential is directly available from the folding procedure. %the same approach is used for calculating, both in the incident and the exit channels, the two contributions.

As briefly addressed in a recent paper \cite{LenskeUniverse}, we are fully aware of the changes of the T-matrix to be expected in the low-energy region. Results of a recent investigation show that below $T_{lab}=$ 50 MeV the strength of the real parts of the (anti-symmetrized) spin-scalar, isoscalar and isovector components of the NN T-matrix remains almost constant, while the imaginary parts increase on a moderate level. A comparison of the total neutron-proton cross section to data and to the one obtained with the CD Bonn potential in the full Lippmann-Schwinger formalism shows that the strong increase of the cross section towards low energies is mainly driven by kinematical effects. %\gre{
The NN T-matrix parameters, as obtained for the central interactions at $T_{lab} = 15$ MeV in the relevant channels for the ion-ion potentials, {\it{i.e.}} the spin - scalar ($S = 0$) and the isospin $T = 0, 1$ channels, are given in Table \ref{tab:tnn}. %}
\begin{table}[t]
\centering
\caption{\label{tab:tnn} Nucleon-nucleon T-matrix interaction strengths at $T_{lab} = 15$ MeV in the spin - scalar ($S = 0$) and the isospin $T = 0, 1$ channels. The ranges are expressed in fm, all other quantities are in MeV.}
\begin{tabular}{*{6}{c}}
\toprule
    Range   &      SE      &      TE      &      SO      &      TO  	\\
\midrule
\midrule
\multicolumn{5}{c}{Real} \\		
     0.25  &   6048.76  &   7972.38  & -51218.10  &   2091.19  \\
     0.40  &  -1754.49  &  -2235.67  &   5541.75  &   -641.32  \\
     1.40  &    -10.50  &    -10.50  &     31.50  &      3.50  \\
\midrule
\midrule     
\multicolumn{5}{c}{Imaginary} \\	
     0.25  &  -1328.78  &  20805.12  &  31381.92  &  16175.65  \\
     0.40  &   -382.85  &  -6968.75  &  -4100.90  &  -2485.94  \\
\bottomrule
\end{tabular}
\end{table}

%An important %approximation 
%assumption in the folding of the DFOL and SPP optical potentials, for both projectile and target isotopes, is %related to the implicitly assumed 
%the spherical shape of their density profiles.
Although the folding procedure is easily extended to cover deformed matter distributions of projectile and target, in the present case by numerical reasons we use spherical density profiles in the calculations of the DFOL and SPP optical potentials. In a more extended approach, reaction observables need to be averaged anyway over the orientations of the deformed ions. To a good approximation the resulting net effects can be simulated by slightly changing the geometrical parameters of the density distributions~\cite{SpaPRC19}. % However, since 
In particular, since the g.s. quadrupole moment of projectile and target are %anything but 
not negligible~\cite{Pri2016, Ram2001}, one may infer that, especially in the case of $^{20}$Ne, the nucleus is significantly deformed, as also confirmed %in turn 
by the large experimental values of the quadrupole deformations ($\beta_2^C = 0.721$ for $^{20}$Ne, $\beta_2^C = 0.194$ for $^{116}$Cd).
An effective way to take %these arguments 
this aspect into account in building the optical potential is thus a change of the density profiles, %used in the folding of the NN interactions, 
as also described in Ref.~\cite{SpaPRC19}, with the requirement to keep constant the volume integral of the nuclear densities to fix the number of nucleons. In Ref.~\cite{SpaPRC19}, where a different reaction %system although 
involving the same projectile nucleus was analyzed, %with the aim to reproduce the slope of the experimental data beyond the grazing angle, 
an effective geometrical modification in the %structure of the 
nuclear densities was %thus 
introduced, %when 
increasing by 5\% the radius of the %nuclear 
density profiles, while renormalizing the central density parameter. For sake of consistency, the same prescription is here adopted for the projectile $^{20}$Ne, while a smaller modification (3\%) %turns out to be required 
is used for the target, %\gre{, 
to achieve the best possible agreement between data and calculations. %}.
The root mean square radius and volume integral per nucleon, as obtained for the real and the imaginary parts of both potentials adopted here, are listed in Tab.~\ref{tab:optical}. One observes that, even though the root mean square radii of the real parts practically coincide, a %quite large 
significant difference exists among the volume integrals of the two optical potentials, % here considered, 
although both values are still compatible with the typical ones expected from the systematics \cite{SatcPR79}.

It is worthwhile to remind that, for the imaginary part, a scaling factor has to be then introduced, depending on the reaction framework adopted, to account for missing couplings to %continuum 
states not explicitly considered here. 

%\gre{
Although it is not obvious that the same scaling factors should be assumed for different nuclear systems, the most reasonable choice would be to adopt a consistent approach with the analysis of some previous works~\cite{CarPRC17, PaePRC17, ErmPRC17, ErmPRC16}. %}
%Anyway, in 
%In both cases, f
For both DFOL and SPP optical potentials, %following the prescriptions given in previous works~\cite{FraneyLove}, %, a scaling factor for the strength of the imaginary part is needed to %also 
%account for missing couplings to %continuum 
%states not explicitly considered here. Such a factor, 
%as suggested by the systematics, 
%\gre{such a factor} is \gre{then} 
such a factor is then set to 0.78 in case of optical model (OM) calculations and 0.6 whenever CC calculations are performed. 

\begin{table}[t]
\centering
\caption{\label{tab:optical} Real $J_V$ and imaginary $J_W$ part of the volume integral per nucleon (in MeV fm$^{3}$) and root mean square radii (in fm) for the real $\sqrt{\langle R^2 \rangle_V}$ and imaginary $\sqrt{\langle R^2 \rangle_W}$ part of DFOL and SPP potentials adopted.}
\begin{tabular}{*{5}{c}}
\toprule
  &  $J_V$ &  $J_W$ &  $\sqrt{\langle R^2 \rangle_V}$  &  $ \sqrt{\langle R^2 \rangle_W}$ \\
\midrule
DFOL & -437.56 & -408.94 & 5.884 & 5.770 \\
SPP & -331.68 & -331.68 & 5.883 & 5.883 \\
\bottomrule
\end{tabular}
\end{table}

A comparison with the %available 
measured quasi-elastic scattering data in terms of the corresponding Rutherford cross section $\sigma_{Ruth}$ %$\sigma_$\sigma / \sigma_{Ruth}$ representation 
is shown in Fig.~\ref{fig:elastic} %\textcolor{magenta}{(a)}
(a). Actually, as already explained in the previous section, the experimental values for the quasi-elastic channel %unavoidably 
include also the inelastic scattering to the (2$_1^+$) target state at 513 keV. %\st{, whose contribution cannot \textcolor{magenta}{could not} be isolated from the total cross section}. 
Therefore, the OM (for the elastic) + DWBA (for the inelastic) calculations plotted in Fig.~\ref{fig:elastic} %\textcolor{magenta}{(a)} 
(a) take into account the incoherent sum of the two different contributions, namely the genuine elastic cross section and the DWBA cross section %related 
of inelastic scattering to the first excited state of $^{116}$Cd.

Despite the differences existing in the %more 
internal region of the two optical potentials, there is no significant difference between the cross sections obtained by the two calculations. This fact confirms that the theoretical description of elastic scattering is not strongly dependent on the choice of %the optical potential 
these optical potentials since the strong absorption confines the reaction source on the surface of the colliding systems, where the two potentials have a similar behavior. 

Heavy ion reactions and therefore elastic scattering and peripheral inelastic reactions are indeed mainly sensitive to the nuclear surface regions of the interacting nuclei and, as a result, the incoming flux is mostly absorbed into a multitude of reaction channels, as demonstrated also by the %huge 
large total reaction cross sections.  %usually characterizing this kind of reactions. 

%Some oscillations are observed in the data in the region %much 
%beyond the grazing angle, where the theoretical 
One observes that theoretical OM+DWBA calculations %do not 
fail to 
reproduce the 
slope exhibited by the 
experimental %results
data beyond the grazing angle, whatever optical %model 
potentials is employed, including the two considered here. Since the behavior around the grazing angle is properly reproduced, the present case is similar to the one discussed in~\cite{SpaPRC19}, where the observed discrepancies with respect to the experimental data were attributed to the coupling of the elastic channel with possible inelastic excitations to the %due to the %strong coupled 
%strongly populated 
2$_1^+$ states of both projectile and target. %contributions. %\textcolor{magenta}{.}
In Fig.~\ref{fig:elastic} (a) %\textcolor{magenta}{(a)} 
CC calculations performed with both DFOL and SPP optical potentials are therefore also included. 

Both one-step DWBA and CC calculations are performed within the rotational model, %by 
following the same prescription adopted in some recent works \cite{CarUNI2021, CavFRO2021}, despite the low-lying states in $^{116}$Cd nuclei might appear to be of vibrational nature. Reduced transition probabilities B(E2;0$^{+} \to 2^{+}$) = 0.0333 $e^2b^2$ for $^{20}$Ne and B(E2;0$^{+} \to 2^{+}$)= 0.58 $e^2b^2$ for $^{116}$Cd are taken from \cite{Pri2016, Pri2017, Kib2002} and used to describe the strength of Coulomb deformation of both projectile and target. 

Nuclear deformations %of the latter 
are described in terms of first-order derivative of the OM potential U(r)
\begin{equation}
V (r) = - \frac{\delta_2}{\sqrt{4 \pi}} \frac{dU(r)}{dr}
\end{equation}
where the strength of the deformation is embedded in the deformation length $\delta_2$ \cite{Sat83}:
\begin{equation}
\delta_2 = \beta_2 R = \frac{4 \pi}{3 Z} \frac{\sqrt{B(E2;0^{+} \to 2^{+})}}{R_V}
\label{eq:delta}
\end{equation}
In Eq.~\eqref{eq:delta}, $\beta_2$ is the deformation parameter characterizing the transition of the given nucleus of charge Z, $R = 1.2 A^{1/3}$ %its 
is the radius and $R_V$ is the root mean square radius of the real part of the adopted optical potential (see Tab.~\ref{tab:optical}). The %values of the 
deformation lenghts obtained %when 
following such a prescription are $\delta_2 = 1.299$ fm and $\delta_2 = 1.130$ fm, for projectile and target, respectively. Exploratory calculations changing the method for the determination of $R_V$ give however similar results. Moreover, the same radial form factors are assumed also for the imaginary coupling potentials.
 %, extracted from the elastic cross section data analysis. 

One observes that the CC calculations %tend to very 
well reproduce the slope exhibited by the %\textcolor{magenta}{
quasi-elastic %} 
experimental %values 
data also at larger angles. A better agreement is thus %deduced 
observed with respect to %previous DWBA curves, 
OM+DWBA, especially in the region of larger momentum transfer, confirming the relevant role played by these rotational (collective) 2$_1^{+}$ states, whose couplings with the g.s. have to be explicitly included. By alternatively switching off the contributions of the %va\textcolor{magenta}{(r}ious 
various inelastic excitations, we found moreover a leading role played by the projectile $2_1^{+}$ state, which is strongly coupled to the elastic channel because of its large deformation parameter. The best agreement is however achieved only when the combined effect with the target inelastic excitation is included. 

%\gre{
The overall outcome is moreover not significantly influenced by the small changes introduced in the target and projectile nuclear density radii. Indeed, we checked that, when ignoring such  modifications, the total cross section in the elastic channel differ at most of 3\% with respect to the results shown before, both in the OM and in the CC approach, in the case of SPP potential. %} 

%\textcolor{magenta}{Moreover/Finally, a
A further %explicit 
validation of the %adopted description 
ingredients 
adopted in the description of the scattering channel analysis is reported in Fig. \ref{fig:elastic} (b). Indeed, a %the good results 
nice agreement is obtained between experimental data and theoretical predictions when disentangling the elastic and the first inelastic transitions in the %available 
angular range considered, especially when the CC calulations are performed. In Fig. \ref{fig:elastic} (b), only the SPP potential is considered, although an analogous result is obtained in the case of DFOL potential. Despite the limitedness of the angular range considered, such a result strengthens all the conclusions drawn from the quasi-elastic analysis. In particular, the matching between the experimental and theoretical elastic distribution confirms the reliability of the prescriptions adopted for the optical potentials, whereas the first inelastic transition provides another proof of the model assumed to describe the coupling with the inelastic excitations considered. In the following sections, we will explicitly take into account the effect of these inelastic excitations, looking also at their influence on the transfer channels investigated within our study. 

\subsection{One-proton transfer $^{116}$Cd($^{20}$Ne,$^{19}$F)$^{117}$In reaction}

\begin{figure}
\centering
\includegraphics[width=\columnwidth]{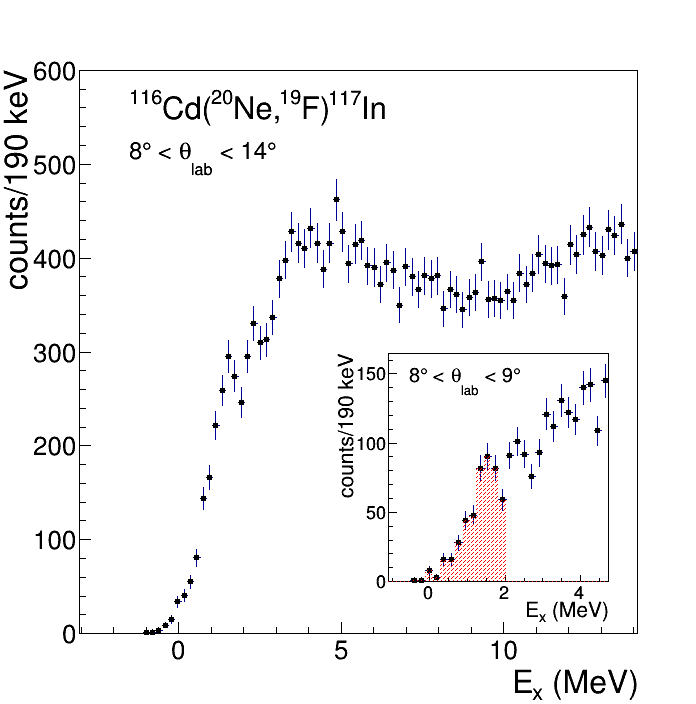}   %spettroadue.png 
\caption{Excitation energy spectrum of the one-proton transfer channel in the 8° $< \theta_{lab} <$ 14° angular range. %A global bump is observed around 5 MeV corresponding %at 
%to the Brink’s energy matching value. 
In the inset, the spectrum up to $\sim$ 4.5 MeV for the interval 8° $< \theta_{lab} <$ 9° is reported. %\st{It is reproduced as the sum (red line) of a % by a %\textcolor{magenta}{
%smooth curve %}
% (in green) and three Gaussian functions (in light blue, magenta and yellow) introduced in the fit procedure as described in the text}. \textcolor{magenta}{
The red hatched area corresponds to the analyzed region %} 
as described in the text. }
\label{fig:spe_onep}
\end{figure}

\begin{figure}
\includegraphics[width=\columnwidth]{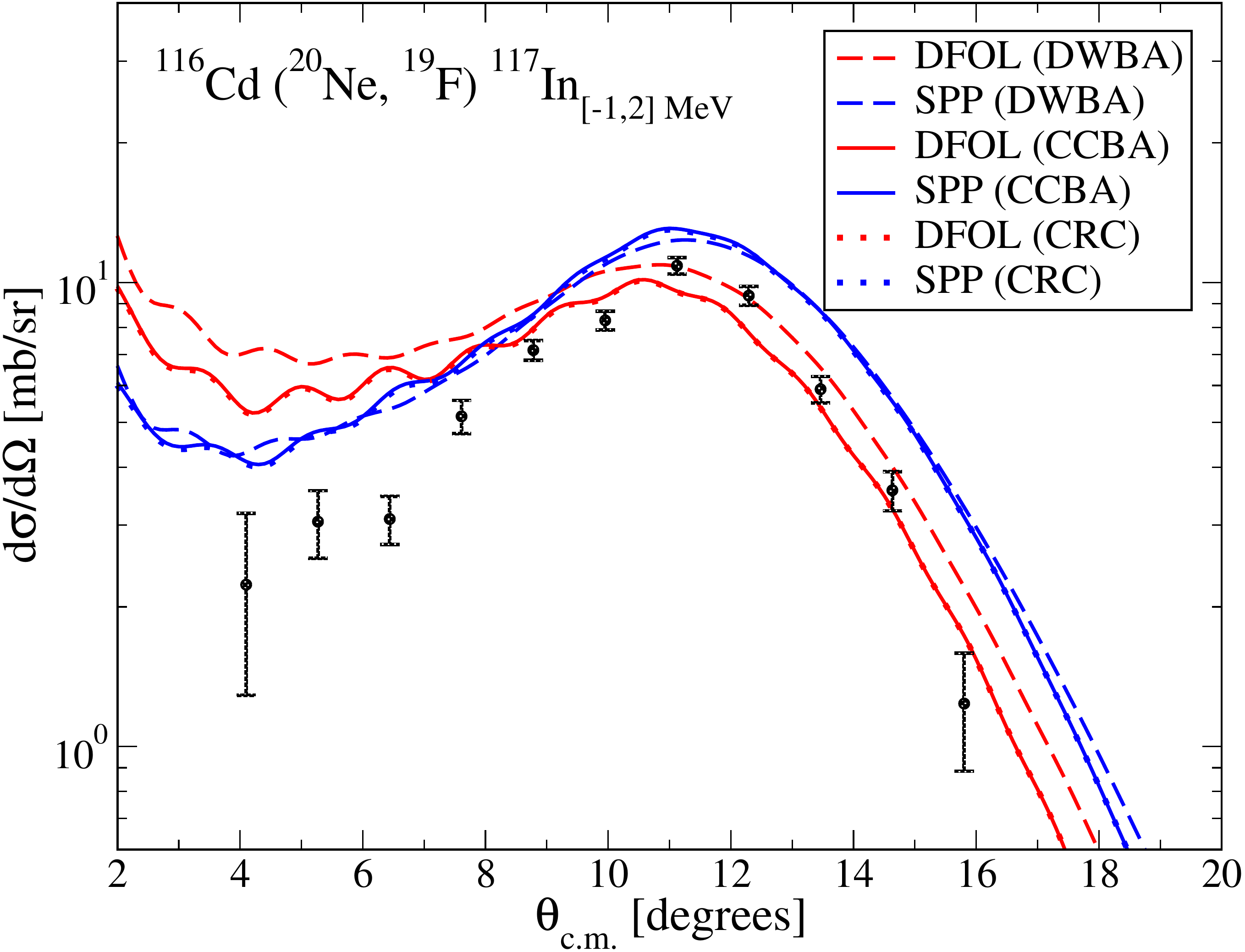}
\caption{Experimental angular distribution of the differential cross section for the one-proton transfer reaction $^{116}$Cd($^{20}$Ne,$^{19}$F)$^{117}$In, as obtained in the %[0.0, 1.0] 
[-1, 2] MeV excitation energy range. Theoretical results refer to three different reaction schemes: DWBA, CCBA and CRC. Both DFOL and SPP optical potentials are employed. %and experimental data from the NUMEN project are also reported.
}
\label{fig:1ptransfer_CRC_CCBA_DWBA}
\end{figure}

The excitation energy spectrum measured for the ($^{20}$Ne,$^{19}$F) one-proton transfer reaction is presented in Fig. \ref{fig:spe_onep}. In agreement with Brink’s transfer matching conditions \cite{BRINK197237}, the spectrum is peaked around $E_x^{opt} \sim$5 MeV. In order to extract the differential cross section angular distribution in the low excitation energy region, the spectra have been analyzed %\st{by a fitting procedure} 
in one-degree angular step. The inset of Fig. \ref{fig:spe_onep} shows the %\st{
%analysis performed %} \textcolor{magenta}{
experimental energy distribution %} 
in the 8° $< \theta_{lab} <$ 9° interval up to $\sim$ 4.5 MeV. %The cross section trend 
Countings at low excitation energy are %\textcolor{magenta}{
smoothly distributed %} \st{described by a %\textcolor{magenta}{smooth %} 
%function accounting for} \textcolor{magenta}{
due to %} 
the $^{117}$In and $^{19}$F high nuclear level densities %\textcolor{magenta}{, also not presenting} \st{the lack of} 
without exhibiting any clear structure in such region. In fact, even if the ground-state-to-ground-state transition as well as the first excited states cannot be %experimentally resolved \textcolor{magenta}{from the first excited states???Non mi sembra chiaro, forse toglierei... che ne pensi?} 
distinguished due to the limited %\textcolor{magenta}{
energy %} 
resolution, none of them dominates over the others thus resulting in a continuous %cross section 
yield trend. Beyond 1 MeV, instead, %\textcolor{magenta}{
some bumps are visible %} \st{three Gaussian functions %were are introduced %\textcolor{magenta}{around/at} at $\sim$ 1.5, $\sim$ 2.4 and $\sim$ 4 MeV to better reproduce the spectra, fitting some %observable observed bumps} which %, due to the level densities and the experimental energy resolution, 
which are expected to be the unresolved sum of several different transitions. %\st{Such fit procedure was introduced, in particular, to estimate the contributions resulting from the tails of the visible structures in the %\textcolor{magenta}{low %} excitation energy region, which could be non negligible.} \textcolor{magenta}{
Examining the inset of Fig. \ref{fig:spe_onep}, however, a local minimum is observed around 2 MeV. %} \textcolor{magenta}{Thus,} 
Then the differential cross section angular distribution in the %[0.0, 1.0] 
[-1, 2] MeV excitation energy range was %\st{thus} 
extracted and it is %reported 
shown in Fig. \ref{fig:1ptransfer_CRC_CCBA_DWBA}. We limited to the lowest excitation energy region, where the reproduction of the energy spectrum in terms of single-particle configurations adopted in the following calculations %might be considered 
is expected to be more reliable. It is interesting to note that the angular distribution shape results clearly peaked close to the grazing angle ($\theta_{gr} \sim$ 15°) region. This ``bell-shaped'' behaviour is the typical feature expected in absorptive reactions involving heavy nuclei \cite{PaePRC17, KahANP1977, Diana}. %}

Once the %projectile-target 
initial state interaction has been properly addressed, it would be important to check whether the theoretical framework adopted in our work allows to get a reliable reproduction also of the transfer channels. Direct reaction calculations for the transfer mechanism %are 
have been performed by employing EFR transfer coupling within, at first, the DWBA approximation. %Later, 
%Then, in analogy with 
Following the analysis performed in some recent works \cite{Diana}, it %would be 
is %worthwhile 
constructive to look also at the effect of the coupling with the inelastic excitations of low-lying states in the incoming partition, within CCBA or CRC formalism. 

%Despite of the reaction framework adopted, to 
To perform microscopic transfer reaction calculations, %in addition to the optical potential, %it is %important essential 
one needs to also know the spectroscopic amplitudes %concerning 
associated to the wave function single-particle overlaps. Such amplitudes are derived %, as it is usually done, 
through large scale shell model calculations performed by using the NuShellX code~\cite{NuShellX}.  

For projectile-like isotopes, the effective Zuker-Buck-McGrory ($zbm$) interaction~\cite{zbm} is considered, within a model space which assumes $^{12}$C as a closed core and valence protons and neutrons in the $1p_{1/2}$, $1d_{5/2}$ and $2s_{1/2}$ orbits. We also checked that similar results would be obtained when employing the $psdmod$~\cite{psdmod} interaction, which was also adopted in some recent works~\cite{CarPRC17, cardozo}.

For the target-like isotopes involved here, discrepancies are generally observed in reproducing the single particle spectrum. Nevertheless, the recently introduced effective interaction $88Sr45$~\cite{coraggio2016}, allows for a quite satisfying determination of the excitation energy of the low-lying states in this mass region%of these large systems
~\cite{Diana}. In that case, %once again 
$^{88}$Sr is assumed as a closed core within a model space which %assumes\textcolor{magenta}{-->che ne dici di considers?} 
considers valence protons in the $2d_{5/2}$, $2p_{1/2}$, $1g_{9/2}$ and $1g_{7/2}$ orbits and neutrons in the $2d_{5/2}$, $3s_{1/2}$, $1g_{7/2}$ and $2d_{3/2}$ orbits. 

\begin{figure}
\centering
\includegraphics[width=\columnwidth]{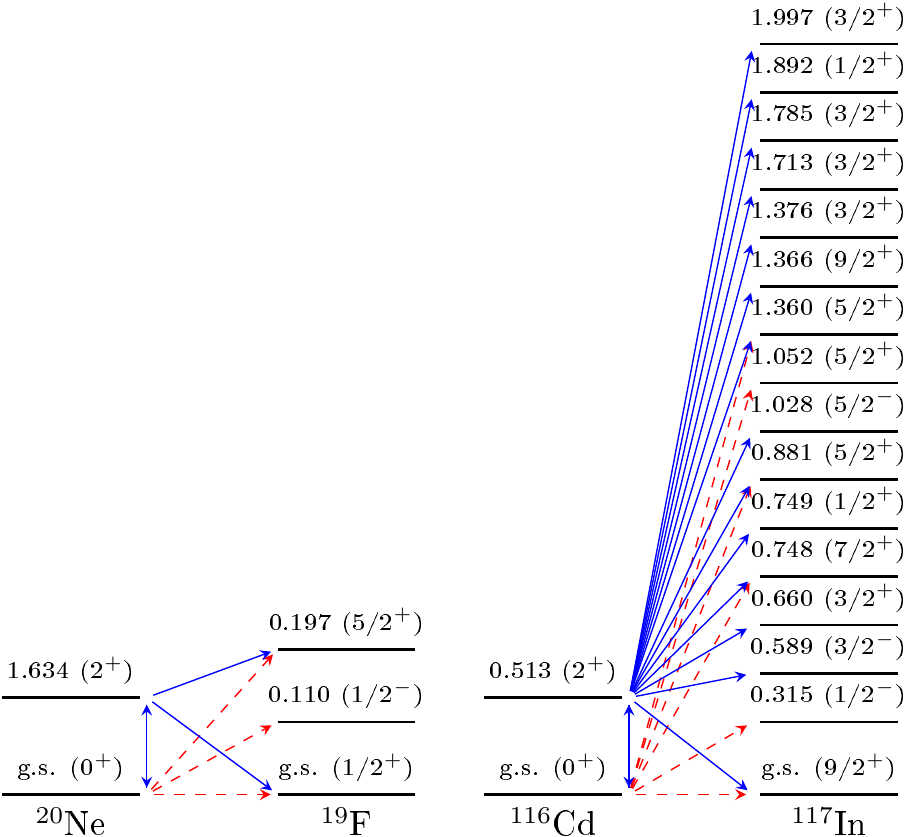} 
\caption{Coupling schemes of the projectile and target overlaps used in the calculations of the $^{116}$Cd($^{20}$Ne,$^{19}$F)$^{117}$In one-proton transfer reaction, within the considered excitation energy range. %[0.0, 1.0]. 
Red dashed arrows concern the coupling considered within the DWBA approximation, while the blue full arrows indicate the additional couplings introduced within the CCBA reaction scheme. In the CRC framework, not shown in the figure, all arrows have to be considered bidirectional (see text).}
\label{fig:coupling_scheme_1ptransfer}     
\end{figure}

The  couplings and level schemes %, as obtained when 
obtained considering $zbm$ and $88Sr45$ interactions for projectile-like and target-like nuclei, respectively, are sketched in Fig.~\ref{fig:coupling_scheme_1ptransfer}, while the spectroscopic amplitudes of the relevant overlaps involved are listed in Table~\ref{tab:1ptransfer}.

The single-particle wave functions %used in the matrix elements calculations 
representing the overlaps are generated by Woods-Saxon binding potentials, whose depths are varied to reproduce the experimental one-proton separation energies. The same procedure is adopted in the following %when considering 
for the one-neutron transfer, where the depths are varied in order to reproduce the experimental one-neutron separation energies. 
The %theoretical 
calculations have been performed by setting the reduced radii and diffuseness parameters of the binding potentials to 1.23 fm and 0.65 fm for %target-like cores 
$^{116}$Cd + p and to 1.26 fm and 0.7 fm for $^{19}$F + p. %projectile-like cores.  
Such values are compatible with the systematics usually adopted %return 
%\st{
and allow for %an %optimal accurate 
the reproduction of the experimental data %} 
\cite{Diana}. %, even though they slightly differ from the parameters of the target-like binding potentials considered in a recent work \cite{Diana}.

The angular distribution of the differential cross section for the %one-proton transfer reaction 
$^{116}$Cd($^{20}$Ne,$^{19}$F)$^{117}$In reaction is shown in Fig. \ref{fig:1ptransfer_CRC_CCBA_DWBA}, as obtained within three different theoretical approaches: %reaction schemes: 
DWBA, CCBA and CRC. In all cases, the post representation is %generally 
adopted and full complex remnant terms are %moreover 
considered. Moreover, the same prescriptions used in OM calculations are employed to determine the corresponding core-core potential. Both DFOL and SPP optical potentials are employed and %experimental data from the NUMEN project are also reported. 
the corresponding results are compared with the experimental data discussed above. 

Regardless of the reaction framework considered, one may observe that, while the DFOL potential tends to overestimate the contribution at smaller angles, the SPP potential returns a satisfactory agreement of the typical bell-shape exhibited by the experimental data around the grazing angle. %It is worthwhile to notice here that 
As shown by the figure, the shape of the angular distribution in the angular range covered by the experimental %values 
data is %strongly 
significantly dependent on the details of the optical potentials, especially in the surface region. From a qualitative point of view, we verified that the same shape would be preserved indeed even when performing the calculations by assuming unitary values for the spectroscopic amplitudes for all the states considered in our coupling scheme of Fig. \ref{fig:coupling_scheme_1ptransfer} or when assuming a different effective interaction in our shell model calculations. 
So, the result of Fig. \ref{fig:1ptransfer_CRC_CCBA_DWBA} confirms the reliability of the potentials adopted to describe the projectile-target interaction, %at least when the SPP potential is concerned.
especially for the SPP parameterization. Nevertheless, it provides some hints to distinguish among different options for the optical potentials, which equally well reproduce the experimental results in the elastic channel, supporting the preferable choice of the SPP potential.

Moreover, %the reproduction 
not only %of 
the shape but also the reproduction of the %\st{absolute value} \textcolor{magenta}{
order of magnitude %} 
of the differential cross section is observed in Fig. \ref{fig:1ptransfer_CRC_CCBA_DWBA}. %\textcolor{magenta}{
Such result, reached %} 
without the need of %to resort to 
any arbitrary scaling factor or any significant modification of the structure parameters of the binding potentials, demonstrates the trustworthiness of the model space and of the effective interaction employed in our shell model calculations, despite some existing and hard to quantify uncertainties. 

\begin{table}[h]
\centering
\caption{\label{tab:1ptransfer}One-proton spectroscopic amplitudes (S.A.) adopted in DWBA and CCBA calculation of the $^{116}$Cd($^{20}$Ne,$^{19}$F)$^{117}$In one-proton transfer reaction. These amplitudes are obtained through shell model calculations by employing $zbm$ interaction for projectile-like nuclei and $88Sr45$ interaction for target-like ones. The column nl$_j$ indicates the principal quantum number, the orbital and the total angular momentum of the single valence proton, respectively, through the usual spectroscopic notation.}
\begin{tabular}{*{4}{c}}
\textbf{Initial state} & \textbf{Final state} & \textbf{nl$_j$} & \textbf{S.A.} \\
\toprule
\multirow{3}*{$^{20}$Ne$_{\textup{g.s.}}$ (0$^{+}$)} &  $^{19}$F$_{\textup{g.s.}}$ (1/2$^{+}$)  &  (2s$_{1/2}$)  &  -0.8584  \\
\cline{2-4}
					   	     &  $^{19}$F$_{0.110}$ (1/2$^{-}$)          &  (1p$_{1/2}$)  &  -1.2702  \\
\cline{2-4}
					   	     &  $^{19}$F$_{0.197}$ (5/2$^{+}$)  	&  (1d$_{5/2}$)  &   1.1741  \\
\midrule
\multirow{3}*{$^{20}$Ne$_{1.634}$ (2$^{+}$)} &  $^{19}$F$_{\textup{g.s.}}$ (1/2$^{+}$)          &  (1d$_{5/2}$)  &   0.6712  \\
\cline{2-4}
					     &  \multirow{2}*{$^{19}$F$_{0.197}$ (5/2$^{+}$)}   &  (2s$_{1/2}$)  &  -0.6922  \\
					     & 							&  (1d$_{5/2}$)  &  -0.6416  \\
\midrule
\midrule
\multirow{6}*{$^{116}$Cd$_{\textup{g.s.}}$ (0$^{+}$)} &  $^{117}$In$_{\textup{g.s.}}$ (9/2$^{+}$)  &  (1g$_{9/2}$)  &  -0.4066  \\
\cline{2-4}
					              &  $^{117}$In$_{0.315}$ (1/2$^{-}$) 	   &  (2p$_{1/2}$)  &  -0.3847  \\
\cline{2-4}
						      &  $^{117}$In$_{0.748}$ (7/2$^{+}$) 	   &  (1g$_{7/2}$)  &   0.0098  \\
\cline{2-4}
						      &  $^{117}$In$_{0.881}$ (5/2$^{+}$) 	   &  (2d$_{5/2}$)  &  -0.1593  \\
\cline{2-4}
						      &  $^{117}$In$_{1.052}$ (5/2$^{+}$) 	   &  (2d$_{5/2}$)  &  -0.2670  \\
\cline{2-4}
						      &  $^{117}$In$_{1.360}$ (5/2$^{+}$) 	   &  (2d$_{5/2}$)  &  -0.7192  \\
\midrule
\multirow{23}*{$^{116}$Cd$_{0.513}$ (2$^{+}$)} &  \multirow{3}*{$^{117}$In$_{\textup{g.s.}}$ (9/2$^{+}$)}  &  (1g$_{9/2}$)  &  -0.6092  \\
					       & 	 						   &  (1g$_{7/2}$)  &   0.0073  \\
					       &  							   &  (2d$_{5/2}$)  &   0.1182  \\
\cline{2-4}
					       &  $^{117}$In$_{0.589}$ (3/2$^{-}$) 			   &  (2p$_{3/2}$)  &  -0.3319  \\
\cline{2-4}
					       &  \multirow{2}*{$^{117}$In$_{0.660}$ (3/2$^{+}$)} 	   &  (1g$_{7/2}$)  &  -0.0157  \\
					       & 	 						   &  (2d$_{5/2}$)  &  -0.1502  \\

\cline{2-4}
					       &  \multirow{3}*{$^{117}$In$_{0.748}$ (7/2$^{+}$)}  	   &  (1g$_{9/2}$)  &  -0.2232  \\
					       & 	 						   &  (1g$_{7/2}$)  &   0.0170  \\
					       &  							   &  (2d$_{5/2}$)  &   0.0674  \\
\cline{2-4}
					       &  $^{117}$In$_{0.749}$ (1/2$^{+}$) 			   &  (2d$_{5/2}$)  &  -0.2796  \\
\cline{2-4}
					       &  \multirow{3}*{$^{117}$In$_{0.881}$ (5/2$^{+}$)} 	   &  (1g$_{9/2}$)  &   0.4308  \\
					       & 	 						   &  (1g$_{7/2}$)  &   0.0201  \\
					       &  							   &  (2d$_{5/2}$)  &  -0.1781  \\
\cline{2-4}
					       &  $^{117}$In$_{1.028}$ (5/2$^{-}$)		 	   &  (2p$_{1/2}$)  &   0.3306  \\
\cline{2-4}
					       &  \multirow{2}*{$^{117}$In$_{1.052}$ (5/2$^{+}$)} 	   &  (2d$_{5/2}$)  &   0.5693  \\
 					       & 	 						   &  (1g$_{9/2}$)  &  -0.1326  \\
\cline{2-4}
					       &  $^{117}$In$_{1.360}$ (5/2$^{+}$)		 	   &  (1g$_{9/2}$)  &  -0.1813  \\
\cline{2-4}
					       &  $^{117}$In$_{1.366}$ (9/2$^{+}$) 	   		   &  (2d$_{5/2}$)  &  -0.4463  \\
\cline{2-4}
					       &  $^{117}$In$_{1.376}$ (3/2$^{+}$) 	  		   &  (2d$_{5/2}$)  &  -0.2946  \\
\cline{2-4}
					       &  $^{117}$In$_{1.713}$ (3/2$^{+}$)		 	   &  (2d$_{5/2}$)  &  -0.1521  \\
\cline{2-4}
					       &  $^{117}$In$_{1.785}$ (3/2$^{+}$)		 	   &  (2d$_{5/2}$)  &   0.1214  \\
\cline{2-4}
					       &  \multirow{2}*{$^{117}$In$_{1.892}$ (1/2$^{+}$)} 	   &  (2d$_{5/2}$)  &  -0.6646  \\
					       & 	 						   &  (1g$_{7/2}$)  &   0.1382  \\
\bottomrule
\end{tabular}
\end{table}

In Fig.~\ref{fig:1ptransfer_CRC_CCBA_DWBA}, the effect of the coupling with inelastic states in the entrance partition %is also emphasized within 
can be inferred from the comparison between the different reaction schemes adopted. Regardless of the optical potential employed, some general features may be discussed. Indeed, similarly to what was %it has been 
observed in our recent work~\cite{Diana}, the inclusion of couplings with low-lying collective inelastic excitations %in the system of coupled equations 
moderately changes the shape of the angular distribution. In particular, it slightly modifies the oscillating behavior of the transfer cross section at small angles, %\textcolor{magenta}{. This is} 
owing to the different orbital angular momentum transferred, when passing through the intermediate states involved in the inelastic excitations. %\textcolor{magenta}{
%which %} 
%introduces 
Moreover, a tiny shift is introduced on the bell-shaped peak, due to %around the peak, in view of 
the larger spatial distribution of the collective wave functions adopted in the coupled channel calculations. %wave function of these excited states involved. 
As would be expected, a larger cross section %an increase 
above the grazing angle is %in particular 
observed in CCBA compared to DWBA. %when comparing the corresponding CCBA calculations with respect to the %previous 
%DWBA ones. 
Such behavior is attributed to couplings with the target 2$_1^{+}$ state, as one observes selectively switching off the contributions of the %va\textcolor{magenta}{(r}ious 
various inelastic excitations. %For what it concerns the coupling with the projectile 2$^{+}$ state, 
As far as the coupling with the projectile $2_1^+$ state is concerned, its inclusion does not imply any change in the shape of the differential cross section, except only for a small global reduction, which suggests a destructive interference with the 0$^{+}$ state contribution. In Fig.~\ref{fig:1ptransfer_CRC_CCBA_DWBA}, CRC results are also displayed. Whatever the optical potential considered, the curve related to CRC calculation is always practically superimposed on the one corresponding to the CCBA results, in the explored angular region. We remind here that, in CRC calculations, the couplings related to the single-nucleon transfer transitions are implemented iteratively, %In all the upcoming theoretical calcu-lations by fresco we have used iterative method for fullCRC calculations 
till the absolute difference between successive S-matrix elements becomes less than 0.01\%, so even including the back-coupling. %at all orders. 
Therefore, our result demonstrates that accounting for the effect of the latter is safely negligible and this holds not only for the transfer but also for both elastic and inelastic channels. %\textcolor{magenta}{da dove si vede? mostriamo l'elastico che viene dal transfer?!} 
%and transfer channels. 
Indeed, in all cases the difference between CRC and CCBA calculations on the total cross section is smaller than 0.2\%.

On the other hand, the small differences found between DWBA and CCBA calculations %the small differences found they can be effectively incorporated in the transition operators. This is especially true when looking..... allows one to justify %thus 
%the use of simplified coupling schemes when studying such a reaction. At the same time, it demonstrates %that the importance of these couplings turns out to be rather reduced, when focusing
demonstrate a mild influence of the high-order couplings between transfer and scattering channels %, which are typical of CRC calculations, 
at least in the present angular window. % we are looking at. 
%As a result, %taking explicitly into account 
The role of the inelastic excitations might be thus disregarded at a first level of accuracy or effectively embedded in the dynamic polarization potential. %transition operators. 
This is especially true when looking at other quantities, such as the (angular-integrated) total cross section, which is obviously less sensitive to the details of the diffraction pattern. 

To summarize, Figs.~\ref{fig:elastic} and \ref{fig:1ptransfer_CRC_CCBA_DWBA} show that %It is shown that, 
with a unified choice of the optical potential and with a rather simple coupling scheme, we are able to reproduce at once the angular distribution of both the quasi-elastic and one-proton transfer channel.  Of course, one might in principle consider more sophisticated reaction frameworks or approximations and more complex coupling schemes. %, such as %again CRC or 
%continuum discretized coupled channels (CCDC). N
Nevertheless, since this simplified approach already demonstrates an acceptable agreement with the experimental %values, 
data, resorting to more complex reaction schemes reveals unnecessary, as long as one does not want to describe more channels. Moreover, it allows to test the basic ingredients of our calculations, validating our choice of the optical potential adopted, as well as the model space and interaction combination considered in the %structure 
shell model calculations.

\subsection{Contribution of sequential neutron-proton transfer on the SCE}
\begin{figure}[t]
    \centering
    \includegraphics[width=\columnwidth]{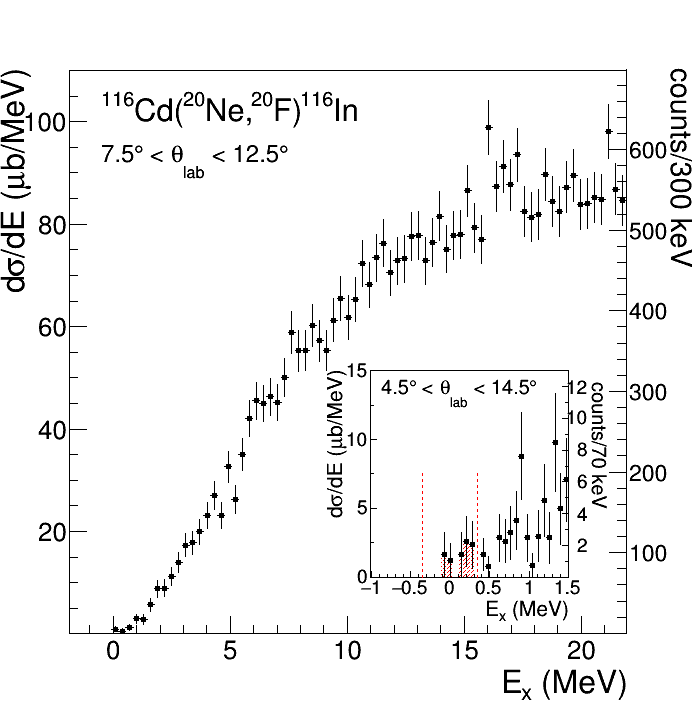}  %scex_aylin_diana
    \caption{Excitation energy cross section spectrum of the single charge exchange channel in the 7.5° $< \theta_{lab} <$ 12.5° angular range. In the inset, the spectrum up to 1.5 MeV in the full angular interval 4.5° $< \theta_{lab} <$ 14.5° is reported. The dashed lines and the red hatched area highlight the selected integration energy region discussed in the text.}
    \label{fig:spe_sce}
\end{figure}

The ($^{20}$Ne,$^{20}$F) %single charge exchange 
SCE reaction absolute cross section energy spectrum is shown in Fig. \ref{fig:spe_sce}. In the inset, a zoom for the excitation energy region up to 1.5 MeV is reported. The continuum shape is a %cooperative 
combined effect of the high level density of the involved ejectile and residual nuclei and the limited energy resolution. Such conditions, %joined to 
together with the poor statistics, do not allow further experimental analysis. Therefore, cross section values %can be 
were extracted only by integrating on selected energy ranges, thus including the contributions from transitions to several states of the final partition. In order to compare the experimental results with the theoretical ones a small excitation energy interval has been considered, namely the one between %\textcolor{magenta}{
-0.35 %} 
and 0.35 MeV. The list of the expected states in the selected energy interval for the ejectile and the residual nuclei and the corresponding experimental integrated cross section value are %reported 
listed in Tab. \ref{tab:20Ne116Cd_integ_cross_section}. %\ref{tab:sce_states}. %}

In the present work, we focus on the competing multi-nucleon transfer mechanisms which end up in the same exit channel as the one reached in the direct SCE process. %, which is not %here addressed 
%addressed here. %reaction.
Hence, in our approach, all excited states in either the projectile- or the target-like outgoing nuclei is thus populated %without any direct isovector excitation, so only resorting to a 
only through the proper rearrangement of neutrons and protons.

\begin{figure}[t]
\centering
\includegraphics[width=\columnwidth]{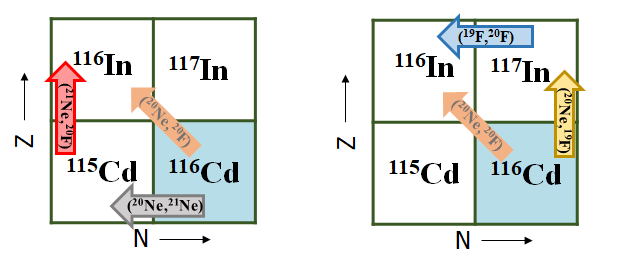}  %scex_aylin_diana
\caption{Schematic representation of the sequential neutron-proton transfer processes, discussed before in Section II. %on transfer mechanisms considered here, in competition with the ($^{20}$Ne, $^{20}$F) SCE channel, not addressed in the present work. 
The one-neutron pick-up + one-proton stripping (path1, Eq. \eqref{eq:np1}) and the one-proton stripping + one-neutron pick-up (path2, Eq. \eqref{eq:np2}) routes are depicted on the left and right, respectively. For completeness, also the direct SCE channel is indicated.}
\label{fig:paths}
\end{figure}

%As a difference with respect to the direct SCE reaction, %which is not %here addressed addressed here, 
%in the sequential transfer mechanism, the two nucleons are transferred one by one, through two different intermediate partitions: ($^{21}$Ne + $^{115}$Cd) or ($^{19}$F + $^{117}$In), i.e. the same final states may be reached by performing either a one-neutron pick-up reaction followed by a one-proton stripping (hereafter labeled as path1) or viceversa (hereafter labeled as path2). 
A schematic representation of the two different paths considered is sketched in Fig. \ref{fig:paths}: the same final states may then be reached by performing either a one-neutron pick-up reaction followed by a one-proton stripping (hereafter labeled as path1) or viceversa (hereafter labeled as path2). 

In the following, the two different reaction paths will be first considered separately and thereafter they will be included within the same scheme, in order to evaluate the interference between the two processes. The prior-post representation is %generally 
used and full complex remnant terms are %moreover 
considered. 

Before embarking in the full calculations, %it would be nice to check the reliability of our numerical results, for example by checking 
we also checked the equivalence between the different (prior/post) representations adopted, once the non-orthogonality terms are included \cite{ThomBCS2013}. %For sake of simplicity, we show here the results just for one specific combination of projectile-like and target-like states of the final partition, namely $^{20}$F$_{\textup{g.s.}}$ (2$^{+}$) and $^{116}$In$_{0.223}$ (4$^{+}$) states, as obtained considering the coherent sum of the contributions 
%along the two different paths described above.
We compare thus the prior-post calculation for which non-orthogonality terms identically vanish \cite{Sat83}, with two other possible combinations, where non-orthogonality corrections are included. As a result, %Fig.~\ref{fig:20Ne116Cd_1nptransfer_path12_testpriorpost} shows 
%when looking at the ratio between the angle differential cross sections, as obtained under the prior-prior or the post-post representation with respect to the prior-post one, a 
a very nice agreement is generally observed. Such an agreement, even though is never worse than 3\%, slightly deteriorates for larger angles (beyond 15 degrees), where however the differential cross section has a steep decrease and its contribution to the total integrated value practically vanishes. Such a result confirms in turn the good numerical convergence of our calculations.
In the second transfer
steps of path1 and path2, contributions 
of spectroscopic strength less than 0.1 are neglected.

%\begin{figure}[t]
%\includegraphics[width=\columnwidth]{./figures/figures_new/20Ne116Cd_1nptransfer_DWBA_4+_SPP_testpriorpost_ratio.eps}
%\caption{Ratio R (see text) for the $^{116}$Cd ($^{20}$Ne, $^{20}$F$_{\textup{g.s.}}$ (2$^{+}$)) $^{116}$In$_{0.223}$ (4$^{+}$) SCE reaction differential cross section obtained through different prior/post combinations even including, whenever non-vanishing, the corresponding non-orthogonality terms.  
%The SCE reaction is described in terms of a two-step transfer mechanism, by assuming the DWBA approximation. The SPP optical potential is employed and only the coherent sum of the two paths is plotted.} 
%\label{fig:20Ne116Cd_1nptransfer_path12_testpriorpost}
%\end{figure}

\begin{figure*}[t]
\begin{center}
    \begin{tikzpicture}[
      scale=0.7,
      level/.style={thick},
      virtual/.style={thick,densely dashed},
      trans/.style={red,thin,dashed,->,shorten >=2pt,shorten <=2pt,>=stealth},
      trans2/.style={blue,thin,->,shorten >=2pt,shorten <=2pt,>=stealth},
      classical/.style={thin,double,<->,shorten >=8pt,shorten <=8pt,>=stealth}
    ]
    % Draw the energy levels.
    \draw[level] (0cm,-2em) -- (2cm,-2em) node[midway,below] {$^{20}$Ne} node[midway,above] {\scriptsize g.s. (0$^{+}$)};
    \draw[level] (0cm,2em) -- (2cm,2em) node[midway,above] {\scriptsize 1.634 (2$^{+}$)};
    \draw[level, white] (4cm,-2em) -- (6cm,-2em) node[midway,below, black] {$^{21}$Ne};
    \draw[level] (4cm,-1em) -- (6cm,-1em) node[midway,above] {\scriptsize g.s. (3/2$^{+}$)};
    \draw[level] (4cm,1em) -- (6cm,1em) node[midway,above] {\scriptsize 0.351 (5/2$^{+}$)};
    \draw[level] (4cm,3em) -- (6cm,3em) node[midway,above] {\scriptsize 1.746 (7/2$^{+}$)};
%	\draw[level] (4cm,3em) -- (6cm,3em) node[midway,above] {\scriptsize 2.789 (1/2$^{-}$)};
	\draw[level] (4cm,5em) -- (6cm,5em) node[midway,above] {\scriptsize 2.794 (1/2$^{+}$)};
	\draw[level] (4cm,7em) -- (6cm,7em) node[midway,above] {\scriptsize 3.736 (5/2$^{+}$)};
    \draw[level] (8cm,-2em) -- (10cm,-2em) node[midway,below] {$^{20}$F} node[midway,above] {\scriptsize g.s. (2$^{+}$)};
%    \draw[trans] (2.1cm,-2em) -- (4cm,-1em) node[midway,left] {};\draw[trans] (6.1cm,-1em) -- (8cm,-2em) node[midway,right] {};
    \draw[trans2] (2.1cm, 2em) -- (4cm,-1em) node[midway,left] {};\draw[trans2] (6.1cm,-1em) -- (8cm,-2em) node[midway,right] {};
    \draw[trans2] (2.1cm,2em) -- (4cm, 1em) node[midway,left] {};	
    \draw[trans] (2.1cm,-2em) -- (4cm, 1em) node[midway,left] {};\draw[trans2] (6.1cm,1em) -- (8cm,-2em) node[midway,right] {};\draw[trans] (6.1cm,1em) -- (8cm,-2em) node[midway,right] {};
%    \draw[trans] (2.1cm,-2em) -- (4cm,3em) node[midway,left] {};\draw[trans] (6.1cm,3em) -- (8cm,-2em) node[midway,right] {};
    \draw[trans2] (2.1cm, 2em) -- (4cm,3em) node[midway,left] {};\draw[trans2] (6.1cm,3em) -- (8cm, -2em) node[midway,right] {};
    \draw[trans2] (2.1cm, 2em) -- (4cm,5em) node[midway,left] {};       
    \draw[trans2] (2.1cm, 2em) -- (4cm,7em) node[midway,left] {};
    \draw[trans] (2.1cm,-2em) -- (4cm,5em) node[midway,left] {};\draw[trans2] (6.1cm,5em) -- (8cm,-2em) node[midway,right] {};\draw[trans] (6.1cm,5em) -- (8cm,-2em) node[midway,right] {};
    \draw[trans] (2.1cm,-2em) -- (4cm,7em) node[midway,left] {};\draw[trans2] (6.1cm,7em) -- (8cm,-2em) node[midway,right] {};\draw[trans] (6.1cm,7em) -- (8cm,-2em) node[midway,right] {};
    \draw[trans2] (2.1cm,-2em) -- (2.1cm,2em) node[midway,left] {};%\draw[trans] (6.1cm,4em) -- (8cm,-2em) node[midway,right] {};
    \draw[trans2] (2.1cm,2em) -- (2.1cm,-2em) node[midway,left] {};%\draw[trans] (6.1cm,4em) -- (8cm,-2em) node[midway,right] {};

    \draw[level] (12cm,-2em) -- (14cm,-2em) node[midway,below] {$^{116}$Cd} node[midway,above] {\scriptsize g.s. (0$^{+}$)};
    \draw[level] (12cm, 2em) -- (14cm, 2em) node[midway,above] {\scriptsize 0.513 (2$^{+}$)};
    \draw[level] (16cm,-2em) -- (18cm,-2em) node[midway,below] {$^{115}$Cd} node[midway,above] {\scriptsize g.s. (1/2$^{+}$)};
	\draw[level] (16cm,0em) -- (18cm,0em) node[midway,above] {\scriptsize 0.229 (3/2$^{+}$)};
	\draw[level] (16cm,2em) -- (18cm,2em) node[midway,above] {\scriptsize 0.361 (5/2$^{+}$)};
	\draw[level] (16cm,4em) -- (18cm,4em) node[midway,above] {\scriptsize 0.389 (7/2$^{+}$)};
	\draw[level] (16cm,6em) -- (18cm,6em) node[midway,above] {\scriptsize 0.389 (9/2$^{+}$)};
	\draw[level] (16cm,8em) -- (18cm,8em) node[midway,above] {\scriptsize 0.473 (3/2$^{+}$)};
	\draw[level] (16cm,10em) -- (18cm,10em) node[midway,above] {\scriptsize 0.473 (5/2$^{+}$)};
	\draw[level] (16cm,12em) -- (18cm,12em) node[midway,above] {\scriptsize 0.507 (3/2$^{+}$)};
	\draw[level] (16cm,14em) -- (18cm,14em) node[midway,above] {\scriptsize 0.507 (5/2$^{+}$)};
	\draw[level] (16cm,16em) -- (18cm,16em) node[midway,above] {\scriptsize 0.649 (1/2$^{+}$)};
	\draw[level] (16cm,18em) -- (18cm,18em) node[midway,above] {\scriptsize 1.062 (7/2$^{+}$)};
   \draw[level] (20cm,-2em) -- (22cm,-2em) node[midway,below] {$^{116}$In} node[midway,above] {\scriptsize g.s. (1$^{+}$)};
    \draw[level] (20cm,0em) -- (22cm,0em) node[midway,above] {\scriptsize 0.128 (5$^{+}$)};
	\draw[level] (20cm,2em) -- (22cm,2em) node[midway,above] {\scriptsize 0.223 (4$^{+}$)};
	\draw[level] (20cm,4em) -- (22cm,4em) node[midway,above] {\scriptsize 0.273 (2$^{+}$)};
	\draw[level] (20cm,6em) -- (22cm,6em) node[midway,above] {\scriptsize 0.313 (4$^{+}$)};
    \draw[trans2] (14.1cm,-2em) -- (14.1cm,2em) node[midway,left] {};%\draw[trans] (6.1cm,4em) -- (8cm,-2em) node[midway,right] {};
    \draw[trans2] (14.1cm,2em) -- (14.1cm,-2em) node[midway,left] {};%\draw[trans] (6.1cm,4em) -- (8cm,-2em) node[midway,right] {};

%    \draw[trans2] (14.1cm,2em) -- (16cm,-2em) node[midway,left] {};							         \draw[trans2] (18.1cm,-2em) -- (20cm,0em) node[midway,right] {}; 
%    \draw[trans2] (18.1cm,-2em) -- (20cm,2em) node[midway,right] {};\draw[trans2] (18.1cm,-2em) -- (20cm,4em) node[midway,right] {};\draw[trans2] (18.1cm,-2em) -- (20cm,6em) node[midway,right] {};

    \draw[trans2] (14.1cm,2em) -- (16cm,0em) node[midway,left] {};\draw[trans2] (18.1cm,0em) -- (20cm,-2em) node[midway,right] {};\draw[trans2] (18.1cm,0em) -- (20cm,0em) node[midway,right] {};
    \draw[trans2] (18.1cm,0em) -- (20cm,2em) node[midway,right] {};

    \draw[trans2] (14.1cm,2em) -- (16cm,2em) node[midway,left] {};								  \draw[trans2] (18.1cm,2em) -- (20cm,0em) node[midway,right] {};
    \draw[trans2] (18.1cm,2em) -- (20cm,2em) node[midway,right] {};\draw[trans2] (18.1cm,2em) -- (20cm,4em) node[midway,right] {};

    \draw[trans2] (14.1cm,2em) -- (16cm,4em) node[midway,left] {};\draw[trans2] (18.1cm,4em) -- (20cm,-2em) node[midway,right] {};\draw[trans2] (18.1cm,4em) -- (20cm,0em) node[midway,right] {};
    \draw[trans2] (18.1cm,4em) -- (20cm,2em) node[midway,right] {};\draw[trans2] (18.1cm,4em) -- (20cm,4em) node[midway,right] {};

    \draw[trans2] (14.1cm,2em) -- (16cm,6em) node[midway,left] {};\draw[trans2] (18.1cm,4em) -- (20cm,-2em) node[midway,right] {};\draw[trans2] (18.1cm,6em) -- (20cm,0em) node[midway,right] {};
    \draw[trans2] (18.1cm,6em) -- (20cm,2em) node[midway,right] {};\draw[trans2] (18.1cm,4em) -- (20cm,4em) node[midway,right] {};

    \draw[trans2] (14.1cm,2em) -- (16cm,8em) node[midway,left] {};\draw[trans2] (18.1cm,6em) -- (20cm,-2em) node[midway,right] {};\draw[trans2] (18.1cm,8em) -- (20cm,0em) node[midway,right] {};
    \draw[trans2] (18.1cm,8em) -- (20cm,2em) node[midway,right] {};\draw[trans2] (18.1cm,6em) -- (20cm,4em) node[midway,right] {};\draw[trans2] (18.1cm,8em) -- (20cm,6em) node[midway,right] {};

    \draw[trans2] (14.1cm,2em) -- (16cm,10em) node[midway,left] {};\draw[trans2] (18.1cm,8em) -- (20cm,-2em) node[midway,right] {};\draw[trans2] (18.1cm,10em) -- (20cm,0em) node[midway,right] {};
    \draw[trans2] (18.1cm,10em) -- (20cm,2em) node[midway,right] {};\draw[trans2] (18.1cm,8em) -- (20cm,4em) node[midway,right] {};

    \draw[trans2] (14.1cm,2em) -- (16cm,12em) node[midway,left] {};\draw[trans2] (18.1cm,10em) -- (20cm,-2em) node[midway,right] {};\draw[trans2] (18.1cm,12em) -- (20cm,0em) node[midway,right] {};

    \draw[trans2] (14.1cm,2em) -- (16cm,14em) node[midway,left] {};								   \draw[trans2] (18.1cm,14em) -- (20cm,0em) node[midway,right] {};
    \draw[trans2] (18.1cm,14em) -- (20cm,2em) node[midway,right] {};\draw[trans2] (18.1cm,12em) -- (20cm,4em) node[midway,right] {};

    \draw[trans2] (14.1cm,2em) -- (16cm,16em) node[midway,left] {};								    \draw[trans2] (18.1cm,16em) -- (20cm,0em) node[midway,right] {};
    \draw[trans2] (18.1cm,16em) -- (20cm,2em) node[midway,right] {};								    \draw[trans2] (18.1cm,16em) -- (20cm,6em) node[midway,right] {};

    \draw[trans2] (14.1cm,2em) -- (16cm,18em) node[midway,left] {};\draw[trans2] (18.1cm,16em) -- (20cm,-2em) node[midway,right] {};\draw[trans2] (18.1cm,18em) -- (20cm,0em) node[midway,right] {};
    \draw[trans2] (18.1cm,18em) -- (20cm,2em) node[midway,right] {};\draw[trans2] (18.1cm,14em) -- (20cm,4em) node[midway,right] {};

    \draw[trans] (14.1cm,-2em) -- (16cm,-2em) node[midway,left] {};							         \draw[trans] (18.1cm,-2em) -- (20cm,0em) node[midway,right] {}; 
    \draw[trans] (18.1cm,-2em) -- (20cm,2em) node[midway,right] {};								 \draw[trans] (18.1cm,-2em) -- (20cm,6em) node[midway,right] {};

    \draw[trans] (14.1cm,-2em) -- (16cm,0em) node[midway,left] {};\draw[trans] (18.1cm,0em) -- (20cm,-2em) node[midway,right] {};\draw[trans] (18.1cm,0em) -- (20cm,0em) node[midway,right] {};
    \draw[trans] (18.1cm,0em) -- (20cm,2em) node[midway,right] {};

    \draw[trans] (14.1cm,-2em) -- (16cm,2em) node[midway,left] {};								\draw[trans] (18.1cm,2em) -- (20cm,0em) node[midway,right] {};
    \draw[trans] (18.1cm,2em) -- (20cm,2em) node[midway,right] {};\draw[trans] (18.1cm,2em) -- (20cm,4em) node[midway,right] {};

    \draw[trans] (14.1cm,-2em) -- (16cm,4em) node[midway,left] {};\draw[trans] (18.1cm,4em) -- (20cm,-2em) node[midway,right] {};\draw[trans] (18.1cm,4em) -- (20cm,0em) node[midway,right] {};
    \draw[trans] (18.1cm,4em) -- (20cm,2em) node[midway,right] {};\draw[trans] (18.1cm,4em) -- (20cm,4em) node[midway,right] {};

%    								  \draw[trans] (18.1cm,6em) -- (20cm,-2em) node[midway,right] {};\draw[trans] (18.1cm,6em) -- (20cm,0em) node[midway,right] {};
%    \draw[trans] (18.1cm,6em) -- (20cm,2em) node[midway,right] {};\draw[trans] (18.1cm,6em) -- (20cm,4em) node[midway,right] {};

    \draw[trans] (14.1cm,-2em) -- (16cm,8em) node[midway,left] {};\draw[trans] (18.1cm,8em) -- (20cm,-2em) node[midway,right] {};\draw[trans] (18.1cm,8em) -- (20cm,0em) node[midway,right] {};
    \draw[trans] (18.1cm,8em) -- (20cm,2em) node[midway,right] {};\draw[trans] (18.1cm,8em) -- (20cm,4em) node[midway,right] {};\draw[trans] (18.1cm,8em) -- (20cm,6em) node[midway,right] {};

    \draw[trans] (14.1cm,-2em) -- (16cm,10em) node[midway,left] {};\draw[trans] (18.1cm,10em) -- (20cm,-2em) node[midway,right] {};\draw[trans] (18.1cm,10em) -- (20cm,0em) node[midway,right] {};
    \draw[trans] (18.1cm,10em) -- (20cm,2em) node[midway,right] {};\draw[trans] (18.1cm,10em) -- (20cm,4em) node[midway,right] {};

    \draw[trans] (14.1cm,-2em) -- (16cm,12em) node[midway,left] {};\draw[trans] (18.1cm,12em) -- (20cm,-2em) node[midway,right] {};\draw[trans] (18.1cm,12em) -- (20cm,0em) node[midway,right] {};

    \draw[trans] (14.1cm,-2em) -- (16cm,14em) node[midway,left] {};								  \draw[trans] (18.1cm,14em) -- (20cm,0em) node[midway,right] {};
    \draw[trans] (18.1cm,14em) -- (20cm,2em) node[midway,right] {};\draw[trans] (18.1cm,14em) -- (20cm,4em) node[midway,right] {};

    \draw[trans] (14.1cm,-2em) -- (16cm,16em) node[midway,left] {};							         \draw[trans] (18.1cm,16em) -- (20cm,0em) node[midway,right] {};
    \draw[trans] (18.1cm,16em) -- (20cm,2em) node[midway,right] {};								 \draw[trans] (18.1cm,16em) -- (20cm,6em) node[midway,right] {};

%    \draw[trans] (14.1cm,-2em) -- (16cm,18em) node[midway,left] {};							            \draw[trans] (18.1cm,18em) -- (20cm,0em) node[midway,right] {};
%    \draw[trans] (18.1cm,18em) -- (20cm,2em) node[midway,right] {};\draw[trans] (18.1cm,18em) -- (20cm,4em) node[midway,right] {};\draw[trans] (18.1cm,18em) -- (20cm,6em) node[midway,right] {};

    \end{tikzpicture}

\caption{Coupling schemes of the projectile and target overlaps used in the sequential two-nucleon transfer reaction calculations, along path 1. The $zbm$ and $88Sr45$ interactions are employed in shell model calculations, for projectile-like and target-like nuclei, respectively. Red arrows concern the coupling considered within the DWBA approximation, while the blue arrows indicate the additional couplings introduced within the CCBA reaction scheme. }
\label{fig:coupling_scheme_path1} 
\end{center}
\end{figure*}
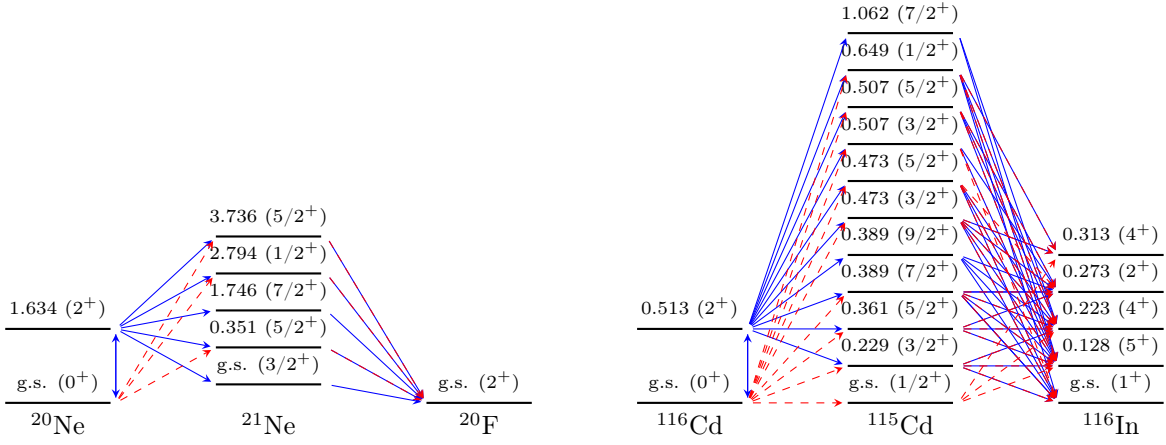

\paragraph{Path1.}
Let us consider the path1, whose couplings and level schemes of the involved nuclei are sketched in Fig.~\ref{fig:coupling_scheme_path1}. The values of the spectroscopic amplitudes extracted by the shell model calculation, when employing $zbm$ and $88Sr45$ interaction, are listed in Table~\ref{tab:amplitudes_path1_proj} for projectile-like overlaps and in Table~\ref{tab:amplitudes_path1_targ1} and~\ref{tab:amplitudes_path1_targ2} for target-like overlaps involved in the first and in the second transfer step, respectively. %In the second transfer step, only the overlaps whose corresponding spectroscopic amplitude comes out to be larger than 0.1 have been included. %Moreover, the 
%The same prescription also applies in case of path2. 

As one may observe in Fig.~\ref{fig:coupling_scheme_path1}, there is a large number of states in the intermediate partition which may be coupled, especially when target-like excitation levels are considered. Moreover, the coupling between two different states may involve even more overlaps, as indicated in Table~\ref{tab:amplitudes_path1_targ1} and~\ref{tab:amplitudes_path1_targ2}. %\textcolor{magenta}{.} 
In principle, one should consider in fact all the possible states of the intermediate partition and, therefore, extend as much as possible the number of couplings %\textcolor{magenta}{s} 
included. However, at least when taking into account transitions which involve low-lying states of projectile and target %in both the initial and final partition, as %like as 
in the initial partition as in this case, the coupling with states having larger excitation energy should be generally reduced. %In fact, %due to the large value of Q_opt and L_opt the transition from the initial 0+ ground states to the intermediate high-lying and high angular momentum are quite strong. The point is that in the second step, due to the same Q and L matching mechanism, larger Q and L are preferred tus suppressing the transtion to the final 0+ ground states.

Moreover, in view of the comparison with the experimental data, which may isolate the contribution to the low-lying states also of the final partition, %i.e. in the range [-0., 0.35] MeV, 
a further reduction of the weight of high-energy contributions to the total cross section is expected. 
Nevertheless, these states may play a role within CCBA formalism, whereas the coupling with the $2_1^+$ states of the initial partition is explicitly taken into account. However, in that case, their contribution is expected to be suppressed by the larger number of steps involved in the whole process. In any case, complete CCBA calculations have been also performed to get a quantitative estimation of these contributions. %\textcolor{magenta}{s}. 

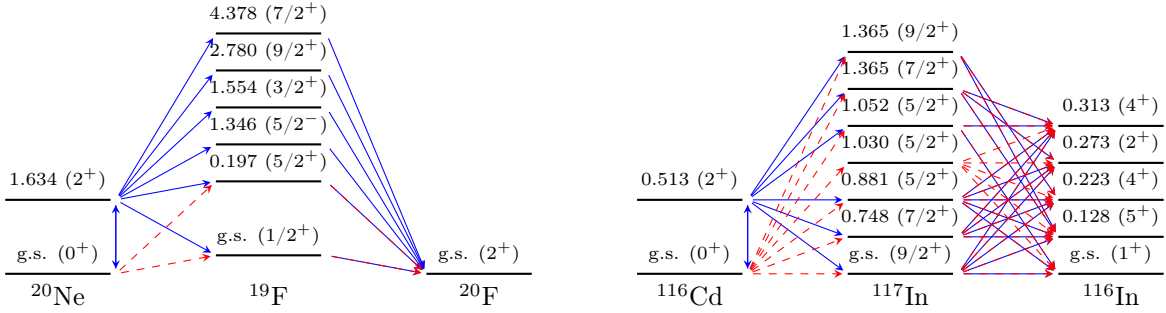
\begin{figure*}[t]
\begin{center}
    \begin{tikzpicture}[
      scale=0.7,
      level/.style={thick},
      virtual/.style={thick,densely dashed},
      trans/.style={red,thin,dashed,->,shorten >=2pt,shorten <=2pt,>=stealth},
      trans2/.style={blue,thin,->,shorten >=2pt,shorten <=2pt,>=stealth},
      classical/.style={thin,double,<->,shorten >=8pt,shorten <=8pt,>=stealth}
    ]
    % Draw the energy levels.
    \draw[level] (0cm,-2em) -- (2cm,-2em) node[midway,below] {$^{20}$Ne} node[midway,above] {\scriptsize g.s. (0$^{+}$)};
    \draw[level] (0cm,2em) -- (2cm,2em) node[midway,above] {\scriptsize 1.634 (2$^{+}$)};
    \draw[level, white] (4cm,-2em) -- (6cm,-2em) node[midway,below, black] {$^{19}$F};
    \draw[level] (4cm,-1em) -- (6cm,-1em) node[midway,above] {\scriptsize g.s. (1/2$^{+}$)};
    \draw[level] (4cm,3em) -- (6cm,3em) node[midway,above] {\scriptsize 0.197 (5/2$^{+}$)};
    \draw[level] (4cm,5em) -- (6cm,5em) node[midway,above] {\scriptsize 1.346 (5/2$^{-}$)};
    \draw[level] (4cm,7em) -- (6cm,7em) node[midway,above] {\scriptsize 1.554 (3/2$^{+}$)};
    \draw[level] (4cm,9em) -- (6cm,9em) node[midway,above] {\scriptsize 2.780 (9/2$^{+}$)};
    \draw[level] (4cm,11em) -- (6cm,11em) node[midway,above] {\scriptsize 4.378 (7/2$^{+}$)};
    \draw[level] (8cm,-2em) -- (10cm,-2em) node[midway,below] {$^{20}$F} node[midway,above] {\scriptsize g.s. (2$^{+}$)};
    \draw[trans2] (2.1cm,2em) -- (4cm,-1em) node[midway,left] {};\draw[trans2] (6.1cm,-1em) -- (8cm,-2em) node[midway,right] {};
    \draw[trans2] (2.1cm,2em) -- (4cm,3em) node[midway,left] {};\draw[trans2] (6.1cm,3em) -- (8cm,-2em) node[midway,right] {};
  \draw[trans2] (2.1cm,2em) -- (4cm,5em) node[midway,left] {};\draw[trans2] (6.1cm,5em) -- (8cm,-2em) node[midway,right] {};
    \draw[trans2] (2.1cm,2em) -- (4cm,7em) node[midway,left] {};\draw[trans2] (6.1cm,7em) -- (8cm,-2em) node[midway,right] {};
    \draw[trans2] (2.1cm,2em) -- (4cm,9em) node[midway,left] {};\draw[trans2] (6.1cm,9em) -- (8cm,-2em) node[midway,right] {};
    \draw[trans2] (2.1cm,2em) -- (4cm,11em) node[midway,left] {};\draw[trans2] (6.1cm,11em) -- (8cm,-2em) node[midway,right] {};
    \draw[trans] (2.1cm,-2em) -- (4cm,-1em) node[midway,left] {};\draw[trans] (6.1cm,-1em) -- (8cm,-2em) node[midway,right] {};
    \draw[trans] (2.1cm,-2em) -- (4cm,3em) node[midway,left] {};\draw[trans] (6.1cm,3em) -- (8cm,-2em) node[midway,right] {};
    \draw[trans2] (2.1cm,-2em) -- (2.1cm,2em) node[midway,left] {};%\draw[trans] (6.1cm,4em) -- (8cm,-2em) node[midway,right] {};
    \draw[trans2] (2.1cm,2em) -- (2.1cm,-2em) node[midway,left] {};%\draw[trans] (6.1cm,4em) -- (8cm,-2em) node[midway,right] {};

    \draw[level] (12cm,-2em) -- (14cm,-2em) node[midway,below] {$^{116}$Cd} node[midway,above] {\scriptsize g.s. (0$^{+}$)};
    \draw[level] (12cm,2em) -- (14cm,2em) node[midway,above] {\scriptsize 0.513 (2$^{+}$)};
    \draw[level] (16cm,-2em) -- (18cm,-2em) node[midway,below] {$^{117}$In} node[midway,above] {\scriptsize g.s. (9/2$^{+}$)};
	\draw[level] (16cm,0em) -- (18cm,0em) node[midway,above] {\scriptsize 0.748 (7/2$^{+}$)};
	\draw[level] (16cm,2em) -- (18cm,2em) node[midway,above] {\scriptsize 0.881 (5/2$^{+}$)};
	\draw[level] (16cm,4em) -- (18cm,4em) node[midway,above] {\scriptsize 1.030 (5/2$^{+}$)};
	\draw[level] (16cm,6em) -- (18cm,6em) node[midway,above] {\scriptsize 1.052 (5/2$^{+}$)};
	\draw[level] (16cm,8em) -- (18cm,8em) node[midway,above] {\scriptsize 1.365 (7/2$^{+}$)};
	\draw[level] (16cm,10em) -- (18cm,10em) node[midway,above] {\scriptsize 1.365 (9/2$^{+}$)};
   \draw[level] (20cm,-2em) -- (22cm,-2em) node[midway,below] {$^{116}$In} node[midway,above] {\scriptsize g.s. (1$^{+}$)};
    \draw[level] (20cm,0em) -- (22cm,0em) node[midway,above] {\scriptsize 0.128 (5$^{+}$)};
	\draw[level] (20cm,2em) -- (22cm,2em) node[midway,above] {\scriptsize 0.223 (4$^{+}$)};
	\draw[level] (20cm,4em) -- (22cm,4em) node[midway,above] {\scriptsize 0.273 (2$^{+}$)};
	\draw[level] (20cm,6em) -- (22cm,6em) node[midway,above] {\scriptsize 0.313 (4$^{+}$)};
    \draw[trans2] (14.1cm,-2em) -- (14.1cm,2em) node[midway,left] {};%\draw[trans] (6.1cm,4em) -- (8cm,-2em) node[midway,right] {};
    \draw[trans2] (14.1cm,2em) -- (14.1cm,-2em) node[midway,left] {};%\draw[trans] (6.1cm,4em) -- (8cm,-2em) node[midway,right] {};

   \draw[trans2] (14.1cm,2em) -- (16cm,-2em) node[midway,left] {};\draw[trans2] (18.1cm,-2em) -- (20cm,-2em) node[midway,right] {};\draw[trans2] (18.1cm,-2em) -- (20cm,0em) node[midway,right] {}; 
    \draw[trans2] (18.1cm,-2em) -- (20cm,2em) node[midway,right] {};\draw[trans2] (18.1cm,-2em) -- (20cm,4em) node[midway,right] {};\draw[trans2] (18.1cm,-2em) -- (20cm,6em) node[midway,right] {};

    \draw[trans2] (14.1cm,2em) -- (16cm,0em) node[midway,left] {};								  \draw[trans2] (18.1cm,0em) -- (20cm,0em) node[midway,right] {};
    \draw[trans2] (18.1cm,0em) -- (20cm,2em) node[midway,right] {};\draw[trans2] (18.1cm,0em) -- (20cm,4em) node[midway,right] {};\draw[trans2] (18.1cm,0em) -- (20cm,6em) node[midway,right] {};

    \draw[trans2] (14.1cm,2em) -- (16cm,2em) node[midway,left] {};\draw[trans2] (18.1cm,2em) -- (20cm,-2em) node[midway,right] {};\draw[trans2] (18.1cm,2em) -- (20cm,0em) node[midway,right] {};
    \draw[trans2] (18.1cm,2em) -- (20cm,2em) node[midway,right] {};\draw[trans2] (18.1cm,2em) -- (20cm,4em) node[midway,right] {};\draw[trans2] (18.1cm,2em) -- (20cm,6em) node[midway,right] {};

    \draw[trans2] (14.1cm,2em) -- (16cm,6em) node[midway,left] {};\draw[trans2] (18.1cm,6em) -- (20cm,-2em) node[midway,right] {};
																  \draw[trans2] (18.1cm,6em) -- (20cm,6em) node[midway,right] {};

    \draw[trans2] (14.1cm,2em) -- (16cm,8em) node[midway,left] {};								  \draw[trans2] (18.1cm,8em) -- (20cm,0em) node[midway,right] {};
    								  \draw[trans2] (18.1cm,8em) -- (20cm,4em) node[midway,right] {};\draw[trans2] (18.1cm,8em) -- (20cm,6em) node[midway,right] {};

    \draw[trans2] (14.1cm,2em) -- (16cm,10em) node[midway,left] {};\								    \draw[trans2] (18.1cm,10em) -- (20cm,0em) node[midway,right] {};
    \draw[trans2] (18.1cm,10em) -- (20cm,2em) node[midway,right] {};

    \draw[trans] (14.1cm,-2em) -- (16cm,-2em) node[midway,left] {};\draw[trans] (18.1cm,-2em) -- (20cm,-2em) node[midway,right] {};\draw[trans] (18.1cm,-2em) -- (20cm,0em) node[midway,right] {}; 
    \draw[trans] (18.1cm,-2em) -- (20cm,2em) node[midway,right] {};\draw[trans] (18.1cm,-2em) -- (20cm,4em) node[midway,right] {};\draw[trans] (18.1cm,-2em) -- (20cm,6em) node[midway,right] {};

    \draw[trans] (14.1cm,-2em) -- (16cm,0em) node[midway,left] {};								\draw[trans] (18.1cm,0em) -- (20cm,0em) node[midway,right] {};
    \draw[trans] (18.1cm,0em) -- (20cm,2em) node[midway,right] {};\draw[trans] (18.1cm,0em) -- (20cm,4em) node[midway,right] {};\draw[trans] (18.1cm,0em) -- (20cm,6em) node[midway,right] {};

    \draw[trans] (14.1cm,-2em) -- (16cm,2em) node[midway,left] {};\draw[trans] (18.1cm,2em) -- (20cm,-2em) node[midway,right] {};\draw[trans] (18.1cm,2em) -- (20cm,0em) node[midway,right] {};
    \draw[trans] (18.1cm,2em) -- (20cm,2em) node[midway,right] {};\draw[trans] (18.1cm,2em) -- (20cm,4em) node[midway,right] {};\draw[trans] (18.1cm,2em) -- (20cm,6em) node[midway,right] {};

    \draw[trans] (14.1cm,-2em) -- (16cm,4em) node[midway,left] {};\draw[trans] (18.1cm,4em) -- (20cm,-2em) node[midway,right] {};\draw[trans] (18.1cm,4em) -- (20cm,0em) node[midway,right] {};
    \draw[trans] (18.1cm,4em) -- (20cm,2em) node[midway,right] {};\draw[trans] (18.1cm,4em) -- (20cm,4em) node[midway,right] {};\draw[trans] (18.1cm,4em) -- (20cm,6em) node[midway,right] {};

    \draw[trans] (14.1cm,-2em) -- (16cm,6em) node[midway,left] {};\draw[trans] (18.1cm,6em) -- (20cm,-2em) node[midway,right] {};
																\draw[trans] (18.1cm,6em) -- (20cm,6em) node[midway,right] {};

    \draw[trans] (14.1cm,-2em) -- (16cm,8em) node[midway,left] {};								\draw[trans] (18.1cm,8em) -- (20cm,0em) node[midway,right] {};
    								  \draw[trans] (18.1cm,8em) -- (20cm,4em) node[midway,right] {};\draw[trans] (18.1cm,8em) -- (20cm,6em) node[midway,right] {};

    \draw[trans] (14.1cm,-2em) -- (16cm,10em) node[midway,left] {};								  \draw[trans] (18.1cm,10em) -- (20cm,0em) node[midway,right] {};
    \draw[trans] (18.1cm,10em) -- (20cm,2em) node[midway,right] {};
    \end{tikzpicture}

\caption{Coupling schemes of the projectile and target overlaps used in the sequential two-nucleon transfer reaction calculations, along path 2. The $zbm$ and $88Sr45$ interactions are employed in shell model calculations, for projectile-like and target-like nuclei, respectively. Red arrows concern the coupling considered within the DWBA approximation, while the blue arrows indicate the additional couplings introduced within the CCBA reaction scheme. }
\label{fig:coupling_scheme_path2}
\end{center}     
\end{figure*}

\paragraph{Path2.}
Analogously to the analysis of the previous paragraph, %here we perform 
a two-step transfer reaction calculation through the intermediate partition ($^{19}$F + $^{117}$In) has been performed, at first within the DWBA approximation. Couplings and level schemes of the involved nuclei are sketched in Fig.~\ref{fig:coupling_scheme_path2}. The values of the spectroscopic amplitudes extracted by the shell model calculation, when employing $zbm$ and $88Sr45$ interactions, for projectile-like and target-like overlaps, are listed in Table~\ref{tab:amplitudes_path2_proj} and \ref{tab:amplitudes_path2_targ}, respectively.

\begin{figure*}
\centering
{\includegraphics[width=0.95\columnwidth]{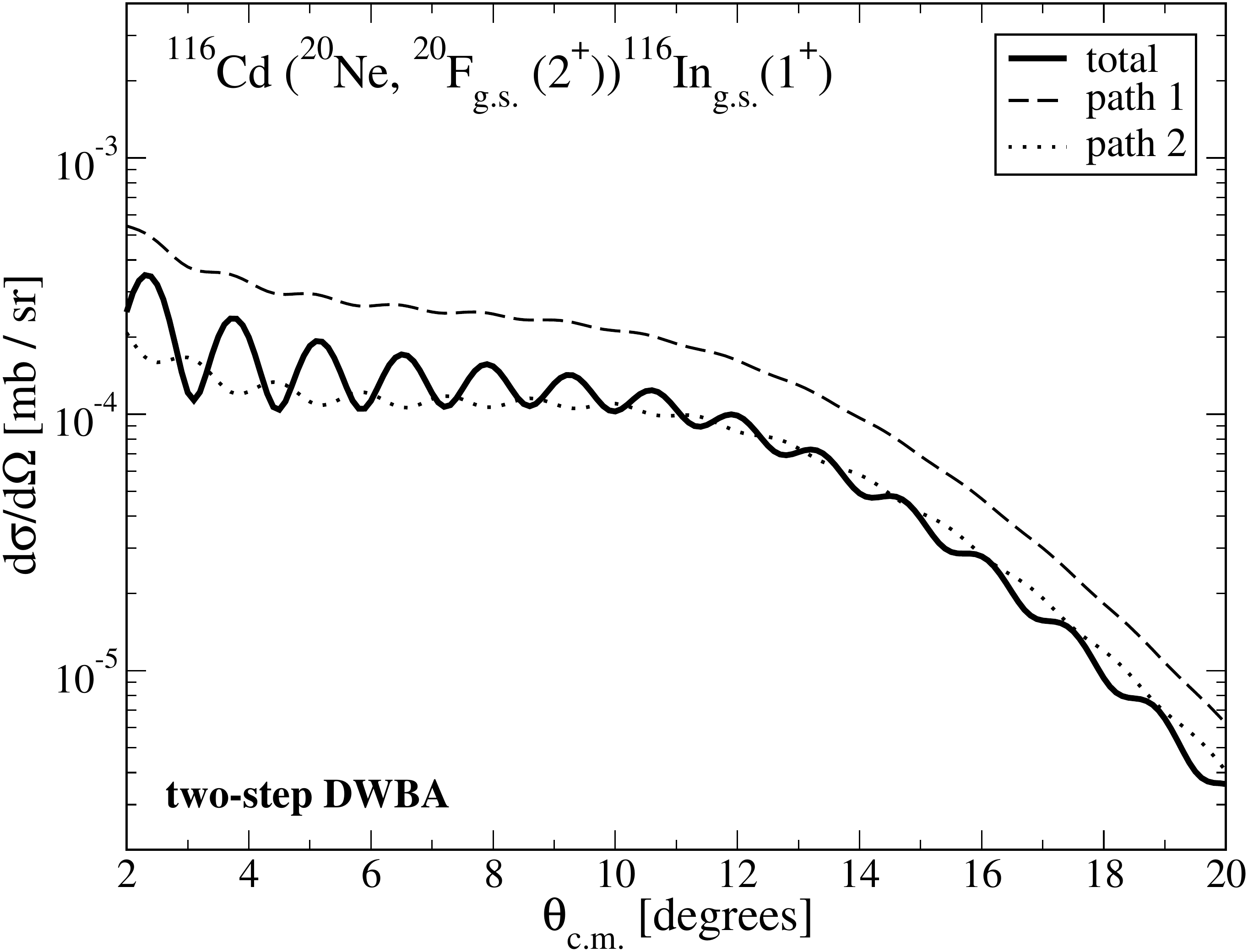}}
{\includegraphics[width=0.95\columnwidth]{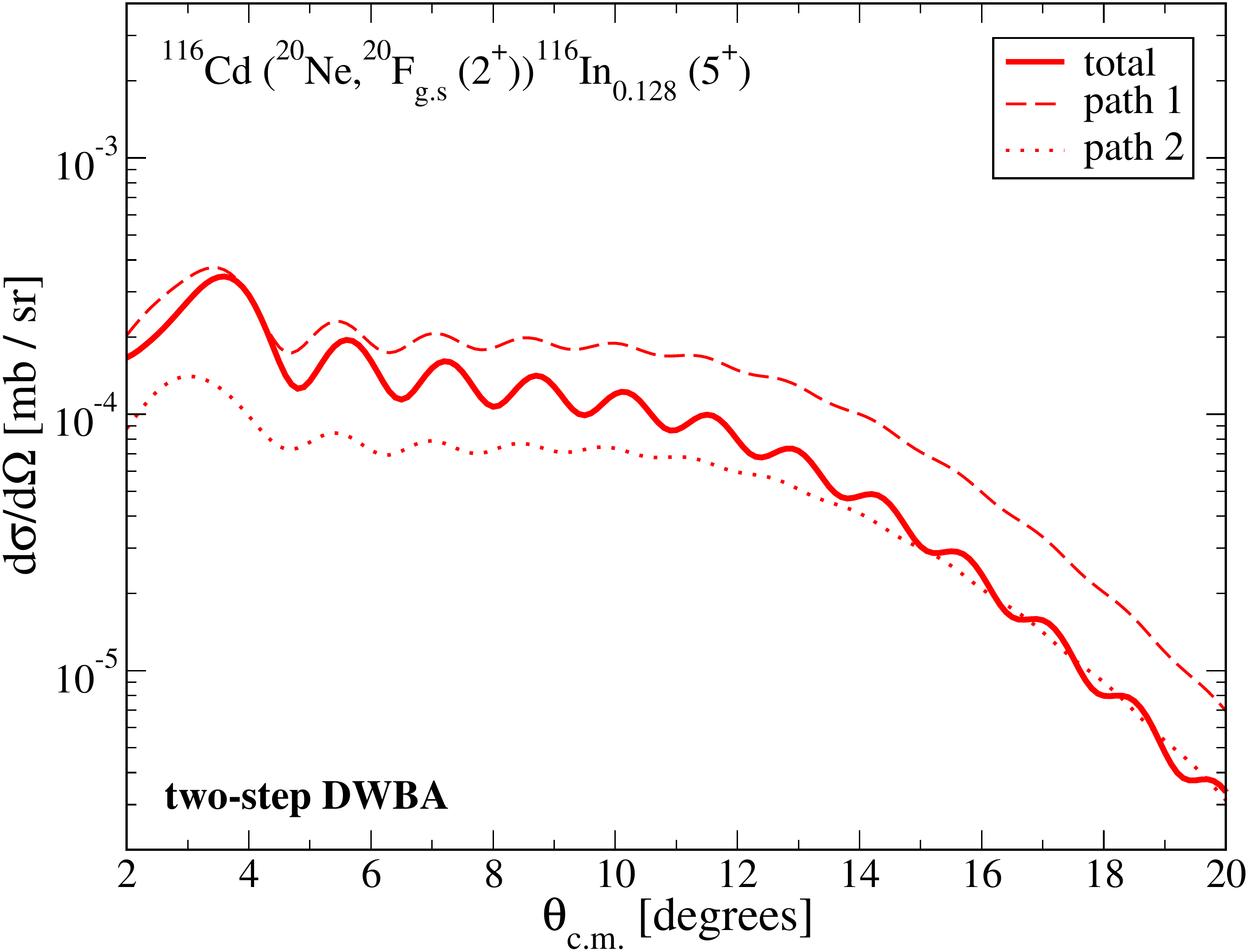}} \\
{\includegraphics[width=0.95\columnwidth]{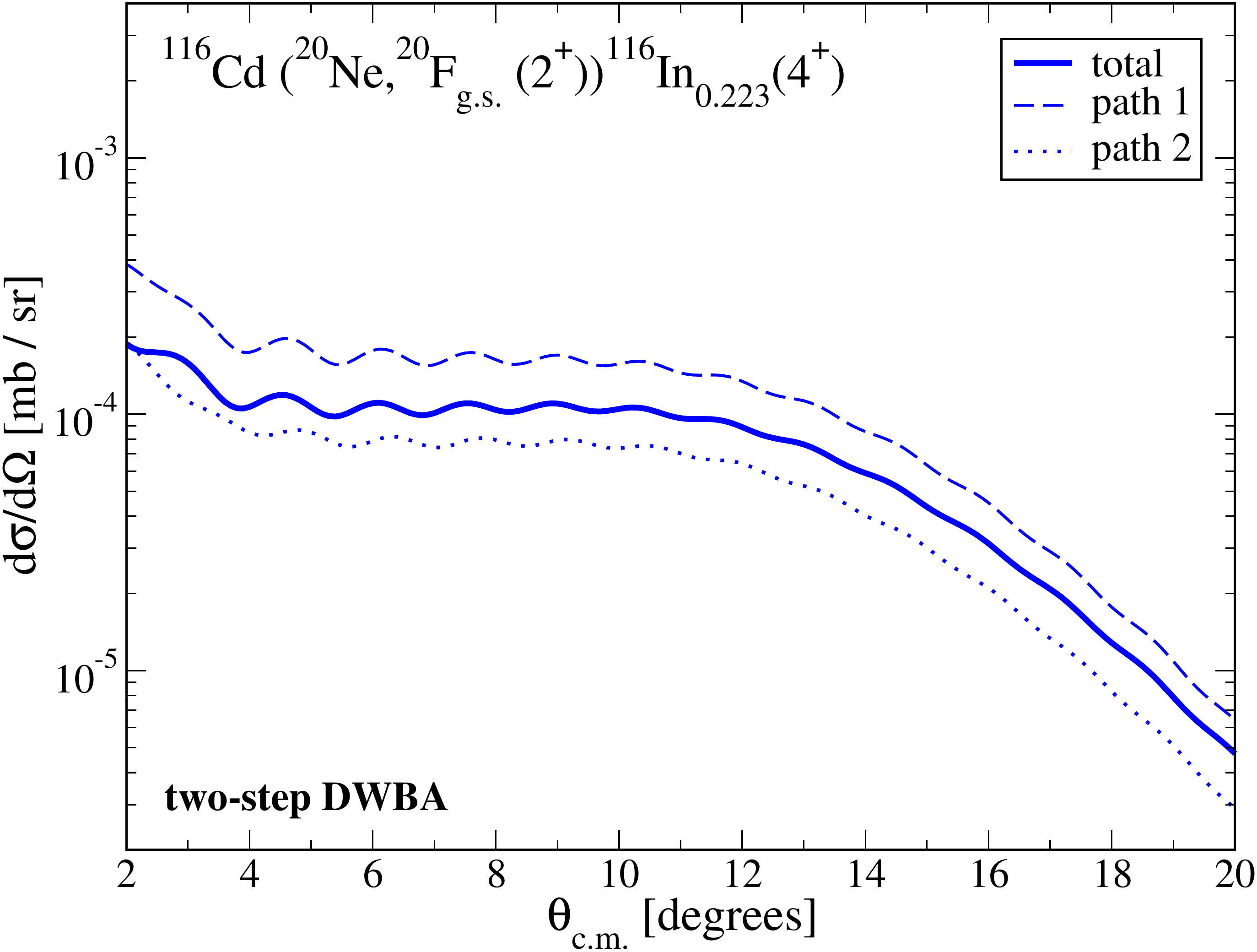}}
{\includegraphics[width=0.95\columnwidth]{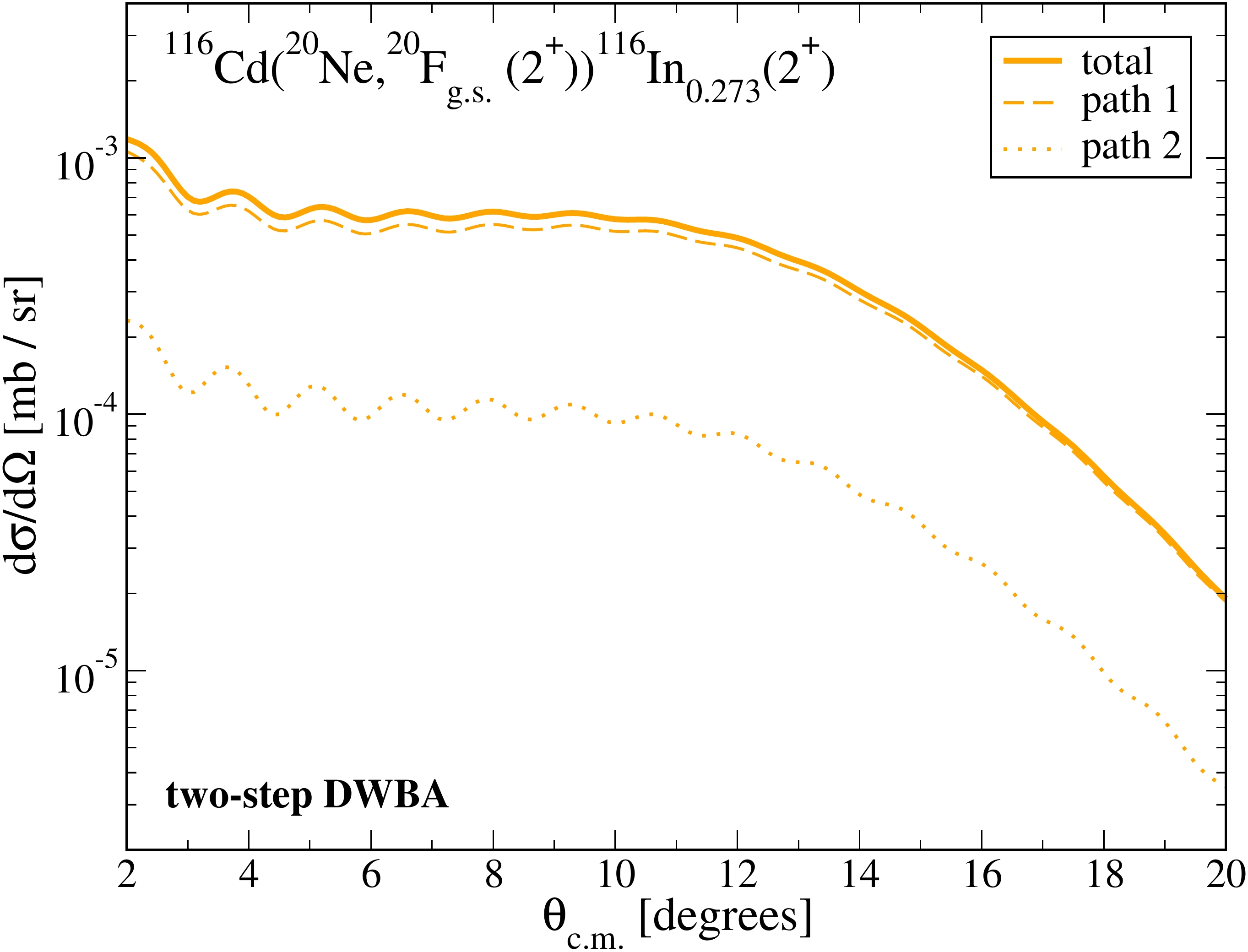}} \\		
{\includegraphics[width=0.95\columnwidth]{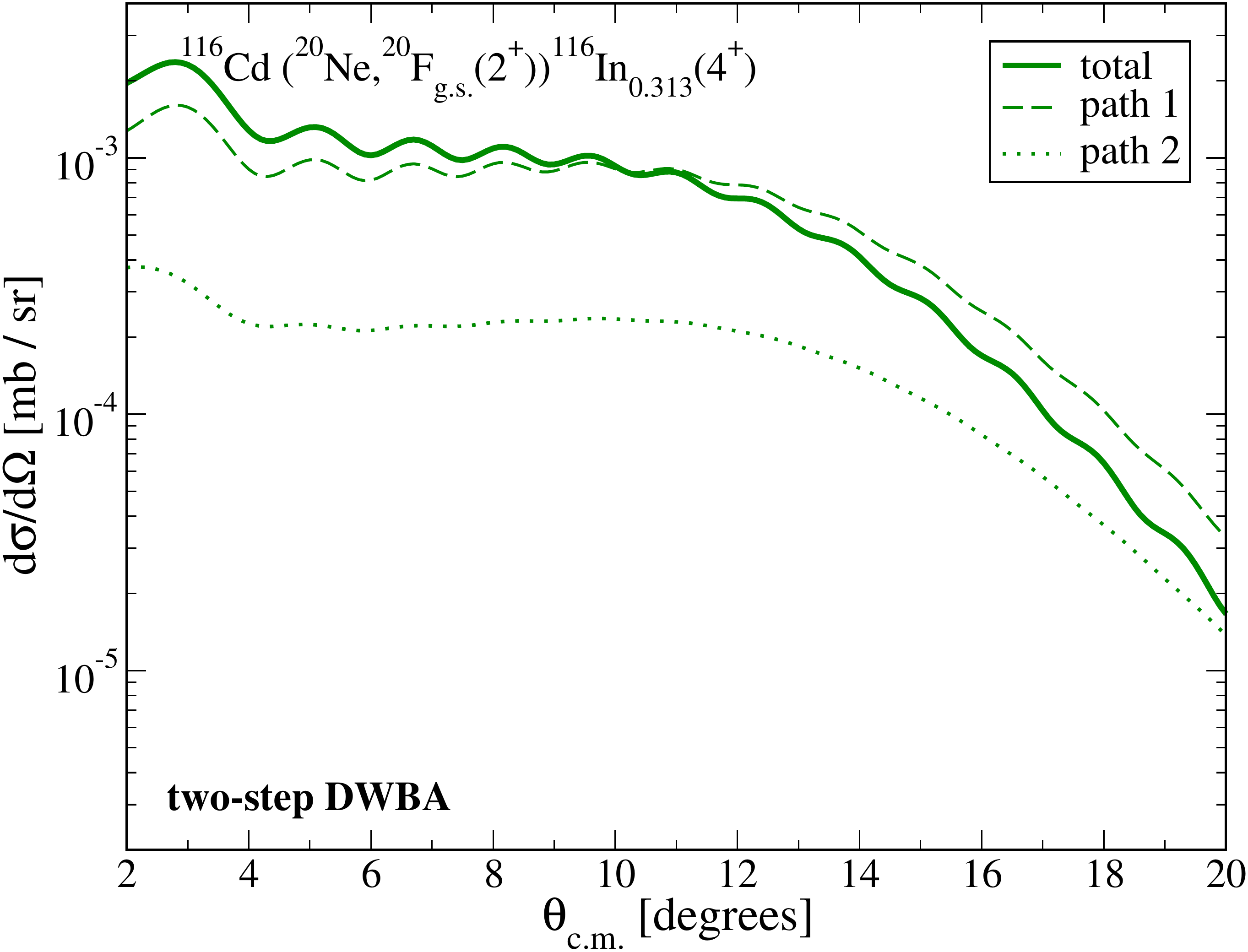}}
\caption{Angular distribution of the differential cross section for the different final states of the $^{116}$Cd($^{20}$Ne,$^{20}$F$_{\textup{g.s.}}$ (2$^{+}$))$^{116}$In SCE reaction, lying within  the considered excitation energy range, %([0.00, 0.35] MeV), 
as obtained when considering only the two-step transfer mechanisms. The two-step DWBA approximation is adopted and the SPP optical potential is employed. %The zbm and 88Sr45 interaction are used for projectile-like and target-like nuclei in shell model calculations. 
The two paths are considered separetely and with their coherent sum.}
\label{fig:1nptrans_DWBA}
\end{figure*}

\paragraph{Coherent sum of two paths.}
With the aim to properly evaluate the role of the two-step transfer mechanisms in determining the total cross section of the $^{116}$Cd($^{20}$Ne,$^{20}$F)$^{116}$In SCE reaction, the contribution of the two different paths illustrated above should be included in a coherent way within the same calculation. The resulting angular behavior of the differential angular distribution cross section, as obtained within the DWBA approximation along path1 or path2 or when including the coherent sum of the two paths, is shown in Fig. \ref{fig:1nptrans_DWBA}, for the different target-like states lying within the excitation energy range considered. %([0.00, 0.35] MeV). 
For the projectile-like states, only the g.s. is taken into account. Only the SPP parameterization is moreover shown for the optical potential, for sake of simplicity.

As a general feature, one observes that the bell-shape exhibited by the angular distribution in the one-proton transfer case is modified in the sequential two-step transfer mechanism. The result is, up to a large extent, a quite %more 
flat angular behavior of the differential cross section, at least at intermediate angles. Moreover, for a given combination of states in the exit channel, the diffraction pattern seems to be practically independent from the considered path while it depends on the angular momentum of the final states involved, i.e. on the total angular momentum transferred in the process. In particular, at least from a qualitative point of view, one notices that the angular position of the first maximum shifts to larger angles when considering target states in the outgoing partition with larger J values. However, this correspondence which is easily recovered in the one-nucleon transfer case, is blended and less clear in the two-step process where an intermediate partition is crossed. As a consequence, some differences arise in the diffraction pattern at small angles even when the angular momentum of the target states in the final partition is the same (see for example the differences existing in the diffraction patterns for the two different 4$^{+}$ states of $^{116}$In shown in Fig. \ref{fig:1nptrans_DWBA}).

A remarkable feature to be noted is that the contribution provided along path1 is systematically larger than the corresponding one obtained along path2. Such a result might be mainly attributed to the  larger level density predicted in the low excitation energy region for the system $^{115}$Cd, with respect to the $^{117}$In, within the model space here adopted. %A possible improvement along this direction is thus envisaged.

%On the other hand, also 
In addition, the role of the coherent interference between the two path turns out to be rather important. The coherent sum of the two contributions may considerably alter indeed the diffraction pattern of the general process, especially when the two paths have a comparable magnitude. As a result, the angular distribution of the differential cross section might appear rather different from the single path components, %although 
whereas the %order of magnitude 
absolute cross section is %not significantly
only partially affected.

\begin{figure}[t]
\includegraphics[width=\columnwidth]{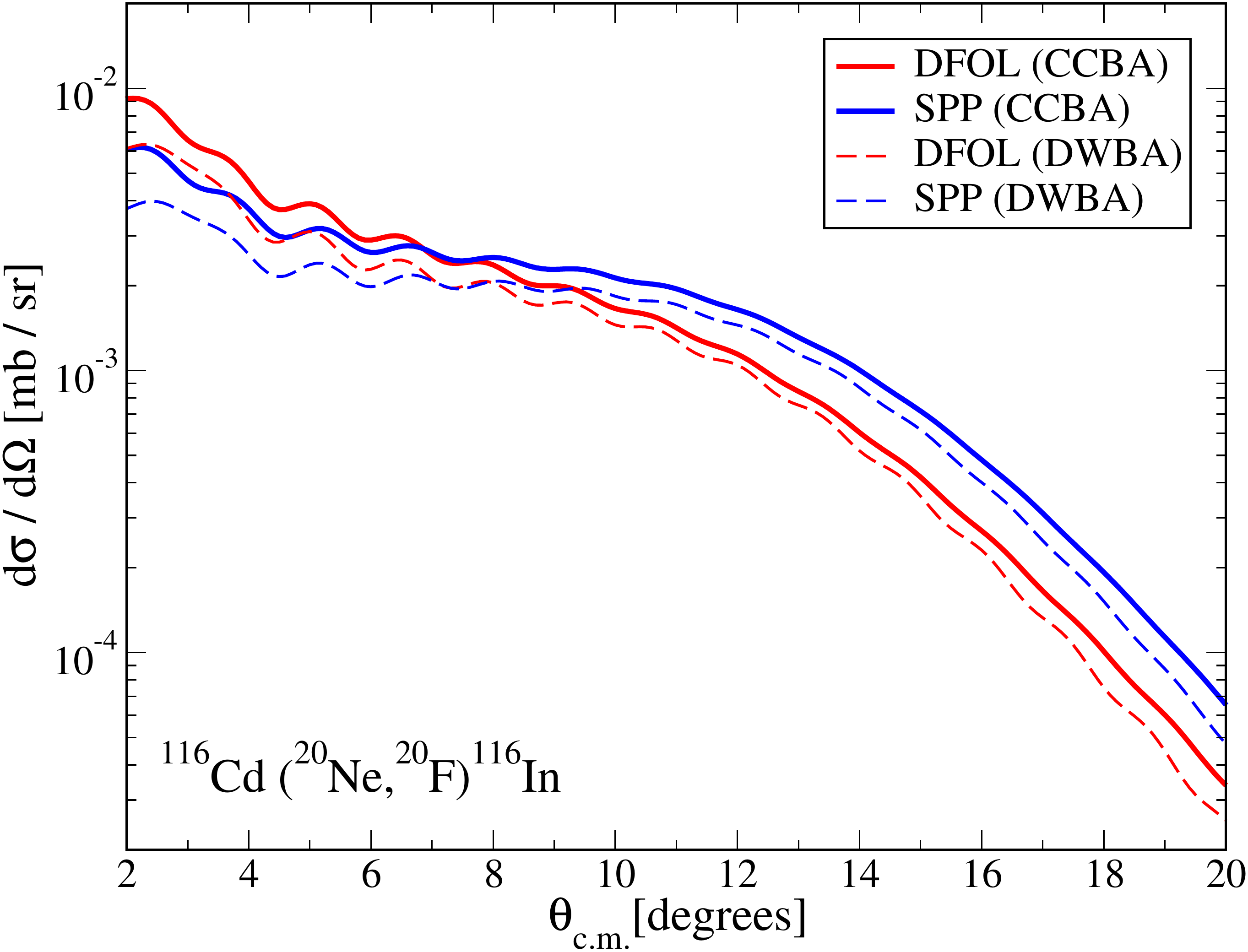}
\caption{Angular distribution of the differential cross section for all the final states of the $^{116}$Cd($^{20}$Ne,$^{20}$F)$^{116}$In SCE reaction, lying within the considered excitation energy range, %([0.00, 0.35] MeV), 
obtained from two-step DWBA (dashed lines) or CCBA (full lines) calculations. Calculation with DFOL and SPP ion-ion potentials are indicated by red and blue lines, respectively. %as obtained when considering the two-step transfer mechanism, by assuming the DWBA approximation and by employing DFOL optical potential. 
%The zbm and 88Sr45 interaction are adopted for projectile-like and target-like nuclei in shell model calculations.
}
\label{fig:1nptransfer_DWBA_CCBA}
\end{figure}

Similar conclusions might be drawn when more complicated coupling schemes are investigated. For this purpose, it could be helpful to compare the DWBA results with the CCBA ones, in order to shed light on the role of the inelastic excitations and on the transfer mechanisms %on 
which they are involved in. The corresponding results are displayed in Fig. \ref{fig:1nptransfer_DWBA_CCBA}, where the differential cross section is plotted as a function of the scattering angle in the center of mass system. Only the total cross section, given by the coherent sum of the two path contributions and the incoherent sum on combinations of projectile and target states lying within the excitation energy range considered %([0.00, 0.35] MeV), 
is reported. As in the one-nucleon transfer case, the CCBA results only slightly impact the diffraction pattern at forward angles, in light of the different combinations of states and of angular momentum transfer involved when exploring the intermediate partition. The results are then slightly sensitive also to the choice of the optical potential, in view of the delicate interplay between the spatial extension of the transition form factor and the surface details of the ion-ion interaction. Indeed, analogously to what was observed also in Fig. \ref{fig:1ptransfer_CRC_CCBA_DWBA} for the one-proton transfer case, DFOL results tend to be larger %overestimate the differential cross section 
at small angles but predict a steeper decrease than SPP with increasing $\theta_{c.m.}$ scattering angle. 

The results discussed above allow thus to finally achieve a quantitative estimation of the total SCE cross section, at least when describing this process as only given by the %sequential succession
sequence of two-nucleon transfer mechanisms and inelastic excitations in the initial partition. Although the genuine direct SCE transition driven by the direct isovector nuclear interaction is not included in such a scheme, it is thus certainly worthwhile to provide a first comparison with the experimental results. %experimentally obtained. 
In Table \ref{tab:20Ne116Cd_integ_cross_section}, we list %therefore 
the theoretical cross section, integrated in the angular range [4.5°, 14.5°] of the laboratory system, for all the states lying within the considered excitation energy range. %[0.00, 0.35] MeV. 
In Table \ref{tab:20Ne116Cd_integ_cross_section}, we also reported the total sum, to be compared with the corresponding experimental value. %Only the SPP potential is considered for sake of simplicity. 

\begin{table}[t]
\centering
\caption{SCE cross section (in $\mu b$) as obtained with the SPP and the DFOL potential through two-step transfer mechanisms, when integrating in the angular range [4.5°, 14.5°] of the laboratory system and for all the combination of projectile and target states lying within the considered excitation energy range. % [0.00, 0.35] MeV. 
The integrated experimental value is also reported. } 
\begin{tabular}{*{9}{c}}
	   &                        &           &   \multicolumn{2}{c}{DFOL} & &\multicolumn{2}{c}{SPP} &	\\ 
\cline{4-5}\cline{7-8}
 $J^\pi_p$ & $J^\pi_t$ & $E_t^*$ & $\sigma_{\textup{DWBA}}$  & $\sigma_{\textup{CCBA}}$ & & $\sigma_{\textup{DWBA}}$  & $\sigma_{\textup{CCBA}}$ &  $\sigma_{\textup{exp}}$ \\ %& $\sigma_{\textup{exp}}$ [$\mu$b]  \\
\hline
 2$^+_{\textup{g.s.}}$  &   1$^+$   &     0.000     & 0.015 & 0.024 & & 0.020 & 0.033 &  \\ %& \\ %&  2.031 $\cdot$ 10$^{-2}$  \\
 2$^+_{\textup{g.s.}}$  &   5$^+$   &     0.128     & 0.017 & 0.043 & & 0.019 & 0.049 &\\ %& \\ %&  1.871 $\cdot$ 10$^{-2}$  \\
 2$^+_{\textup{g.s.}}$  &   4$^+$   &     0.223     & 0.014 & 0.021 & & 0.018 & 0.028 & \\ %& \\ %&  1.831 $\cdot$ 10$^{-2}$  \\
 2$^+_{\textup{g.s.}}$  &   2$^+$   &     0.273     & 0.078 & 0.078 & & 0.099 & 0.100 &\\ %& \\ %&  9.586 $\cdot$ 10$^{-2}$  \\
 2$^+_{\textup{g.s.}}$  &   4$^+$   &     0.313     & 0.128 & 0.129 & & 0.153 & 0.156 & \\ %& \\ %& 14.206 $\cdot$ 10$^{-2}$  \\
\hline
 total       &           &               &  	      0.252           &         0.295      & &  	      0.309           &         0.366  & 0.7 $\pm$ 0.3\\ 
\end{tabular}
%\begin{tabular}{*{7}{c}}
%        $J^\pi_{E_p^*}$      & $J^\pi_{E_t^*}$   & $\sigma_{\textup{DWBA}}^{\textup{DFOL}}$  & $\sigma_{\textup{CCBA}}^{\textup{DFOL}}$ & $\sigma_{\textup{DWBA}}^{\textup{SPP}}$  & $\sigma_{\textup{CCBA}}^{\textup{SPP}}$  & $\sigma_{\textup{exp}}$ \\ %& $\sigma_{\textup{exp}}$ [$\mu$b]  \\
%\hline
% 2$^+_{\textup{g.s.}}$ &   1$^+_{\textup{g.s.}}$   & 0.020 & 0.033 & 0.020 & 0.033 \\ %& \\ %&  2.031 $\cdot$ 10$^{-2}$  \\
% 2$^+_{\textup{g.s.}}$ &   5$^+_{\textup{0.128}}$  & 0.019 & 0.049 & 0.019 & 0.049 \\ %& \\ %&  1.871 $\cdot$ 10$^{-2}$  \\
% 2$^+_{\textup{g.s.}}$ &   4$^+_{\textup{0.223}}$  & 0.018 & 0.028 & 0.018 & 0.028 \\ %& \\ %&  1.831 $\cdot$ 10$^{-2}$  \\
% 2$^+_{\textup{g.s.}}$ &   2$^+_{\textup{0.273}}$  & 0.099 & 0.100 & 0.099 & 0.100 \\ %& \\ %&  9.586 $\cdot$ 10$^{-2}$  \\
% 2$^+_{\textup{g.s.}}$ &   4$^+_{\textup{0.313}}$  & 0.153 & 0.156 & 0.153 & 0.156 \\ %& \\ %& 14.206 $\cdot$ 10$^{-2}$  \\
%\hline
%	 total 	       &            &  	      0.309           &         0.366       & 0.309           &         0.366 & 0.7 $\pm$ 0.3\\ 
%\end{tabular}
\label{tab:20Ne116Cd_integ_cross_section}
\end{table}

In light of the results of Table \ref{tab:20Ne116Cd_integ_cross_section}, one can assess that, when describing the $^{116}$Cd($^{20}$Ne,$^{20}$F)$^{116}$In reaction only in terms of two successive transfer processes, the order of magnitude of the %total reaction 
cross section is nicely recovered in our theoretical calculations. Moreover, this is achieved already within the simplified DWBA framework, although the role of the coupling with the inelastic excitations of the states which belong to the initial partition seems to be not completely negligible.

It is also worth noting that, due to the finite experimental energy resolution, the set interval upper limit rules out part of the high-energy contributions of the considered transitions. On the other hand, the present theoretical predictions of the sequential proton-neutron transfer processes fully account for such contributions, leading to an expected overestimation of the relative amount in the total reaction cross section attributed to the multi-step mechanism. Nonetheless, this does not affect the validity of this first study, since the order of magnitude of the transfer processess would be in any case preserved.

We would like to remind here that our result has been obtained by employing as theoretical ingredients an optical potential, which has demonstrated to provide a reliable reproduction of the elastic channel, and a well-defined model space with its corresponding shell model effective interaction, which returns a satisfying spectroscopic description of the nuclear systems involved here and allows for %\st{a very good} \textcolor{magenta}{reasonable} 
reasonable agreement with experimental results of the one-proton transfer process. We notice, by the way, that the same ingredients have been already successfully applied also in the description of two-neutron and two-proton transfer reactions, %which start from 
having in common the same initial partition \cite{Diana}.

Of course, our result %might be still considered to be partially affected
might still be considered affected by a certain degree of model dependence, as suggested by the small sensitivity here highlighted with respect to the optical potential adopted and to the fine details of the structure ingredients. %Nevertheless, owing to the variety of different processes involved and to the uncertainties existing in the %structure 
%description of the %\textcolor{magenta}{considered} 
%considered nuclei, this 
This unavoidable model dependence, which might be ascribable to the variety of different processes involved and to the uncertainties existing in the %structure 
description of the %\textcolor{magenta}{considered} 
considered nuclei, is however strongly mitigated in our work by the multi-channel experimental constraints and the fully consistent theoretical approach employed. %this dependence is however practically unavoidable. 
%In any case, at least from a qualitative point of view, 
%Therefore, we may consider our result fully satisfying and our finding well-founded.
Moreover, our study suggests that SPP might be globally considered a preferrable choice with respect to DFOL. %Despite 
Although they both give a good reproduction of the scattering data, SPP gives indeed a better description of the transfer channels, in one-proton transfer as well as in the two-neutron and two-proton channels considered in Ref. \cite{Diana}.

The analysis performed here does not prevent us thus %thus to come out 
to draw %\red{remarkable} \gre{important} 
important conclusions.
%Indeed, on 
On the one hand, our work demonstrates that the two-step transfer mechanisms provide a non negligible contribution to the total cross section of the $^{116}$Cd($^{20}$Ne,$^{20}$F)$^{116}$In reaction channel. Such a result, in the energy/mass range investigated in the present work, was not so obvious. Although in some previous works (Refs.~\cite{Oertzen, Winfield, Len89, DasPRC1986}), which mainly refer to lighter systems, a primary role of the multi-step transfer contribution was already assessed, in some other light-ion induced studies (Ref.~\cite{CapLenske04}) a different behaviour was observed, thus highlighting a not universal scenario. Moreover, in this work, different kinematical conditions, which are of crucial importance in determining the magnitude of the transfer mechanisms, have been generally explored.
On the other hand, our finding suggests that a relevant fraction of this total cross section is still missing within our picture. Such a fraction should be provided by the direct mechanism, namely from the direct SCE process, which has not been addressed here,%\gre{, 
owing to unavoidable uncertainties which would affect the results for a such heavy system, both from the experimental and theoretical perspectives. %}. 
Such a statement might be then definitely validated only when confirmed by a consistent calculation taking into account the direct charge-exchange contribution. Nevertheless, for the time being, we highlight that this result seems to be in line with recent findings discussed in Ref. \cite{CavFRO2021}, where a significant contribution from direct isovector exchange mechanism is highlighted from the analysis of the $^{40}$Ca($^{18}$O,$^{18}$F)$^{40}$K SCE reaction, at the same bombarding energy per nucleon. %\textcolor{magenta}{at the same bombarding energy per nucleon}.

The competition between the two processes is however anything but trivial, since its fully consistent analysis would require all the structure ingredients needed for describing both mechanisms being extracted within the same coherent theoretical framework. Such a competition is currently under investigation and will be the topic of a forthcoming paper. 

However, since the transfer mechanisms %seem to be not suppressed, %by orders of magnitude, 
do not seem to be suppressed, the present study already suggests that %if one wants 
in order to extract structure information %concerning 
on the NME %of the CE process 
from the SCE %reaction 
cross section, a careful estimation of the competing transfer processes is needed. These higher order processes are instead expected to be suppressed when further increasing the number of steps involved, for example when investigating DCE reactions, and preliminary attempts along this direction seem to confirm such %a statement 
an expectation \cite{LayBurrello, Diana}. 

\section{Conclusions and outlooks}

In the present work, the $^{116}$Cd($^{20}$Ne,$^{20}$F)$^{116}$In reaction has been studied at an incident energy of 306 MeV, through a multi-channel analysis, involving also the elastic scattering and the one-proton transfer $^{116}$Cd($^{20}$Ne,$^{19}$F)$^{117}$In reaction.
%A large number of different channels is indeed populated in this kind of heavy ion collisions above the Coulomb barrier and a multi-channel approach should be recommended if one wants to assess the reliability of the ingredients usually adopted in the description of the complex reaction mechanisms involved. 
Such a wide collection of experimental data was indeed instrumental to put more stringent constraints on the nuclear models that are also adopted for the calculations of the SCE %NME and the related beta-decay ones. 
direct process, the corresponding NME and the related beta-decay ones. %In this framework, the reaction investigated in this work deserves also a special attention, since the $^{116}$Cd nucleus involved is a candidate for the elusive 0$\nu$2$\beta$ decay. 
In perspective, studying CE reactions on $^{116}$Cd will contribute to encircle and to understand the expected $0\nu 2\beta$-decay properties of that nucleus.   

The experimental data for multi-nucleon transfer have been then compared with the theoretical predictions obtained by performing (sequential) two-step distorted wave Born approximation and higher-order coupled reaction channel calculations. Two different optical potentials were employed for modeling the initial and final state interactions and a large-scale shell model approach was adopted to evaluate the spectroscopic amplitudes for the single-particle transitions characterizing one-nucleon transfer processes.

Through a combination of structure and reaction studies and a multi-channel comparison with the experimental data, we validated the choice of the optical potentials adopted and we checked the %trustworthiness 
reliability of the model space and interactions assumed in the shell model calculations. Moreover, we shed light on the influence of the high-order couplings between transfer and scattering channels, bringing to light a bland influence of the inelastic excitations in the initial partition both on the one-proton transfer process and on the SCE reaction.

Finally, within the same unified scheme, our calculations were able to reproduce also the order of magnitude of the experimental cross section, by describing the $^{116}$Cd($^{20}$Ne,$^{20}$F)$^{116}$In reaction only in terms of successive transfer processes. %We obtained this result 
This result was obtained already within the simplified two-step DWBA framework, proving that resorting to more complex reaction schemes, at a first level of accuracy, reveals unnecessary. It is %wortwhile to notice 
worth noting that the robustness of our finding is corroborated by the fact that %, despite a rather intrinsic model dependence of the theoretical framework adopted, 
no parameter has been pratically adjusted with respect to the systematics. 

However, the sequence of second or higher order transfer or inelastic processes considered in our study does not entirely exhaust the total cross section. The origin of the remaining difference between the theoretical and experimental results is not easy to access at this stage, although being for a large extent ascribable to the direct SCE mechanism, which is not included within the present picture. Our conclusion is thus that room is left for a %\red{significant} \gre{possible relevant} 
possible %relevant role played by 
contribution of the direct isovector exchange mechanism in the explored SCE transitions. 

%\red{As a perspective, the} \gre{The} 
The genuine contribution of the latter mechanism might be isolated when a fully consistent analysis, taking into account the competition among all the possible processes involved, including the direct SCE mechanism, is carried out within the same theoretical framework. %\gre{
As a perspective, the study of the competition among the different charge exchange reaction mechanisms demands the exploration of new set of data. To this aim, the analysis of simple systems, in which transitions to individual states can be isolated, is particularly helpful for untangling direct and multi-step contributions and comparing with the predictions of theoretical models. Studies of heavy-ion induced charge exchange transitions on $^{12}$C target are already in progress \cite{spatafora}, taking advantage of the low level density in the light reaction products and of the well known spectroscopic properties of the involved nuclei. In this view, the analysis of new charge exchange data at different beam energies is also interesting and it is among the research lines of the NUMEN and NURE projects~\cite{EPJA2018}. %} 
Work is in progress along this direction, with the aim of extending this analysis also to the DCE 
%cross section,
channel, of particular interest for the possibility to extract information on the NME of 2$\beta$-decay. %, according to the main purposes of the NUMEN and NURE projects. %to disentangle these competing mechanisms, looking also at the DCE cross section, to hopefully extract information of the NME of 2$\beta$-decay, according to the main purposes of NUMEN and NURE projects.

\appendix

\begin{table}[tp!]
\caption{\label{tab:amplitudes_path1_proj} Spectroscopic amplitudes (S.A.) obtained through shell model $zbm$ interaction and adopted in CCBA calculations for all the projectile-like nuclei involved along the path1 (see text) of the $^{116}$Cd($^{20}$Ne,$^{20}$F)$^{116}$In SCE reaction, whose final states lie within the excitation energy range here considered. % ([0.00, 0.35] MeV). 
The column nl$_j$ indicates the principal quantum number, the orbital and the total angular momentum of the single valence proton, respectively, through the usual spectroscopic notation.}
\centering
\begin{tabular}{*{4}{c}}
\textbf{Initial state}				       & \textbf{Final state} 				 	& \textbf{nl$_j$} & \textbf{S.A.} \\
\toprule
\multirow{3}*{$^{20}$Ne$_{\textup{g.s.}}$ (0$^{+}$)}   &  $^{21}$Ne$_{0.351}$ (5/2$^{+}$)   		 	&  (1d$_{5/2}$)   &   0.7696  \\
%\cline{2-4}
%					   	       &  $^{21}$Ne$_{2.789}$ (1/2$^{-}$)  	         	&  (1p$_{1/2}$)   &   0.3935  \\
\cline{2-4}	
					   	       &  $^{21}$Ne$_{2.794}$ (1/2$^{+}$)   		        &  (2s$_{1/2}$)   &  -0.7355  \\
\cline{2-4}
						       &  $^{21}$Ne$_{3.736}$ (5/2$^{+}$)   		 	&  (1d$_{5/2}$)   &  -0.0891  \\
\midrule	
\multirow{8}*{$^{20}$Ne$_{\textup{1.634}}$ (2$^{+}$)} &  \multirow{2}*{$^{21}$Ne$_{\textup{g.s.}}$ (3/2$^{+}$)}  	&  (2s$_{1/2}$)   &   0.2077  \\
						       &   						 	&  (1d$_{5/2}$)   &  -1.0262  \\
\cline{2-4}
					               &  \multirow{2}*{$^{21}$Ne$_{0.351}$ (5/2$^{+}$)} 	&  (2s$_{1/2}$)   &   0.3181  \\
					               &  							&  (1d$_{5/2}$)   &   0.1650  \\
\cline{2-4}
					   	       &  $^{21}$Ne$_{1.746}$ (7/2$^{+}$)  	        	&  (1d$_{5/2}$)   &  -0.9131  \\
\cline{2-4}
					   	       &  $^{21}$Ne$_{2.794}$ (1/2$^{+}$)   		 	&  (1d$_{5/2}$)   &  -0.6516  \\
%\cline{2-4}
%					   	       &  $^{21}$Ne$_{2.867}$ (1/2$^{+}$)   		 	&  (1d$_{5/2}$)   &   0.6234  \\
%\cline{2-4}
%					   	       &  $^{21}$Ne$_{3.664}$ (3/2$^{-}$)   		 	&  (1p$_{1/2}$)   &  -0.3631  \\
\cline{2-4}
						       &  \multirow{2}*{$^{21}$Ne$_{3.736}$ (5/2$^{+}$)} 	&  (2s$_{1/2}$)   &  -0.7344  \\
						       &  						 	&  (1d$_{5/2}$)   &   0.5534  \\
%\cline{2-4}
%					   	       &  $^{21}$Ne$_{3.884}$ (5/2$^{-}$)   		 	&  (1p$_{1/2}$)   &  -0.3353  \\
\midrule
\midrule
\multirow{2}*{$^{21}$Ne$_{\textup{g.s.}}$ (3/2$^{+}$)} & \multirow{8}*{$^{20}$F$_{\textup{g.s.}}$ (2$^{+}$)}    &  (2s$_{1/2}$)   &  -0.6366  \\
						       & 				 		        &  (1d$_{5/2}$)   &   0.5209  \\
\cline{1-1}\cline{3-4}
\multirow{2}*{$^{21}$Ne$_{0.351}$ (5/2$^{+}$)}         & 						        &  (2s$_{1/2}$)   &   0.0656  \\
						       & 				 		        &  (1d$_{5/2}$)   &  -0.6412  \\
\cline{1-1}\cline{3-4}
$^{21}$Ne$_{1.746}$ (7/2$^{+}$)		               & 				                        &  (1d$_{5/2}$)   &  -0.4159  \\
\cline{1-1}\cline{3-4}
$^{21}$Ne$_{2.794}$ (1/2$^{+}$)		               & 				                        &  (1d$_{5/2}$)   &  -0.0431  \\
\cline{1-1}\cline{3-4}
\multirow{2}*{$^{21}$Ne$_{3.736}$ (5/2$^{+}$)}         & 				                        &  (2s$_{1/2}$)   &  -0.0331  \\
					               & 						        &  (1d$_{5/2}$)   &  -0.2296  \\
\bottomrule
\end{tabular}
\end{table}

\begin{table}[tp!]
\caption{\label{tab:amplitudes_path1_targ1} Spectroscopic amplitudes (S.A.) obtained through shell model $88Sr45$ interaction and adopted in CCBA calculations for all the target-like nuclei involved along the first step of path1 (see text) of the $^{116}$Cd($^{20}$Ne,$^{20}$F)$^{116}$In SCE reaction. The column nl$_j$ indicates the principal quantum number, the orbital and the total angular momentum of the single valence proton, respectively, through the usual spectroscopic notation.}
\centering
\begin{tabular}{*{4}{c}}
\textbf{Initial state}				       & \textbf{Final state} 				 	& \textbf{nl$_j$} & \textbf{S.A.} \\
\toprule
\multirow{9}*{$^{116}$Cd$_{\textup{g.s.}}$ (0$^{+}$)}  &  $^{115}$Cd$_{\textup{g.s.}}$ (1/2$^{+}$) 	    &  (3s$_{1/2}$)  &   0.8913  \\
\cline{2-4}
						       &  $^{115}$Cd$_{0.229}$ (3/2$^{+}$) 	  	    &  (2d$_{3/2}$)  &  -1.2814  \\
\cline{2-4}
						       &  $^{115}$Cd$_{0.361}$ (5/2$^{+}$) 	 	    &  (2d$_{5/2}$)  &  -0.0927  \\
\cline{2-4}
						       &  $^{115}$Cd$_{0.389}$ (7/2$^{+}$) 	   	    &  (1g$_{7/2}$)  &   1.8526  \\
\cline{2-4}
						       &  $^{115}$Cd$_{0.473}$ (3/2$^{+}$) 	            &  (2d$_{3/2}$)  &   0.0356  \\
\cline{2-4}
						       &  $^{115}$Cd$_{0.473}$ (5/2$^{+}$) 	            &  (2d$_{5/2}$)  &   0.3192  \\
\cline{2-4}
						       &  $^{115}$Cd$_{0.507}$ (3/2$^{+}$) 	  	    &  (2d$_{3/2}$)  &  -0.0219  \\
\cline{2-4}
						       &  $^{115}$Cd$_{0.507}$ (5/2$^{+}$) 	 	    &  (2d$_{5/2}$)  &   0.0950  \\
\cline{2-4}
						       &  $^{115}$Cd$_{0.649}$ (1/2$^{+}$) 	            &  (3s$_{1/2}$)  &   0.9347  \\ 
%\cline{2-4}
%						       &  $^{115}$Cd$_{1.062}$ (7/2$^{+}$) 	  	    &  (1g$_{7/2}$)  &  -0.0062  \\
\midrule
\multirow{24}*{$^{116}$Cd$_{\textup{0.513}}$ (2$^{+}$)}&  \multirow{3}*{$^{115}$Cd$_{0.229}$ (3/2$^{+}$)}   &  (3s$_{1/2}$)  &   0.1381  \\
						       &  				 	 	    &  (2d$_{3/2}$)  &   0.9889  \\
						       &  				 	 	    &  (1g$_{7/2}$)  &  -0.1855  \\
\cline{2-4}
						       &  \multirow{3}*{$^{115}$Cd$_{0.361}$ (5/2$^{+}$)}   &  (3s$_{1/2}$)  &  -0.5688  \\
						       &  				 	 	    &  (2d$_{3/2}$)  &   0.5895  \\
						       &  				 	 	    &  (1g$_{7/2}$)  &   0.3823  \\
\cline{2-4}
						       &  \multirow{2}*{$^{115}$Cd$_{0.389}$ (7/2$^{+}$)}   &  (2d$_{1/2}$)  &  -0.1917  \\
						       &  				 	 	    &  (1g$_{3/2}$)  &   0.5723  \\
\cline{2-4}
						       &  \multirow{2}*{$^{115}$Cd$_{0.389}$ (9/2$^{+}$)}   &  (2d$_{5/2}$)  &   0.1209  \\
						       &  				 	 	    &  (1g$_{3/2}$)  &  -0.5996  \\
\cline{2-4}
						       &  \multirow{2}*{$^{115}$Cd$_{0.473}$ (3/2$^{+}$)}   &  (3s$_{1/2}$)  &  -0.4137  \\ 
						       &  				 	 	    &  (1g$_{7/2}$)  &  -0.2472  \\
\cline{2-4}
						       &  \multirow{2}*{$^{115}$Cd$_{0.473}$ (5/2$^{+}$)}   &  (3s$_{1/2}$)  &   0.1866  \\
						       &  						    &  (1g$_{7/2}$)  &   0.5306  \\
\cline{2-4}
						       &  \multirow{3}*{$^{115}$Cd$_{0.507}$ (3/2$^{+}$)}   &  (3s$_{1/2}$)  &   0.1228  \\ 
						       &  				 	 	    &  (2d$_{5/2}$)  &   0.1089  \\
						       &  				 	 	    &  (1g$_{7/2}$)  &  -0.4713  \\ 
\cline{2-4}
						       &  \multirow{3}*{$^{115}$Cd$_{0.507}$ (5/2$^{+}$)}   &  (3s$_{1/2}$)  &  -0.1367   \\ 
						       &  				 	 	    &  (2d$_{3/2}$)  &   0.2446   \\
						       &  				 	 	    &  (1g$_{7/2}$)  &  -0.2707   \\ 
\cline{2-4}
						       &  $^{115}$Cd$_{0.649}$ (1/2$^{+}$) 	            &  (2d$_{3/2}$)  &  -0.2863  \\
						       &						    &  (2d$_{5/2}$)  &   0.1247  \\
\cline{2-4}
						       &  \multirow{2}*{$^{115}$Cd$_{1.062}$ (7/2$^{+}$)}   &  (2d$_{3/2}$)  &   0.5040  \\
						       &  				 	 	    &  (1g$_{7/2}$)  &   0.4286  \\

\bottomrule
\end{tabular}
\end{table}

\begin{table}[tp!]
\caption{\label{tab:amplitudes_path1_targ2} Spectroscopic amplitudes (S.A.) obtained through shell model $88Sr45$ interaction and adopted in CCBA calculations for all the target-like nuclei involved along the second step of path1 (see text) of the $^{116}$Cd($^{20}$Ne,$^{20}$F)$^{116}$In SCE reaction and whose final states lie within the excitation energy range here considered. % ([0.00, 0.35] MeV). 
The column nl$_j$ indicates the principal quantum number, the orbital and the total angular momentum of the single valence proton, respectively, through the usual spectroscopic notation.}
\centering
\begin{tabular}{*{4}{c}}
\textbf{Initial state}				       & \textbf{Final state} 				 	& \textbf{nl$_j$} & \textbf{S.A.} \\
\toprule
$^{115}$Cd$_{0.229}$ (3/2$^{+}$)        	     & \multirow{8}*{$^{116}$In$_{\textup{g.s.}}$ (1$^{+}$)}  	&  (2d$_{5/2}$)  &  -0.1978  \\
\cline{1-1}\cline{3-4}
$^{115}$Cd$_{0.389}$ (7/2$^{+}$)  	 	     & 				                      		&  (1g$_{9/2}$)  &  -0.9798  \\
\cline{1-1}\cline{3-4}
$^{115}$Cd$_{0.389}$ (9/2$^{+}$)  	 	     & 				                      		&  (1g$_{9/2}$)  &   0.3486  \\
\cline{1-1}\cline{3-4}
$^{115}$Cd$_{0.473}$ (3/2$^{+}$)  	 	     & 				                      		&  (2d$_{5/2}$)  &  -0.1457  \\
\cline{1-1}\cline{3-4}
$^{115}$Cd$_{0.473}$ (5/2$^{+}$)  	 	     & 				                      		&  (2d$_{5/2}$)  &   0.1498  \\
\cline{1-1}\cline{3-4}
$^{115}$Cd$_{0.507}$ (3/2$^{+}$)  	 	     & 				                      		&  (2d$_{5/2}$)  &  -0.1540  \\
\cline{1-1}\cline{3-4}
\multirow{2}*{$^{115}$Cd$_{1.062}$ (7/2$^{+}$)}      & 				                      		&  (2d$_{5/2}$)  &   0.1505  \\
						     & 				                      		&  (1g$_{9/2}$)  &   0.2514  \\
\midrule
$^{115}$Cd$_{\textup{g.s.}}$ (1/2$^{+}$)  	     & 	\multirow{11}*{$^{116}$In$_{0.128}$ (5$^{+}$)}		&  (1g$_{9/2}$)  &  -0.3145  \\
\cline{1-1}\cline{3-4}
$^{115}$Cd$_{0.229}$ (3/2$^{+}$)        	     & 						 	 	&  (1g$_{9/2}$)  &   0.5385  \\
\cline{1-1}\cline{3-4}
$^{115}$Cd$_{0.361}$ (5/2$^{+}$)        	     & 						 	 	&  (1g$_{9/2}$)  &   0.1772  \\
\cline{1-1}\cline{3-4}
$^{115}$Cd$_{0.389}$ (7/2$^{+}$)  	 	     & 				                      		&  (1g$_{9/2}$)  &  -0.2004  \\
\cline{1-1}\cline{3-4}
$^{115}$Cd$_{0.389}$ (9/2$^{+}$)  	 	     & 				                      		&  (2d$_{5/2}$)  &  -0.1084  \\
\cline{1-1}\cline{3-4}
$^{115}$Cd$_{0.473}$ (3/2$^{+}$)  	 	     & 				                      		&  (1g$_{9/2}$)  &  -0.2877  \\
\cline{1-1}\cline{3-4}
$^{115}$Cd$_{0.473}$ (5/2$^{+}$)  	 	     & 				                      		&  (1g$_{9/2}$)  &  -0.2605  \\
\cline{1-1}\cline{3-4}
$^{115}$Cd$_{0.507}$ (3/2$^{+}$)  	 	     & 				                      		&  (1g$_{9/2}$)  &   0.1585  \\
\cline{1-1}\cline{3-4}
$^{115}$Cd$_{0.507}$ (5/2$^{+}$)  	 	     & 				                      		&  (1g$_{9/2}$)  &  -0.2633  \\
\cline{1-1}\cline{3-4}
$^{115}$Cd$_{0.649}$ (1/2$^{+}$)  	 	     & 				                      		&  (1g$_{9/2}$)  &   0.1831  \\
\cline{1-1}\cline{3-4}
$^{115}$Cd$_{1.062}$ (7/2$^{+}$)	             & 				                      		&  (1g$_{9/2}$)  &  -0.8241  \\
\midrule
$^{115}$Cd$_{\textup{g.s.}}$ (1/2$^{+}$)  	     & 	\multirow{12}*{$^{116}$In$_{0.223}$ (4$^{+}$)}		&  (1g$_{9/2}$)  &  -0.4179  \\
\cline{1-1}\cline{3-4}
$^{115}$Cd$_{0.229}$ (3/2$^{+}$)        	     & 						 	 	&  (1g$_{9/2}$)  &  -0.4774  \\
\cline{1-1}\cline{3-4}
\multirow{2}*{$^{115}$Cd$_{0.361}$ (5/2$^{+}$)}	     & 						 	 	&  (2d$_{5/2}$)  &  -0.1412  \\
						     & 						 	 	&  (1g$_{9/2}$)  &   0.5945  \\
\cline{1-1}\cline{3-4}
$^{115}$Cd$_{0.389}$ (7/2$^{+}$)  	 	     & 				                      		&  (1g$_{9/2}$)  &   0.1464  \\
\cline{1-1}\cline{3-4}
\multirow{2}*{$^{115}$Cd$_{0.389}$ (9/2$^{+}$)}	     & 						 	 	&  (2d$_{5/2}$)  &   0.1171  \\
						     & 						 	 	&  (1g$_{9/2}$)  &  -0.8536  \\
\cline{1-1}\cline{3-4}
$^{115}$Cd$_{0.473}$ (3/2$^{+}$)  	 	     & 				                      		&  (1g$_{9/2}$)  &   0.1176  \\
\cline{1-1}\cline{3-4}
$^{115}$Cd$_{0.473}$ (5/2$^{+}$)  	 	     & 				                      		&  (1g$_{9/2}$)  &  -0.4535  \\
\cline{1-1}\cline{3-4}
$^{115}$Cd$_{0.507}$ (5/2$^{+}$)  	 	     & 				                      		&  (1g$_{9/2}$)  &  -0.2989  \\
\cline{1-1}\cline{3-4}
$^{115}$Cd$_{0.649}$ (1/2$^{+}$)  	 	     & 				                      		&  (1g$_{9/2}$)  &   0.2337  \\
\cline{1-1}\cline{3-4}
$^{115}$Cd$_{1.062}$ (7/2$^{+}$)	             & 				                      		&  (1g$_{9/2}$)  &   0.5216  \\
\midrule
$^{115}$Cd$_{0.361}$ (5/2$^{+}$) 	    	     & 	\multirow{9}*{$^{116}$In$_{0.273}$ (2$^{+}$)}		&  (2d$_{5/2}$)  &  -0.1858  \\
\cline{1-1}\cline{3-4}
\multirow{2}*{$^{115}$Cd$_{0.389}$ (7/2$^{+}$)}      & 						 	 	&  (2d$_{5/2}$)  &  -0.1401  \\
						     & 						 	 	&  (1g$_{9/2}$)  &   0.5919  \\
\cline{1-1}\cline{3-4}
$^{115}$Cd$_{0.389}$ (7/2$^{+}$)	             & 						 	 	&  (1g$_{9/2}$)  &  -0.4539  \\
\cline{1-1}\cline{3-4}
$^{115}$Cd$_{0.473}$ (3/2$^{+}$)  	 	     & 				                      		&  (2d$_{5/2}$)  &   0.1242  \\
\cline{1-1}\cline{3-4}
$^{115}$Cd$_{0.473}$ (5/2$^{+}$)  	 	     & 				                      		&  (1g$_{9/2}$)  &  -0.1142  \\
\cline{1-1}\cline{3-4}
$^{115}$Cd$_{0.507}$ (5/2$^{+}$)  	 	     & 				                      		&  (2d$_{5/2}$)  &   0.1230 \\
\cline{1-1}\cline{3-4}
\multirow{2}*{$^{115}$Cd$_{1.062}$ (7/2$^{+}$)}      & 				                      		&  (2d$_{5/2}$)  &  -0.1611  \\
						     & 				                      		&  (1g$_{9/2}$)  &  -0.1623  \\
\midrule
$^{115}$Cd$_{\textup{g.s.}}$ (1/2$^{+}$)  	     & 	\multirow{3}*{$^{116}$In$_{0.313}$ (4$^{+}$)}		&  (1g$_{9/2}$)  &  -0.1678  \\
\cline{1-1}\cline{3-4}
$^{115}$Cd$_{0.473}$ (3/2$^{+}$)  	 	     & 				                      		&  (1g$_{9/2}$)  &   0.1308  \\
\cline{1-1}\cline{3-4}
$^{115}$Cd$_{0.649}$ (1/2$^{+}$)  	 	     & 				                      		&  (1g$_{9/2}$)  &  -0.3518  \\
\bottomrule
\end{tabular}
\end{table}

\begin{table}[tp!]
\caption{\label{tab:amplitudes_path2_proj} Spectroscopic amplitudes (S.A.) obtained through shell model $zbm$ interaction and adopted in CCBA calculations for all the projectile-like nuclei involved along the path2 (see text) of the $^{116}$Cd($^{20}$Ne,$^{20}$F)$^{116}$In SCE reaction and whose final states lie within the excitation energy range here considered. % ([0.00, 0.35] MeV). 
The column nl$_j$ indicates the principal quantum number, the orbital and the total angular momentum of the single valence proton, respectively, through the usual spectroscopic notation.}
\centering
\begin{tabular}{*{4}{c}}
\textbf{Initial state} & \textbf{Final state} & \textbf{nl$_j$} & \textbf{S.A.} \\
\toprule
\multirow{2}*{$^{20}$Ne$_{\textup{g.s.}}$ (0$^{+}$)}  &  $^{19}$F$_{\textup{g.s.}}$ (1/2$^{+}$)	        &  (2s$_{1/2}$)  &  -0.8584  \\
%\cline{2-4}
%					   	      &  $^{19}$F$_{0.110}$ (1/2$^{-}$)  		&  (1p$_{1/2}$)  &  -1.2702  \\
\cline{2-4}
					   	      &  $^{19}$F$_{0.197}$ (5/2$^{+}$)  		&  (1d$_{5/2}$)  &   1.1741  \\
\midrule
\multirow{8}*{$^{20}$Ne$_{\textup{1.634}}$ (2$^{+}$)} &  $^{19}$F$_{\textup{g.s.}}$ (1/2$^{+}$) 	&  (1d$_{5/2}$)  &   0.6712  \\
%\cline{2-4}
%					   	     &  $^{19}$F$_{0.110}$ (1/2$^{-}$)  		&  (1p$_{1/2}$)  &  -1.2702  \\
\cline{2-4}
					   	     &  \multirow{2}*{$^{19}$F$_{0.197}$ (5/2$^{+}$)}  	&  (2s$_{1/2}$)  &  -0.6922  \\
						     &  					 	&  (1d$_{5/2}$)  &  -0.6416  \\
\cline{2-4}
						     & 	$^{19}$F$_{1.346}$ (5/2$^{-}$)			&  (1p$_{1/2}$)  &  -0.9779  \\
%\cline{2-4}
%						     & 	$^{19}$F$_{1.459}$ (3/2$^{-}$)			&  (1p$_{1/2}$)  &   0.8157  \\
\cline{2-4}
					   	     &  \multirow{2}*{$^{19}$F$_{1.554}$ (3/2$^{+}$)}  	&  (2s$_{1/2}$)  &   0.3767  \\
						     &  					 	&  (1d$_{5/2}$)  &   0.2923  \\
\cline{2-4}
						     & 	$^{19}$F$_{2.780}$ (9/2$^{+}$)			&  (1d$_{5/2}$)  &  -0.6248  \\
\cline{2-4}
						     & 	$^{19}$F$_{4.378}$ (7/2$^{+}$)			&  (1d$_{5/2}$)  &  -0.1995  \\
\midrule
\midrule
$^{19}$F$_{\textup{g.s.}}$ (1/2$^{+}$)        	     & \multirow{7}*{$^{20}$F$_{\textup{g.s.}}$ (2$^{+}$)}  	&  (1d$_{5/2}$)  &  -0.2291  \\
\cline{1-1}\cline{3-4}
\multirow{2}*{$^{19}$F$_{0.197}$ (5/2$^{+}$)}  	     & 				                      		&  (2s$_{1/2}$)  &   0.1215  \\
					             & 						      		&  (1d$_{5/2}$)  &  -0.8891  \\
\cline{1-1}\cline{3-4}
$^{19}$F$_{\textup{1.346}}$ (5/2$^{-}$)              & 						      		&  (1p$_{1/2}$)  &  -0.1475  \\
\cline{1-1}\cline{3-4}
$^{19}$F$_{\textup{1.554}}$ (3/2$^{+}$)              & 							  	&  (1d$_{5/2}$)  &  -0.4445  \\
\cline{1-1}\cline{3-4}
$^{19}$F$_{\textup{2.780}}$ (9/2$^{+}$)              &  							&  (1d$_{5/2}$)  &  -0.5137  \\
\cline{1-1}\cline{3-4}
$^{19}$F$_{\textup{4.378}}$ (7/2$^{+}$)              & 								&  (1d$_{5/2}$)  &  -0.5312  \\
\bottomrule
\end{tabular}
\end{table}

\begin{table}[tp!]
\caption{\label{tab:amplitudes_path2_targ} Spectroscopic amplitudes (S.A.) obtained through shell model $88Sr45$ interaction and adopted in CCBA calculations for all the target-like nuclei involved along the path2 (see text) of the $^{116}$Cd($^{20}$Ne,$^{20}$F)$^{116}$In SCE reaction and whose final states lie within the excitation energy range here considered. % ([0.00, 0.35] MeV). 
The column nl$_j$ indicates the principal quantum number, the orbital and the total angular momentum of the single valence proton, respectively, through the usual spectroscopic notation.}
\centering
\begin{tabular}{*{4}{c}}
\textbf{Initial state}				       & \textbf{Final state} 				 	& \textbf{nl$_j$} & \textbf{S.A.} \\
\toprule
\multirow{6}*{$^{116}$Cd$_{\textup{g.s.}}$ (0$^{+}$)}   &  $^{117}$In$_{\textup{g.s.}}$ (9/2$^{+}$)  		    &  (1g$_{9/2}$)  &  -0.4066  \\
%\cline{2-4}
%						        &  $^{117}$In$_{0.315}$ (1/2$^{-}$) 	    		    &  (2s$_{1/2}$)  &  -0.3847  \\
\cline{2-4}
						        &  $^{117}$In$_{0.748}$ (7/2$^{+}$) 	    		    &  (1g$_{7/2}$)  &   0.0098  \\
\cline{2-4}
						        &  $^{117}$In$_{0.881}$ (5/2$^{+}$) 	    		    &  (2d$_{5/2}$)  &  -0.1593  \\
\cline{2-4}
						        &  $^{117}$In$_{1.030}$ (5/2$^{+}$) 	    		    &  (2d$_{5/2}$)  &  -0.2670  \\
\cline{2-4}
						        &  $^{117}$In$_{1.052}$ (5/2$^{+}$) 	    		    &  (2d$_{5/2}$)  &  -0.7192  \\
\cline{2-4}
						        &  $^{117}$In$_{1.365}$ (7/2$^{+}$) 	    		    &  (1g$_{7/2}$)  &  -0.0027  \\
\cline{2-4}
						        &  $^{117}$In$_{1.365}$ (9/2$^{+}$) 	    		    &  (1g$_{9/2}$)  &   0.0383  \\
\midrule
\multirow{10}*{$^{116}$Cd$_{\textup{0.513}}$ (2$^{+}$)} &  \multirow{2}*{$^{117}$In$_{\textup{g.s.}}$ (9/2$^{+}$)}  &  (2d$_{5/2}$)  &   0.1182  \\
						        &  					   		    &  (1g$_{9/2}$)  &  -0.6092  \\
%\cline{2-4}
%						        &  $^{117}$In$_{0.589}$ (3/2$^{-}$) 	   		    &  (2p$_{3/2}$)  &  -0.3319  \\
%\cline{2-4}
%						        &  $^{117}$In$_{0.660}$ (3/2$^{+}$) 	   		    &  (2d$_{5/2}$)  &  -0.1502  \\
\cline{2-4}
						        &  $^{117}$In$_{0.748}$ (7/2$^{+}$) 	    		    &  (1g$_{9/2}$)  &  -0.2232  \\
%\cline{2-4}
%						        &  $^{117}$In$_{0.749}$ (1/2$^{+}$) 	    		    &  (2d$_{5/2}$)  &  -0.2796  \\
%\cline{2-4}
%						        &  $^{117}$In$_{0.780}$ (5/2$^{-}$) 	    		    &  (2p$_{3/2}$)  &   0.3306  \\
\cline{2-4}
						        &  \multirow{2}*{$^{117}$In$_{0.881}$ (5/2$^{+}$)} 	    &  (2d$_{5/2}$)  &  -0.1781  \\
							&  					   		    &  (1g$_{9/2}$)  &   0.4308  \\
\cline{2-4}
						        &  \multirow{2}*{$^{117}$In$_{1.052}$ (5/2$^{+}$)} 	    &  (2d$_{5/2}$)  &   0.5693  \\
							&  					   		    &  (1g$_{9/2}$)  &  -0.1326  \\
%\cline{2-4}
%						        &  $^{117}$In$_{1.066}$ (11/2$^{+}$) 	   		    &  (1g$_{9/2}$)  &  -0.2847  \\
%\cline{2-4}
%						        &  $^{117}$In$_{1.235}$ (13/2$^{+}$) 	   		    &  (1g$_{9/2}$)  &  -0.3737  \\
\cline{2-4}
						        &  $^{117}$In$_{1.365}$ (7/2$^{+}$) 	   		    &  (2d$_{5/2}$)  &   0.1889  \\
\cline{2-4}
						        &  $^{117}$In$_{1.365}$ (9/2$^{+}$) 	   		    &  (1g$_{9/2}$)  &  -0.1813  \\
%\cline{2-4}
%						        &  $^{117}$In$_{1.609}$ (5/2$^{-}$) 	   		    &  (2p$_{1/2}$)  &   0.3306  \\
\midrule
\midrule
$^{117}$In$_{\textup{g.s.}}$ (9/2$^{+}$) 		&  \multirow{4}*{$^{116}$In$_{\textup{g.s.}}$ (1$^{+}$)}   &  (1g$_{7/2}$)  &  -0.5008  \\
\cline{1-1}\cline{3-4}
$^{117}$In$_{0.881}$ (5/2$^{+}$)			&  					           	   &  (2d$_{3/2}$)  &  -0.2335  \\
\cline{1-1}\cline{3-4}
$^{117}$In$_{1.030}$ (5/2$^{+}$)			&  					           	   &  (2d$_{3/2}$)  &  -0.4548  \\
\cline{1-1}\cline{3-4}
$^{117}$In$_{1.052}$ (5/2$^{+}$)			&  					           	   &  (2d$_{3/2}$)  &   0.2233  \\
\midrule
\multirow{3}*{$^{117}$In$_{\textup{g.s.}}$ (9/2$^{+}$)} &  \multirow{10}*{$^{116}$In$_{\textup{0.128}}$ (5$^{+}$)}  &  (1g$_{7/2}$)  &  -0.1534  \\
							&	  					           &  (2d$_{3/2}$)  &  -0.9107  \\
							&  					           	   &  (3s$_{1/2}$)  &  -0.1969  \\
\cline{1-1}\cline{3-4}
\multirow{3}*{$^{117}$In$_{0.748}$ (7/2$^{+}$)}  	&  							   &  (1g$_{7/2}$)  &   0.1283  \\
							&	  					           &  (2d$_{3/2}$)  &   0.7751  \\
\cline{1-1}\cline{3-4} 
$^{117}$In$_{0.881}$ (5/2$^{+}$)			&  					           	   &  (2d$_{5/2}$)  &   0.1249  \\
\cline{1-1}\cline{3-4}
$^{117}$In$_{1.030}$ (5/2$^{+}$)			&  					           	   &  (1g$_{7/2}$)  &   0.1178  \\
\cline{1-1}\cline{3-4}
$^{117}$In$_{1.365}$ (7/2$^{+}$)			&  					           	   &  (2d$_{3/2}$)  &  -0.4105  \\
\cline{1-1}\cline{3-4}
\multirow{2}*{$^{117}$In$_{1.365}$ (9/2$^{+}$)}		&  					           	   &  (1g$_{7/2}$)  &  -0.1519  \\
							&  					           	   &  (2d$_{3/2}$)  &  -0.4675  \\
\midrule
\multirow{2}*{$^{117}$In$_{\textup{g.s.}}$ (9/2$^{+}$)} &  \multirow{7}*{$^{116}$In$_{\textup{0.223}}$ (4$^{+}$)} &  (2d$_{3/2}$)  &   0.6852  \\
							&  					           	   &  (3s$_{1/2}$)  &  -0.2581  \\
\cline{1-1}\cline{3-4}
\multirow{2}*{$^{117}$In$_{0.748}$ (7/2$^{+}$)}  	&  							   &  (1g$_{7/2}$)  &   0.1143  \\
							&	  					           &  (2d$_{3/2}$)  &   0.2821  \\
\cline{1-1}\cline{3-4} 
$^{117}$In$_{0.881}$ (5/2$^{+}$)			&  					           	   &  (2d$_{3/2}$)  &   0.9649  \\
\cline{1-1}\cline{3-4}
$^{117}$In$_{1.030}$ (5/2$^{+}$)			&  					           	   &  (2d$_{3/2}$)  &  -0.4573  \\
\cline{1-1}\cline{3-4}
$^{117}$In$_{1.365}$ (9/2$^{+}$)			&  					           	   &  (3s$_{1/2}$)  &  -0.1604  \\
\midrule
$^{117}$In$_{\textup{g.s.}}$ (9/2$^{+}$)	        &  \multirow{5}*{$^{116}$In$_{\textup{0.273}}$ (2$^{+}$)}  &  (1g$_{7/2}$)  &   0.6212  \\
\cline{1-1}\cline{3-4}
$^{117}$In$_{0.748}$ (7/2$^{+}$)		 	&  							   &  (2d$_{3/2}$)  &  -0.1605  \\
\cline{1-1}\cline{3-4} 
$^{117}$In$_{0.881}$ (5/2$^{+}$)			&  					           	   &  (2d$_{3/2}$)  &   0.1612  \\
\cline{1-1}\cline{3-4}
$^{117}$In$_{1.030}$ (5/2$^{+}$)			&  					           	   &  (2d$_{3/2}$)  &   0.3029  \\
\cline{1-1}\cline{3-4}
$^{117}$In$_{1.365}$ (7/2$^{+}$)			&  					           	   &  (2d$_{3/2}$)  &  -0.3473  \\
\midrule
\multirow{3}*{$^{117}$In$_{\textup{g.s.}}$ (9/2$^{+}$)} &  \multirow{8}*{$^{116}$In$_{\textup{0.313}}$ (4$^{+}$)}  &  (2d$_{5/2}$)  &  -0.1060  \\
							&  					           	   &  (2d$_{3/2}$)  &  -0.2653  \\
							&  					           	   &  (3s$_{1/2}$)  &  -0.8248  \\
\cline{1-1}\cline{3-4}
$^{117}$In$_{0.748}$ (7/2$^{+}$)		  	&  							   &  (2d$_{3/2}$)  &  -0.1569 \\
\cline{1-1}\cline{3-4} 
$^{117}$In$_{0.881}$ (5/2$^{+}$)			&  					           	   &  (2d$_{3/2}$)  &   0.1433  \\
\cline{1-1}\cline{3-4}
$^{117}$In$_{1.030}$ (5/2$^{+}$)			&  					           	   &  (2d$_{3/2}$)  &   0.2034  \\
\cline{1-1}\cline{3-4}
$^{117}$In$_{1.052}$ (5/2$^{+}$)			&  					           	   &  (2d$_{3/2}$)  &  -0.1049  \\
\cline{1-1}\cline{3-4}
$^{117}$In$_{1.365}$ (7/2$^{+}$)			&  					           	   &  (2d$_{3/2}$)  &   0.2134  \\
\bottomrule
\end{tabular}
\end{table}

\bibliographystyle{ieeetr}
\bibliography{references.bib}

\newpage


\begin{thebibliography}{}{}
\bibitem{Osterfeld} F. Osterfeld, Rev. Mod. Phys. {\bf 64} 2, 491-550 (1992). %Nuclear spin and isospin excitations
\bibitem{Ichimura06} M. Ichimura, H. Sakai and T. Wakasa, Prog. Part. Nucl. Phys. {\bf 56}, 446 (2006). %Gamow-Teller Strength Distributions in [...] and Two-Neutrino Double- Decay Nuclear Matrix Elements
\bibitem{Frekers13} D. Frekers et al., Nucl. Phys. A {\bf 916}, 219 (2013). %Gamow–Teller strength extraction from (3He, t) reactions
\bibitem{Brendel:1988mwl} C. Brendel et al., Nucl. Phys. A {\bf 477} 162-188 (1988).
\bibitem{AndersonPRC1991} B. D. Anderson et al., Phys. Rev. C {\bf 43}, 50 (1991).%Gamow-Teller strength in the (p,n) reaction at 136 MeV on 20Ne, 24Mg, and 28Si
\bibitem{blomgren} J. Blomgren et al., Phys. Lett. B {\bf 362}, 34 - 38 (1995). %Search for double Gamow-Teller strength by heavy-ion double charge exchange
\bibitem{YakoPLB2005} K. Yako et al., Phys. Lett. B {\bf 615}, 193-199 (2005). %Determination of the Gamow–Teller quenching factor from charge exchange reactions on 90Zr
\bibitem{FrekersPPNP2006} D. Frekers, Progr. Part. Nucl. Phys. {\bf 57}, 217–22 (2006). %Facets of charge-exchange reactions — from astrophysics to double beta decay
\bibitem{FujitaPRC2007} H. Fujita, Y. Fujita, T. Adachi, A. D. Bacher, G. P. A. Berg, T. Black et al., Phys. Rev. C {\bf 75}, 034310 (2007). %Isospin structure of Jπ=1+ states in 58Ni and 58Cu studied by 58Ni(p,p′) and 58Ni(3He,t)58Cu measurements
\bibitem{FujitaPPNP2011} Y. Fujita, B. Rubio, W. Gelletly, Progr. Part. Nucl. Phys. {\bf 66}, 549-606 (2011). %Spin–isospin excitations probed by strong, weak and electro-magnetic interactions
\bibitem{Douma2020} C. A. Douma et al., Eur. Phys. J. A {\bf 56}, 51 (2020). %Gamow–Teller strength distributions of 116Sb and 122Sb using the (3He,𝑡) charge-exchange reaction
\bibitem{EPJA2018} F. Cappuzzello et al., Eur. Phys. J. A {\bf 54}, 72 (2018). %The NUMEN project: NUclear Matrix Elements for Neutrinoless double beta decay
\bibitem{EPJA2015} F. Cappuzzello et al., Eur. Phys. J. A {\bf 51}, 145 (2015). %Heavy-ion double charge exchange reactions: A tool toward 0vbb nuclear matrix elements
\bibitem{Agodi1} C. Agodi et al., Nucl. Part. Phys. Proceed. {\bf 265}-{\bf 266}, 28-30 (2015).
\bibitem{Agodi2} C. Agodi et al., Universe {\bf 7(3)}, 72 (2021).
\bibitem{sasanoPRC2012} M. Sasano et al., Phys. Rev. C {\bf 85}, 061301(R) (2012).
\bibitem{Kisamori2016} K. Kisamori et al., Phys. Rev. Lett. {\bf 116}, 052501 (2016). %Candidate Resonant Tetraneutron State Populated by the 4He(8He,8Be) Reaction
\bibitem{Miki2017} K. Miki et al., Phys. Lett. B {\bf 769}, 339-344 (2017). %Isovector excitations in 100Nb and their decays by neutron emission studied via the Mo100 (t, He3+ n) reaction at 115 MeV/u
\bibitem{Taddeucci87} T. N. Taddeucci et al., Nucl. Phys. A {\bf 469}, 125 (1987). %The (p, n) reaction as a probe of beta decay strength
\bibitem{Alford} W. P. Alford and B. M. Spicer, Adv. Nucl. Phys. {\bf 24}, 1 (1998).
\bibitem{LenPPNP2019} H. Lenske et al., Progr. Part. Nucl. Phys. {\bf 109}, 103716 (2019). %Heavy ion charge exchange reactions as probes for nuclear β\betaβ-decay
\bibitem{Lenske18} H. Lenske, J. I. Bellone, M. Colonna, J. A. Lay, Phys. Rev. C {\bf 98}, 044620 (2018). %Theory of single-charge exchange heavy-ion reactions
\bibitem{CavFRO2021} M. Cavallaro et al., Front. Astron. Space Sci. {\bf 8}, 659815 (2021).%A constrained analysisof the 40Ca(18O,18F)40K direct charge exchange reaction mechanism at 275 MeV.
\bibitem{BelPLB2020} J. I. Bellone et al., Phys. Lett. B {\bf 807}, 135528 (2020). %Two-step description of heavy ion double charge exchange reactions
\bibitem{LenskeUniverse} Lenske et al., Universe 7, 98 (2021).
\bibitem{SanPRC2018} E. Santopinto, H. Garcia-Tecocoatzi, R. I. Magana Vsevolodovna, J. Ferretti et al., Phys. Rev. C {\bf 98}, 061601(R), (2018). %Heavy-ion double-charge-exchange and its relation to neutrinoless double-β decay
\bibitem{ShimizuPRL2018} N. Shimizu, J. Menéndez, and K. Yako, Phys. Rev. Lett. {\bf 120}, 142502 (2018). %Double Gamow-Teller Transitions and its Relation to Neutrinoless ββ Decay
\bibitem{EngRPP2017} J. Engel, and J. Menéndez, Rep. Progr. Phys. {\bf 80}, 046301 (2017). %Status and future of nuclear matrix elements for neutrinoless double-beta decay: a review
\bibitem{GamPRL2020} D. Gambacurta, M. Grasso, J. Engel, Phys. Rev. Lett. {\bf 125}, 212501 (2020). %Gamow-Teller Strength in 48Ca and 78Ni with the Charge-Exchange Subtracted Second Random-Phase Approximation
\bibitem{CapLenske04} F. Cappuzzello et al., Nucl. Phys. A {\bf 739}, 30-56 (2004). %Analysis of the 11B(7Li, 7Be)11Be reaction at 57 MeV in a microscopic approach
\bibitem{LayBurrello} J. A. Lay et al., Jour. Phys.: Conf. Series {\bf 1056}, 012029 (2018). %Double Charge-Exchange Reactions and the effect of transfer
\bibitem{Oertzen} W. von Oertzen, Nucl. Phys. A {\bf 482}, 357 (1988). %Excitation of isovector modes in heavy ion induced charge exchange reactions
\bibitem{Winfield} J. S. Winfield et al., Phys. Rev. C {\bf 33}, 1333 (1986); Phys. Rev. C {\bf 35}, 1166(E) (1987). %Mechanism of the heavy-ion charge exchange reaction 12C(12C,12N)12B at 35 MeV/nucleon
\bibitem{Len89} H. Lenske, H. H. Wolter and H. G. Bohlen, Phys. Rev. Lett. {\bf 62}, 1457 (1989). %Reaction mechanism of heavy-ion charge-exchange scattering at intermediate energies
\bibitem{DasPRC1986} C. H. Dasso and A. Vitturi, Phys. Rev. C {\bf 34}, 743 (1986). %Mechanism for double-charge exchange in heavy ion reactions
\bibitem{SpaPRC19} A. Spatafora, F. Cappuzzello, D. Carbone, M. Cavallaro, J. A. Lay, L. Acosta, et al., Phys. Rev. C {\bf 100}, 034620 (2019).
\bibitem{Akis} O. Sgouros et al., Phys. Rev. C {\bf 104}, 034617 (2021). %Study of one-proton transfer reaction for the 18O + 48Ti system at 275 MeV
\bibitem{FauPRC21} L. La Fauci et al., Phys. Rev. C {\bf 104}, 054610 (2021). %O18+Se76 elastic and inelastic scattering at 275 MeV
\bibitem{CarUNI2021} D. Carbone et al., Universe {\bf 7}, 58 (2021). %Initial State Interaction for the 20Ne + 130Te and 18O + 116Sn Systems at 15.3 AMeV from Elastic and Inelastic Scattering Measurements
\bibitem{Diana} D. Carbone, J. L. Ferreira, S. Calabrese, F. Cappuzzello, M. Cavallaro, A. Hacisalihoglu et al., Phys. Rev. C {\bf 102}, 044606 (2020). %Analysis of two-nucleon transfer reactions in the 20Ne + 116Cd system at 306 MeV
\bibitem{Jonas} J. L. Ferreira, D. Carbone, M. Cavallaro, N. N. Deshmukh, C. Agodi, G. A. Brischetto et al., Phys. Rev. C {\bf 103}, 054604 (2021). %Analysis of two-proton transfer in the 40Ca(18O,20Ne)38Ar reaction at 270 MeV incident energy
\bibitem{Sat83} G. R. Satchler. Direct Nuclear Reactions, Oxford University Press, Oxford and New York 1983.
\bibitem{TimoPPNP2019} N. K. Timofeyuk and R. C. Jonhson, Prog. Part. Nucl. Phys. {\bf 111}, 103738 (2019).
\bibitem{MontPRC2011} D. Montanari, S. Leoni, L. Corradi, G. Pollarolo, G. Benzoni, N. Blasi et al., Phys. Rev. C {\bf 84}, 054613 (2011).
\bibitem{GasPRC2018} L. R. Gasques  et al., Phys. Rev. C {\bf 97}, 034629 (2018).
\bibitem{Tamura:1974} T. Tamura, Phys. Rep. \textbf{14}, 59-96 (1974). %Compact reformulation of distorted-wave and coupled-channel Born approximations for transfer reactions between nuclei,
\bibitem{KeePRC2020} N. Keeley, K. W. Kemper and K. Rusek, Phys. Rev. C {\bf 102}, 014617 (2020).
\bibitem{Thompson} I. J. Thompson, Comp. Phys. Rep. {\bf 7} 167-212 (1988). 
\bibitem{CAVALLARO2020334} M. Cavallaro et al., Nucl. Instrum Methods Phys. Res. B {\bf 463}, 334-338 (2020). %The MAGNEX magnetic spectrometer for double charge exchange reactions
\bibitem{shima1992} K. Shima, N. Kuno, M. Yamanouchi, and H. Tawara, Atom. Data Nucl. Data Tabl. {\bf 51}, 173-241 (1992). %Equilibrium charge fractions of ions of [...] emerging from a carbon foil",
\bibitem{cavallaro2019} M. Cavallaro et al., Results in Physics {\bf 13}, 102191 (2019). %Charge-state distributions of 20Ne ions emerging from thin foils"
\bibitem{magnex_review} F. Cappuzzello, C. Agodi, D. Carbone, and M. Cavallaro, Eur. Phys. J. A {\bf 52}, 167 (2016). %The MAGNEX spectrometer: results and perspectives
\bibitem{cappuzzello2010} F. Cappuzzello et al., Nucl. Instrum. Methods Phys. Res. A {\bf 621}, 419-423 (2010). %A particle identification technique for large acceptance spectrometers",
\bibitem{cavallaro2012} M. Cavallaro et al., Eur. Phys. J. A {\bf 48}, 59 (2012). %The low-pressure focal plane detector of the MAGNEX spectrometer
\bibitem{TORRESI2021164918} D. Torresi et al., Nucl. Instrum Methods Phys. Res. A {\bf 989}, 164918 (2021). %An upgraded focal plane detector for the MAGNEX spectrometer 
\bibitem{PerPLB09} D. Pereira et al., Phys. Lett. B {\bf 670}, 330 (2009).
\bibitem{PerPLB2012} D. Pereira et al., Phys. Lett. B {\bf 710}, 426 (2012).
\bibitem{FonPRC2019} L. M. Fonseca, R. Linares, V. A. B. Zagatto, F. Cappuzzello, D. Carbone, M. Cavallaro, C. Agodi, J. Lubian, J. R. B. Oliveira et al., Phys. Rev. C {\bf 100}, 014604 (2019).
\bibitem{PerNPA2009} D. Pereira et al., Nucl. Phys. A {\bf 826}, 211 (2009).
\bibitem{ZagPRC2018} V. A. Zagatto et al., Phys. Rev. C {\bf 97}, 054608 (2018). %Important role of projectile excitation in 16O + 60Ni and 16O + 27Al scattering at intermediate energies
\bibitem{AlvNPA2003} M. A. G. Alvarez et al., Nucl. Phys. A {\bf 723}, 93 (2003).
\bibitem{FraneyLove} M. A. Franey and W. G. Love, Phys. Rev. C {\bf 31} 488 (1985).
\bibitem{Ram2001} S. Raman, C. W. Nestor, and P. Tikkanen. Atom. Data Nucl. Data Tabl. {\bf 78}, 1-128, (2001). %Transition probability from the ground to the first-excited 2+ state of even-even nuclides
\bibitem{Pri2016} B. Pritychenko, M. Birch, B. Singh, and M. Horoi, Atom. Data Nucl. Data Tabl. {\bf 107}, 1 (2016).
\bibitem{SatcPR79} G.R. Satchler and W. G. Love., Phys. Rep. {\bf 55}, 183-254 (1979). %Folding model potentials from realistic interactions for heavy-ion scattering.
\bibitem{PaePRC17} B. Paes et al., Phys. Rev. C {\bf 96}, 044612 (2017). %Long-range versus short-range correlations in the two-neutron transfer reaction 64Ni(18O,16O)66Ni
\bibitem{CarPRC17} D. Carbone et al., Phys. Rev. C {\bf 95}, 034603 (2017).
\bibitem{ErmPRC17} M. J. Ermamatov et al., Phys. Rev. C {\bf 96}, 044603 (2017).
\bibitem{ErmPRC16} M. J. Ermamatov, F. Cappuzzello, J. Lubian, M. Cubero, C. Agodi, D. Carbone et al., Phys. Rev. C {\bf 94}, 024610 (2016).
\bibitem{Pri2017} B. Pritychenko, M. Birch, B. Singh, and M. Horoi, Atom. Data Nucl. Data Tabl. {\bf 114}, 371 (2017).
\bibitem{Kib2002} T. Kibedi, R. H. Spear, Atom. Data Nucl. Data Tabl. {\bf 80}, 3582 (2002). 
\bibitem{BRINK197237} D. M. Brink, Phys. Lett. B {\bf 40}, 37-40 (1972). %Kinematical effects in heavy-ion reactions
\bibitem{KahANP1977} S. Kahana and A. J. Baltz, Adv. Nucl. Phys. {\bf 9}, 1 (1977). %One- and Two-Nucleon Transfer Reactions with Heavy Ions
\bibitem{NuShellX} W. D. M. Rae, http://www.garsington.eclipse.co.uk.
\bibitem{zbm} A. P. Zuker, B. Buck, and J. B. McGrory, Phys. Rev. Lett. {\bf 21}, 39 (1968).
\bibitem{psdmod} Y. Utsuno, S. Chiba, Phys. Rev. C {\bf 83}, 021301(R) (2011). %Multiparticle-multihole states around 16O and correlation-energy effect on the shell gap
\bibitem{cardozo} E. N. Cardozo et al., Phys. Rev. C {\bf 97}, 064611 (2018). %Competition between direct and sequential two-neutron transfers in the 18O+28Si collision at 84 MeV
\bibitem{coraggio2016} L. Coraggio, A. Gargano, N. Itaco, Phys. Rev. C {\bf 93}, 064328 (2016). %Double-step truncation procedure for large-scale shell-model calculations
\bibitem{ThomBCS2013} I. J. Thompson in {\it{Fifty Years of Nuclear BCS, Pairing in Finite Systems}}, edited by R. A. Broglia and V. Zelevinsky (World Scientific), 455–467 (2013).
\bibitem{spatafora} A. Spatafora et al., {\it{in preparation}}.
\end{thebibliography}

\section*{Acknowledgments} 
The authors warmly acknowledge the operators of INFN-LNS Accelerator Division for the production and delivery of the $^{20}$Ne beam and their support throughout the experiments. This project has received funding from the European Research Council (ERC) under the European Union's Horizon 2020 Research and Innovation Program (Grant Agreement No. 714625). S. Burrello and J. A. Lay acknowledge support from the Spanish Ministry of Economy and Competitiveness and the European Regional Development Fund (FEDER) under Project Nº FIS2017-88410-P and FIS2014-53448-C2-1-P. E. R. Ch\'{a}vez-Lomel\'{i} acknowledges DGAPA-UNAM IN107820, AG101120 and CONACyT 314857. J.R.B. Oliveira acknowledges "FAPESP proc. 2019/07767-1 and INCT-FNA, project n. 464898/2014-5), Brazil, by JRBO". DGAPA-UNAM IN107820, AG101120 and CONACyT 314857.

\end{document}